\documentclass[aps,prd,a4paper,twocolumn,superscriptaddress,nofootinbib]{revtex4-1}
\usepackage{ulem}
\usepackage{color}
\usepackage{graphicx}
\usepackage{hyperref}   
\usepackage{appendix}
\usepackage{epsfig}
\usepackage{epstopdf}
\usepackage{amsmath}
\usepackage{subfig}
\usepackage{comment}
\usepackage{bm}
\usepackage{slashed}
\usepackage[dvipsnames]{xcolor}
\hypersetup{pdftex,colorlinks=true,linkcolor=blue,citecolor=magenta,menucolor=black,urlcolor=blue,filecolor=blue}

\begin{document}

\title{Responses of quark-antiquark interactions and heavy quark dynamics to magnetic field } 

\author{He-Xia Zhang}
\email{hxzhang@m.scnu.edu.cn}

\affiliation{Key Laboratory of Atomic and Subatomic Structure and Quantum Control (MOE), Guangdong Basic Research Center of Excellence for Structure and Fundamental Interactions of Matter, Institute of Quantum Matter, South China Normal University, Guangzhou 510006, China}
\affiliation{Guangdong-Hong Kong Joint Laboratory of Quantum Matter, Guangdong Provincial Key Laboratory of Nuclear Science, Southern Nuclear Science Computing Center, South China Normal University, Guangzhou 510006, China}

\author{Enke Wang}
\affiliation{Key Laboratory of Atomic and Subatomic Structure and Quantum Control (MOE), Guangdong Basic Research Center of Excellence for Structure and Fundamental Interactions of Matter, Institute of Quantum Matter, South China Normal University, Guangzhou 510006, China}
\affiliation{Guangdong-Hong Kong Joint Laboratory of Quantum Matter, Guangdong Provincial Key Laboratory of Nuclear Science, Southern Nuclear Science Computing Center, South China Normal University, Guangzhou 510006, China}

\begin{abstract}
We investigate the impact of the magnetic field generated by colliding nuclei on heavy quark-antiquark interactions and heavy quark dynamics in the quark-gluon plasma (QGP).
By means of hard-thermal-loop resummation technique combined with dimension-two gluon condensates, the static heavy quark potential and heavy quark momentum diffusion coefficient, which incorporate both perturbative and non-perturbative interactions between heavy quarks and the QGP medium, are computed beyond the lowest Landau level approximation.
We find that the imaginary part of the heavy quark potential in the magnetic field exhibits significant anisotropy. Specifically, the absolute value of the imaginary part is larger when the quark-antiquark separation is aligned perpendicular to the magnetic field direction, compared to when it is aligned parallel to the  magnetic field direction.
The heavy quark momentum diffusion coefficient in the magnetized QGP medium also becomes anisotropic.
As the temperature rises, the influence of higher Landau levels becomes increasingly significant, resulting in a decrease in the anisotropy ratio of the heavy quark momentum diffusion coefficient to values even below 1. At sufficiently high temperatures, this ratio ultimately approaches 1.
The non-perturbative interactions are indispensable for understanding heavy quark dynamics in the low-temperature region. We also study the response of viscous quark matter to the magnetic field and explore its implications  for heavy quark potential, thermal decay widths of quarkonium states, as well as heavy quark momentum diffusion coefficient.
\end{abstract}

\maketitle

\section{Introduction}
In the past two decades, heavy-ion collision (HIC) experiments at the Relativistic Heavy Ion Collider (RHIC) and the Large Hadron Collider (LHC)  have provided convincing evidence to reveal that a deconfined state of quarks and gluons - quark-gluon plasma (QGP) described by quantum chromodynamics (QCD), can be generated at high temperatures~\cite{STAR:2005gfr, PHENIX:2004vcz, ALICE:2008ngc}. During these heavy-ion collisions, the strong transient magnetic field also can be generated in the direction perpendicular to the reaction plane due to the relativistic motion of the positively charged colliding heavy ions. The estimated value of the field strength in the primary stage ($<0.5$~fm)  can reach  $eB\sim m_{\pi}^2$ in Au + Au collisions at the RHIC energies and $eB\sim 15 m_{\pi}^2$ in Pb+Pb collisions at the LHC energies~\cite{magnetic, Bzdak:2011yy, Deng:2012pc, Skokov:2009qp, Kharzeev:2007jp, Voronyuk:2011jd}, where $m_\pi$ is pion mass. 
The existence of such intense magnetic fields in HICs opens a new frontier of high-energy physics and induces novel quantum transport phenomena like chiral magnetic effect~\cite{Voronyuk:2011jd, Fukushima:2008xe, Alver:2010gr}.

Heavy quarks, which are produced from the early stage of HICs via hard scattering processes and difficult to thermalize, can offer valuable insights into the properties of the medium they cross~\cite{Moore:2004tg, Rapp:2009my}. Since the formation times of heavy quarks are comparable to the time scale of the maximum magnetic field, heavy quarks propagating through the QCD medium are profoundly influenced by the magnetic field, which will be  inherited by the final heavy-flavor hadrons. The difference in the directed flow $v_1$ between open charm mesons $D^0$ and $\bar{D}^0$,  arising from the competing Faraday and Hall effects caused by the decreasing magnetic field, can serve as a direct probe of the initial electromagnetic (EM) field created in HICs. 
Theoretical predictions based on the Langevin transport equation within the relativistic hydrodynamics~\cite{Das:2016cwd, Chatterjee:2018lsx} have indicated that the $v_1$ of open charm mesons is larger than that of charged light hadrons, and $\Delta v_1=v_1(D^0)-v_1(\bar{D}^0)$ is nonzero, which is also confirmed by the STAR experimental results at RHIC~\cite{STAR:2019clv}. However, at LHC energies, the Langevin transport model coupled to the hydrodynamics model~\cite{Das:2016cwd} with a constant electrical conductivity extracted from lattice QCD calculation~\cite{Ding:2010ga, Amato:2013naa} predicts a qualitative behavior of $\Delta v_1$  opposite to the experimental result~\cite{ALICE:2019sgg}. 
Recently, the authors of Ref.~\cite{Jiang:2022uoe} have  adopted the EM field evolution model from Ref.~\cite{Sun:2020wkg} instead of the direct solution of the Maxwell equation with a constant electric conductivity to enhance the effect of Lorentz force relative to Coulomb force, thereby obtaining qualitative results  consistent with the experimental results at LHC. Even so, there are many aspects to improve the theoretical calculation. On the one hand, the (viscous) hydrodynamic model coupled with the dynamical electromagnetic fields, such as the relativistic magnetohydrodynamics (RMHD)~\cite{RHMD1, RHMD2, RHMD3, RHMD4}, is anticipated to provide a proper description of the dynamical evolution of the created magnetized QCD matter. So far, there still  remain many challenges to be addressed within the RMHD approach used in relativistic heavy-ion collisions (see~\cite{Hattori:2022hyo} for details). On the other hand, the HQ momentum diffusion coefficient as a vital input parameter in the Langevin transport model, is only affected by the thermal random force, while the EM field effects are treated as external forces in the transport equation. In principle, the HQ momentum diffusion coefficient is also influenced by the EM field through interacting with magnetized light (anti-)quarks, investigating the response of heavy quark dynamics to the magnetic field could offer insight into the transport properties of the QGP.
In the presence of a magnetic field oriented along $z$-axis ($\bm{B}=B\hat{z}$), the dispersion relation for light (anti-)quarks due to the Landau level quantization is obtained as~\cite{Gusynin:1995nb,Akhiezer} 
$E_{k,n}^f$=$\sqrt{k_z^2+m_{f,n}^2}$ with  $ m_{f,n}=\sqrt{m_f^{2}+2n|q_feB|}$, where the index $n=0,1,2,..$ label the quantum number of the Landau levels; $q_fe$ is electric charge of $f$-th flavor quark; $m_f$ is current mass of $f$-th flavor (anti-)quark. 
The estimation of HQ momentum diffusion coefficients in the lowest Landau level (LLL) approximation has been carried out based on hard-thermal-loop (HTL) perturbative theory in the static limit~\cite{Fukushima:2015wck} and beyond the static limit~\cite{Bandyopadhyay:2021zlm}. In Refs.~\cite{Kurian:2020kct, Kurian:2019nna}, the scattering rate of heavy quarks with LLL massless quarks derived in Ref.~\cite{Fukushima:2015wck}, was utilized to explore the effect of higher Landau level (HLL) on HQ dynamics.  Actually, the heavy quark scattering rate with thermal HLL quarks differs significantly from that with  LLL massless quarks. Therefore, more careful calculations of the HQ momentum diffusion coefficient beyond the LLL approximation are necessary for obtaining a more robust qualitative result. Very Recently,  HQ momentum diffusion coefficient has also been estimated for arbitrary Landau level beyond the static limit~\cite{Bandyopadhyay:2023hiv}.

In addition to HQ dynamics, another particular interest regarding the physics of heavy flavor is the heavy quark-antiquark potential, which is the starting point for the non-relativistic approaches to study the in-medium static properties of heavy quarkonia. The real (imaginary) part of in-medium HQ potential determines the binding energy (thermal decay width) of quarkonium states. In Refs.~\cite{Bonati:2016kxj, Bonati:2018uwh}, C. Bonati $et~ al$. utilized the lattice QCD method to investigate the impact of magnetic field on the real part of HQ potential, $\mathrm{Re}V$, in the vacuum and below the pseudocritical temperature. To our best knowledge, there are no lattice QCD studies for the imaginary part of the HQ potential, $\mathrm{Im}V$, in the magnetic field. 
On the other hand, phenomenological studies based on potential models~\cite{Hasan:2020iwa, Khan:2021syq, Singh:2017nfa} have attempted to derive stable qualitative properties of heavy quarkonia in the magnetic fields. In Ref.~\cite{Singh:2017nfa}, the in-medium HQ potential, accommodating both short-range Yukawa (perturbative) interaction and large-range string-like (non-perturbative) interaction, was derived using the dielectric permittivity  that encodes the effects of the deconfined medium in the LLL approximation. Here, the gluon self-energy was obtained using the imaginary-time formalism. The results in Ref.~\cite{Singh:2017nfa} indicated that both the $\mathrm{Re}V$ and the $\mathrm{Im}V$  in  a strong magnetic field exhibit isotropy. Recently, R. Ghosh  $et ~al$. have also explored the $\mathrm{Im }V$ at finite temperature and magnetic field utilizing dielectric permittivity~\cite{Ghosh:2022sxi}, where all Landau level summation has been considered in the gluon self-energy calculation with imaginary-time formalism. The associated results in Ref.~\cite{Ghosh:2022sxi}  showed that the magnetic field leads to anisotropy in the $\mathrm{Im}V$.  Nevertheless, it is noteworthy that in the literature, the contribution of the LLL massless quark-loop to the gluon self-energy was usually overlooked, which might result in incomplete expressions for the imaginary part of the effective gluon propagator and the subsequent $\mathrm{Im}V$.

Furthermore, current computations of HQ potential and HQ momentum diffusion coefficient are mostly performed using the assumption that the collision partners from the QGP medium are in thermal equilibrium. For a more realistic perspective, the QGP medium created in HICs behaves like a fluid with viscosity. Due to the existence of a magnetic field breaking the spatial rotational symmetry, the viscous coefficients of the QGP medium are no longer isotropic, giving rise to anisotropic transport coefficients~\cite{Hattori:2022hyo}. 
As a result of Landau level quantization, the motions of light (anti-)quarks are primarily restricted to the longitudinal direction. In the strong magnetic field limit ($T^2\ll eB $) with the LLL approximation, the viscous effect from quark matter is dominant, and the light (anti-)quarks only contribute to the longitudinal pressure, which leads to a non-zero longitudinal bulk viscosity $\zeta_{\|}$~\cite{Hattori:2017qih}. 
The theoretical calculations of $\zeta_{\|}$ in both the LLL and HLL approximations have indicated that the ratio of longitudinal bulk viscosity to entropy density $\zeta_{\|}/s$ in the magnetic field is much larger than that in the zero magnetic fields~\cite{Hattori:2017qih, Kurian:2018qwb, Rath:2020beo,Ghosh:2020wqx}. Therefore, it is crucial to investigate the impact of these non-equilibrium effects on the HQ potential and HQ transport coefficient to gain a deeper understanding of the properties of the out-of-equilibrium QGP medium.
 
In the present work, the numerical calculations are performed based on weak coupling perturbative QCD theory in the hierarchies of scale $(\alpha_s T^2,~\alpha_s eB )\ll (T^2,~eB)$ with $\alpha_s$ being strong coupling constant. In the one-loop calculation, the typical momenta of thermal quarks and gluons, comprising the internal lines of the gluon self-energy diagrams, are hard scales ($\sim T$ for gluon-loop, $\sim T, \sqrt{eB}$ for quark-loop). The momentum transfers for quarks and gluons via an exchanged gluon, comprising the external lines of the gluon self-energy diagrams, are soft scales  ($\sim\alpha_s eB$ and $\sim\alpha_s T^2$). The inequalities  $(\alpha_s T^2,~\alpha_s eB) \ll T^2$ justify the application of the hard-thermal-loop technique~\cite{Fukushima:2017lvb,Bandyopadhyay:2021zlm,Hattori:2017xoo}. We consider that the magnetic field can be comparable to the square of temperature,  the HLL effects become non-negligible~\cite{Fukushima:2017lvb, Kurian:2020kct}. We first utilize a widely used HQ potential model which is defined through the Fourier transform of the static effective gluon propagator, to investigate the influence of magnetic field on both static HQ potential and  thermal decay widths of quarkonium states in the QGP. Such an effective gluon propagator incorporates both perturbative QCD effects and non-perturbative effects from dimension-two gluon condensates~\cite{Guo:2018vwy, Megias:2007pq, Megias:2005ve}. Both the LLL quark-loop and HLL quark-loop contributions to the gluon self-energy are carefully considered in the effective gluon propagator. Using the imaginary part of the effective gluon propagator mentioned above, we also compute the magnetic field-dependent scattering rate between heavy quark and thermal partons.  Subsequently,  we calculate the HQ momentum diffusion coefficients beyond the LLL approximation. The calculation of the HQ momentum diffusion coefficient is performed in the static limit.  
Furthermore, the response of hot and viscous quark matter to the magnetic field is studied within the kinetic theory under the relaxation time approximation (RTA). 
By employing the derived longitudinal viscous modified distribution function of thermal quarks, we extend the HTL perturbation theory to such a non-equilibrium scenario, aiming to further explore the longitudinal viscous correction on HQ potential, thermal decay width of quarkonium states, and HQ momentum diffusion coefficient.

The layout of this work is as follows. In Sec.~\ref{sec:gluon_self-energy}, we provide a detailed derivation of the gluon self-energy at the magnetic field in the real-time thermal field theory, utilizing the  HTL resummation technique.  The  HTL resummed effective gluon propagator is evaluated in Sec.~\ref{sec:propagator}. In Sec.~\ref{sec:gluon_self-energy_bulk}, we derive the longitudinal bulk viscous modified distribution function of light (anti-)quarks, and subsequently extend the HTL resummation technique to the magnetized viscous QGP. Besides the usual perturbative resummed gluon propagator, a phenomenological gluon propagator that includes the non-perturbative effects from dimension-two gluon condensates, is introduced in Sec.~\ref{sec:HQ_potential}. The HQ potential is constructed by the Fourier transform of the effective gluon propagator in the static limit. Based on the gluon propagators derived in Sec.~\ref{sec:propagator} and Sec.~\ref{sec:gluon_self-energy_bulk},  we present the formulae for anisotropic HQ momentum diffusion coefficients in Sec.~\ref{sec:HQ_diffusion}. In Sec.~\ref{sec:discussions}, we discuss the impacts of temperature, magnetic field, the HLL effect as well as longitudinal bulk viscous correction on the HQ potential, quarkonium state’s thermal decay width, and HQ momentum diffusion coefficients. We summarize our findings in Sec.~\ref{sec:summary}. 

\section{Gluon self-energy in magnetized QGP medium}\label{sec:gluon_self-energy}
To estimate the  HQ potential and leading order HQ scattering rate with thermal partons from the magnetized QGP medium, it is imperative to evaluate the gluon self-energy.
Throughout this work, we employ the real-time formalism of thermal field theory, which is more appropriate when dealing with a non-equilibrium situation and can immediately split the computations of gluon self-energy into a zero-temperature part and a temperature-dependent part, by virtue of the split in real-time propagators. Unless otherwise stated, all calculations are for massless QCD at zero chemical potential. In the Landau level representation, the free quark propagators for the $f$-th flavor can be represented by 2 $\times$ 2 matrices~\cite{Mallik:2009pj,Miransky:2015ava,Kobes:1984vb,Rath:2017fdv}
 \begin{align}\label{eq:Sij}
 S^f
 (k,m_{f,n})=&ie^{-\frac{\bm{k}_{\perp}^2}{|q_f eB|}}\sum_{n=0}^{\infty}(-1)^nD_{n}^f(k)\nonumber\\
&\times \bigg[
 \begin{pmatrix}
 \frac{1}{k_{\|}^2-m_{f,n}^2+i\epsilon} & 0\\
 0 & \frac{-1}{k_{\|}^2-m_{f,n}^2-i\epsilon}
 \end{pmatrix}\nonumber\\
 &+i2\pi \delta({k}_{\|}^2-m_{f,n}^2)\nonumber\\
 &\times\begin{pmatrix}
 X & X-\Theta(-k_0)\\	
 X-\Theta(k_0) &  X
 \end{pmatrix}
 \bigg].
 \end{align}
The metric is defined as $g^{\mu\nu}=g_{\|}^{\mu\nu}+g_{\perp}^{\mu\nu}$, where $g_{\|}^{\mu\nu}=\mathrm{diag}(1,0,0,-1)$ and $g_{\perp}^{\mu\nu}=\mathrm{diag}(0,-1,-1,0)$ are the projectors onto the longitudinal subspace and transverse (with respect to the direction of magnetic field) subspace, respectively. For four-vectors $k^\mu$, its associated notations can be written as $k_{\|}^\mu=(k_0,0,0,k_z)$ and $k_{\perp}^\mu=(0,k_x,k_y,0)$ with $k^2=k_{\|}^2-\bm{k}_{\perp}^2 $, where $k_{\|}^2=k_0^2-k_z^2$ and $\bm{k}_{\perp}^2=k_x^2+k_y^2$. 
The abbreviations $X$ and $Y$ in Eq.~(\ref{eq:Sij}) are respectively given by $X=\Theta(k_0)f_{+}(k_0)+\Theta(-k_0)f_{-}(-k_0)$, where $\Theta(k_0)$ is the Heaviside function. In global equilibrium, $f_{\pm}(x)=f^0_{\pm}(x)=[e^{(|x|\pm\mu)/T}+1]^{-1}$ represents the Fermi-Dirac thermal distribution function, where the subscript $``\pm"$ corresponds to anti-fermions and fermions, separately; $\mu$ is quark chemical potential. The $D_{n}^f(k)$ in Eq.~(\ref{eq:Sij}) can be expressed as
\begin{align}
D_n^f(k)=&2(\slashed{k}_{\|}+m_{f})\big[\mathcal{P}^f_+L_n\bigg(\frac{2\bm{k}_{\perp}^2}{|q_f eB|}\bigg)\nonumber\\
&-\mathcal{P}^f_-L_{n-1}\bigg(\frac{2\bm{k}_{\perp}^2}{|q_f eB|}\bigg)\big]\nonumber\\
&+4\slashed{\bm{k}}_{\perp}L_{n-1}^{1}\bigg(\frac{2\bm{k}_{\perp}^2}{|q_f eB|}\bigg),
\end{align}
with $\mathcal{P}_{\pm}^f=[1\pm i\gamma^x\gamma^y\mathrm{sgn}(q_feB)]/2$ being the projection operator into a state with spin aligning with magnetic field direction. Here, $L_n^{m}(x)$ are the generalized Laguerre polynomials, $L_{n}(x)=L^0_{n}(x) $ and $L_{-1}(x)=0$. Now, we convert to a more convenient representation, the Keldysh representation. 
The retarded, advanced, and symmetric propagators can be obtained from the Keldysh representation (which satisfies $S_{11}-S_{12}-S_{21}+S_{22}=0$) via~\cite{Keldysh:1964ud,Chou:1984es}
\begin{eqnarray}\label{eq:SRAF}
S_{R}=S_{11}-S_{12}, 
S_{A}=S_{11}-S_{21},
S_{F}=S_{11}+S_{22}.
\end{eqnarray}
Similarly, the retarded ($R$), advanced ($A$), and symmetric ($F$) gluon self-energies are given as  
\begin{equation}\label{eq:Pi_{RAF}}
\Pi_R=\Pi_{11}+\Pi_{12},
\Pi_A=\Pi_{11}+\Pi_{21},
\Pi_F=\Pi_{11}+\Pi_{22},
\end{equation} 
and hold $\Pi_{11}+\Pi_{12}+\Pi_{21}+\Pi_{22}=0$.
Furthermore, the symmetric gluon self-energy $\Pi_F$  in an equilibrium QGP within the massless limit also can be obtained through the Kubo-Martin-Schwinger (KMS) condition~\cite{KMS1,KMS2}
\begin{eqnarray}\label{eq:KMS}
\Pi_F(k)=(1+2f^0_B(k_0))\mathrm{sgn}(k_0)(\Pi_R(k)-\Pi_A(k)),
\end{eqnarray}
where $f^0_B(k_0)$ is the Bose-Einstein distribution function. The validity of Eq.~(\ref{eq:KMS}) in the magnetized viscous QGP medium needs further verification. 

\begin{figure}
\subfloat{\includegraphics[scale=0.7]{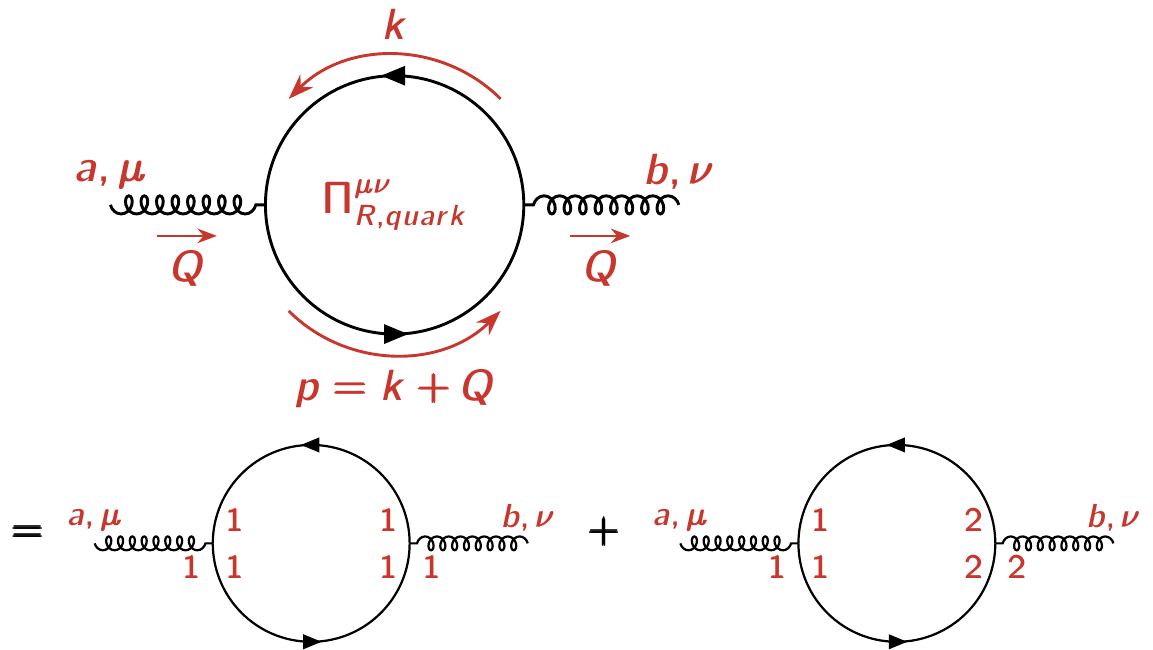}}\hspace{.1 cm}
	\caption{The one-loop diagram for quark contribution to the retarded gluon self-energy $\Pi_{R,\rm quark}^{\mu\nu}$.
 The solid (curly) line represents a quark
(gluon) propagator, where the quark can occupy different Landau levels. $k$ and $p=k+Q$ are the loop momenta. $Q=(q_0, {\bm q})$ is the external momentum and denotes the four-momentum transfer. The below row corresponds to the retarded  gluon self-energy, with indices ``1" and ``2" being two types of vertices. 
$\mu,\nu$ are Lorentz indices, $a,b$ are color indices.
 }
	\label{fig_Feynman_plot}
\end{figure} 
\subsection{Quark-loop contribution to retarded gluon self-energy}
According to the relations in Eq.~(\ref{eq:Pi_{RAF}}) and Fig.~\ref{fig_Feynman_plot}, the one quark-loop contribution to the retarded gluon self-energy tensor in a magnetic field using the standard Feynman rules can be  read off
	\begin{align}\label{eq:Pimunu}
	\Pi^{\mu\nu}_{R,\rm quark}(Q)=&\Pi_{11,\rm quark}^{\mu\nu}(Q)+\Pi_{12,\rm quark}^{\mu\nu}(Q),\nonumber\\
 =&-ig_s^2
	\int\frac{d^4k}{(2\pi)^4}\nonumber\\
 &\times \mathrm{Tr}[t_b\gamma^\mu S_{11}^f(k,m_{f,n})t_a\gamma^\nu S_{11}^f(p,m_{f,l})\nonumber\\
	&-t_b\gamma^\mu S^f_{21}(k,m_{f,n})t_a\gamma^\nu S_{12}^f(p,m_{f,l})],
	\end{align}
where the minus sign in the square bracket originates from the vertex of type-2 field~\cite{Landsman:1986uw}. 
$\mathrm{Tr}(t_at_b)=\frac{\delta_{ab}}{2}$, where $t_{a,b}$ refer to the generators of color group. 
In order to  simplify Eq.~(\ref{eq:Pimunu}), the transverse part can be written in the following form: 
	 \begin{align}\label{eq:Lmunu}
	 L^{\mu\nu}&=\int\frac{d^2\bm{k}_{\perp}}{(2\pi)^2}(-1)^{n+l}e^{\frac{-\bm{k}_{\perp}^2-\bm{p}_{\perp}^2}{|q_f eB|}}\mathrm{Tr}[\gamma^\mu D_{n}^f(k)\gamma^\nu D_{l}^f(p)]\nonumber\\
	 &=\int\frac{d^2\bm{k}_{\perp}}{(2\pi)^2}e^{\frac{-\bm{k}_{\perp}^2-\bm{p}_{\perp}^2}{|q_f eB|}}(-1)^{n+l}\bigg\{8(k_{\|}^\mu p_{\|}^\nu+k_{\|}^\nu p_{\|}^\mu\nonumber\\
	 &-g_{\|}^{\mu\nu}(k_{\|}\cdot p_{\|}-m_{f}^2))\bigg[L_{n}\bigg(\frac{2\bm{k}_{\perp}^2}{|q_feB|}\bigg)L_{l}\bigg(\frac{2\bm{p}_{\perp}^2}{|q_feB|}\bigg)\nonumber\\
	 &+L_{n-1}\bigg(\frac{2\bm{k}_{\perp}^2}{|q_feB|}\bigg)L_{l-1}\bigg(\frac{2\bm{p}_{\perp}^2}{|q_feB|}\bigg)\bigg]\nonumber\\
	 &+64({k}_{\perp}^\mu {p}_{\perp}^\nu+{k}_{\perp}^\nu p_{\perp}^\mu- (k_{\perp}\cdot p_{\perp}) g_{}^{\mu\nu})\nonumber\\
	&\times L_{l-1}^1\bigg(\frac{2\bm{k}_{\perp}^2}{|q_feB|}\bigg)
	 L_{n-1}^1\bigg(\frac{2\bm{p}_{\perp}^2}{|q_feB|}\bigg)\nonumber\\
	& +8g_{\perp}^{\mu\nu}(k_{\|}\cdot p_{\|}-m_{f}^2)\bigg[L_{n}\bigg(\frac{2\bm{k}_{\perp}^2}{|q_feB|}\bigg)L_{l-1}\bigg(\frac{2\bm{p}_{\perp}^2}{|q_feB|}\bigg)\nonumber\\
	&+L_{n-1}\bigg(\frac{2\bm{k}_{\perp}^2}{|q_feB|}\bigg)L_{l}\bigg(\frac{2\bm{p}_{\perp}^2}{|q_feB|}\bigg)\bigg]\nonumber\\
&	+16 (k_{\|}^\mu p_{\perp}^\nu+k_{\|}^\nu p_{\perp}^\mu)\bigg[L_{n}\bigg(\frac{2\bm{k}_{\perp}^2}{|q_feB|}\bigg)L_{l-1}^1\bigg(\frac{2\bm{p}_{\perp}^2}{|q_feB|}\bigg)\nonumber\\
&-L_{n-1}\bigg(\frac{2\bm{k}_{\perp}^2}{|q_feB|}\bigg)L_{l-1}^1\bigg(\frac{2\bm{p}_{\perp}^2}{|q_feB|}\bigg)\bigg]\nonumber\\
	 &+16(p_{\|}^\mu k_{\perp}^\nu+p_{\|}^\nu k_{\perp}^\mu)\bigg[L_{l}\bigg(\frac{2\bm{p}_{\perp}^2}{|q_feB|}\bigg)L_{n-1}^1\bigg(\frac{2\bm{k}_{\perp}^2}{|q_feB|}\bigg)\nonumber\\
	& -L_{l-1}\bigg(\frac{2\bm{p}_{\perp}^2}{|q_feB|}\bigg)L_{n-1}^1\bigg(\frac{2\bm{k}_{\perp}^2}{|q_feB|}\bigg)\bigg]
	 \bigg\},
	 \end{align}
 with $({k}_{\perp}\cdot {p}_{\perp})=k_xp_x+k_yp_y$. By utilizing the integral identities listed in the Appendix, the tensor $L^{\mu\nu}$ can be further simplified as
    $L^{\mu\nu}=(-1)^{n+l}\exp(\frac{-\bm{q}_{\perp}^2}{2|q_feB|})\frac{|q_fe B|}{\pi}
	\bigg[4|q_feB|n\delta_{l-1}^{n-1} g_{\|}^{\mu\nu}+(\delta^n_{l}+\delta^{n-1}_{l-1})[k_{\|}^\mu p_{\|}^\nu+k_{\|}^\nu p_{\|}^\mu-g_{\|}^{\mu\nu}(k_{\|}\cdot p_{\|}-m_{f}^2)]-(\delta_{l}^{n-1}+\delta^n_{l-1})(k_{\|}\cdot p_{\|}-m_{f}^2)g_{\perp}^{\mu\nu}\bigg]$, where $\bm{q}^2_{\perp}$ denotes the squared momentum transfer transverse to the magnetic field direction, and $\exp(\frac{-\bm{q}_{\perp}^2}{2|q_feB|})$ represents the asymmetry factor. 
Since we are working in the soft momentum transfer limit with the hierarchies of scale $m_f^2\ll \alpha_sT^2, \alpha_s eB\ll eB$, the perturbative QCD interactions cannot induce the Landau level transition for light (anti-)quarks because they don't have enough energy to jump across the energy gap separating the Landau levels that is proportional to $ \sqrt {eB}$. Take $n=l$, 
then $L^{\mu\nu}$ is reduced as $L^{\mu\nu}=\exp(\frac{-\bm{q}_{\perp}^2}{2|q_feB|})\frac{q_feB}{\pi}\alpha_{0n}[2n|q_feB|g_{\|}^{\mu\nu}+(k_{\|}^\mu p_{\|}^\nu+k_{\|}^\nu p_{\|}^\mu)-g_{\|}^{\mu\nu}(k_{\|}\cdot p_{\|}-m_{f}^2)]$, with $\alpha_{0n}=(2-\delta_{0,n})$ being the Landau level-dependent spin degeneracy. It is effective to obtain the retarded gluon self-energy from the quark-loop by calculating only the physical ``11"-component.  This component can be written as
\begin{align}\label{eq:general_tensor}
	\Pi_{11,\rm quark}^{\mu\nu}
=& i\sum_{f}\sum_{n=0}^{\infty}\frac{g_s^2}{2}\int\frac{d^2k_\|}{(2\pi)^2}L^{\mu\nu}\biggl\{\bigg[\frac{1}{(k_{\|}^2-m_{f,n}^2+i\epsilon )}\nonumber\\
	&+2\pi i\bigg(\Theta(k_0)f_+(k_0)+\Theta(-k_0)f_{-}(-k_0)\bigg)\nonumber\\
	&\times\delta(k_{\|}^2-m_{f,n}^2)\bigg]\times\bigg[\frac{1}{(p_{\|}^2-m_{f,n}^2+i\epsilon )}\nonumber\\
	&+2\pi i\bigg(\Theta(p_0)f_+(p_0)+\Theta(-p_0)f_{-}(-p_0)\bigg)\nonumber\\
&\times	\delta(p_{\|}^2-m_{f,n}^2)\bigg]
	\biggr\}.
	\end{align}
The strong coupling strength $g_s$ in this work is obtained from~\cite{Ayala:2018wux}
\begin{eqnarray}
\alpha_s(\Lambda^2,eB)=\frac{g_s^2}{4\pi}=\frac{\alpha_s(\Lambda^2)}{1+\frac{11N_c-2N_f}{12\pi}\alpha_s(\Lambda^2)\ln (\frac{\Lambda^2}{\Lambda^2+eB})},
\end{eqnarray}
where the strong coupling constant $\alpha_s$ for vanishing magnetic field is given by $\alpha_s(\Lambda^2)=\frac{12\pi}{11N_c-2N_f}\ln\frac{\Lambda_{\bar{MS}}^2}{\Lambda^2}$ at $\Lambda_{\bar{MS}}=176~\mathrm{MeV}$ for $N_f=N_c=3$~\cite{Bazavov:2012ka}, and the scale $\Lambda$ is chosen as $\Lambda=2\pi T$. 

\subsubsection*{A-1.~ Vacuum LLL quark-loop contribution to retarded  gluon self-energy }
In the magnetic field, the vacuum contribution to gluon self-energy should be maintained even at high temperatures, unlike in the absence of a magnetic field.
In Eq.~(\ref{eq:general_tensor}), the term exclusively related to vacuum quark propagators is given as $\Pi^{\mu\nu}_{11,\mathrm{quark}, T=0}$, which can be written as 
\begin{align}\label{eq:tensor_vac}
\Pi^{\mu\nu}_{11,\mathrm{quark}, T=0}=&\sum_{f}\sum_{n=0}^{\infty}\frac{ig_s^2|q_f eB|}{2\pi}\alpha_{0n}\int \frac{d^2k_\|}{(2\pi)^2} e^ {-\frac{\bm{q}_{\perp}^2}{2|q_feB|}}\nonumber\\
&\times\frac{k_{\|}^\mu p_{\|}^\nu+k_{\|}^\nu p_{\|}^\mu-g_{\|}^{\mu\nu}(k_{\|}\cdot p_{\|}-m_{f,n}^2)}{(k_{\|}^2-m_{f,n}^2+i\epsilon)(p_{\|}^2-m_{f,n}^2+i\epsilon)}.
\end{align}
Using the Feynman's parameterization (see details in Refs.~\cite{Ayala:2018ina,Hattori:2022uzp}), Eq.~(\ref{eq:tensor_vac}) can be rewritten as
\begin{align}\label{eq:Vac}
\Pi^{\mu\nu}_{11,\mathrm{quark},T=0}=&g_s^2\sum_{f}\sum_{n=0}^{\infty}\frac{\alpha_{0n}|q_fe B|}{4\pi^2}
\bigg[\frac{q_{\|}^2g_{\|}^{\mu\nu}-q_{\|}^\mu q_{\|}^\nu}{(q_{0}+i\epsilon)^2-q_z^2}\bigg]\nonumber\\
&\times e^{-\frac{\bm{q}_{\perp}^2}{2|q_feB|}}\underbrace{\int_{0}^{1} dx\frac{x y_{\|}(1-x)}{-1+x(1-x) y_{\|}}}_{I(y_\|)},
\end{align}
with $y_{\|}=q_{\|}^2/m_{f,n}^2$. We can see the light quarks in a magnetic field at zero temperature don't couple to the transverse subspace spanned by $g_{\perp}^{\mu\nu}$ and $q_{\perp}^{\mu}$. 
In the space-like region $q_{\|}^2\leq 0$, the function $I(y_{\|})$ is given as~\cite{Fukushima:2011nu,Hattori:2022uzp},
\begin{eqnarray}\label{eq:I}
    I(y_{\|})=1-\frac{4\arctan(\sqrt{y_{\|}/(4-y_\|)})}{\sqrt{y_{\|}/(4-y_\|)}},
\end{eqnarray}
where the first term corresponds to the massless Schwinger model~\cite{Schwinger:1962tn} and the second term accounts for mass correction. 
The function $I(y_{\|})$ has two limiting behaviors~\cite{Fukushima:2011nu,Hattori:2022uzp}, i.e.,
\begin{eqnarray}\label{eq:I_2}
I(0)=0,~I(\infty)=1.
\end{eqnarray}
When the light quarks occupy the LLL ($n=0$), $y_{\|}\to \infty$,  it results in  $I(y_{\|})=1$.
Note the mass correction ($m_f\neq 0$) in Eq.~(\ref{eq:I}) can safely be neglected due to the HTL approximation with $m_f^2\ll |q_{\|}^2| \ll eB$.
To compute the HQ potential, we need the temporal (``00") component of the gluon self-energy. Specializing the Lorentz indices as $\mu=\nu=0$, the vacuum LLL quark-loop contribution to the real part of the retarded gluon self-energy is computed as 
\begin{align}
 &\mathrm{Re} \Pi^{00}_{R,\mathrm{quark},T=0,n=0} =\mathrm{Re}\Pi^{00}_{11,\mathrm{quark},T=0,n=0}=-\frac{q_z^2}{q_{\|}^2}s_f(\bm{q}_{\perp}),
\label{eq:RePiT=0}
\end{align}
where the abbreviation $s_f(\bm{q}_{\perp})=\alpha_s\frac{|q_feB|}{\pi}\exp(-\frac{\bm{q}_{\perp}^2}{2|q_feB|})$ is used for the reader's convenience. Subsequently, the imaginary part of the retarded gluon self-energy can be obtained by utilizing the relation 
$\mathrm{Im}\Pi^{00}_{R}=\tanh|\frac{q_0}{2T}|\mathrm{Im}\Pi_{11}^{00}$~\cite{Baier:1991gg, Kobes:1985kc, Fujimoto:1985me}. Thus, the vacuum LLL quark-loop contribution to the imaginary part of the retarded gluon self-energy is computed as 
\begin{align}
&\mathrm{Im}\Pi^{00}_{R,\mathrm{quark},T=0,n=0}=\mathrm{Im}\Pi^{00}_{11,\mathrm{quark},T=0,n=0}\nonumber\\
&=\sum_{f}\frac{\pi q_0}{2}s_f(\bm{q}_{\perp})\bigg[\delta(q_0+q_z)+\delta(q_0-q_z)\bigg]. \label{eq:ImPiT=0}
\end{align}
Here, Eq.~(\ref{eq:RePiT=0}) and Eq.~(\ref{eq:ImPiT=0}) are consistent with the results in Ref.~\cite{Fukushima:2015wck}.  
Notice that the delta functions in Eq.~(\ref{eq:ImPiT=0}) are related to the chiral dispersion of LLL with $\delta(q_0\pm q_z)$ corresponding to right-handed and left-handed chiral fermions or massless quarks. These LLL chiral fermions have the linear dispersion relations, $E_{LLL}(k_z)=\pm k_z$, where the sign $``+"$ and the sign $``-"$ correspond to right-handed chirality and left-handed chirality, respectively. 
Since the chirality of chiral fermions cannot be flipped in the presence of perturbative QCD interaction, the four-momentum transfer from the LLL chiral fermions is given as  $(\Delta E,~\Delta k_z)=(q_0,~q_z)$, and must satisfy $q_0=\pm q_z$. In the static limit $q_0\to 0$, the longitudinal momentum transfer $q_z$ is kinematically prohibited and there is no scattering along the magnetic field direction for LLL massless quarks.

\subsubsection*{A-2.~ Vacuum HLL quark-loop contribution to retarded gluon self-energy}
When light quarks occupy the higher Landau levels ($n\geq 1$),  $y_{\|}\approx0$
due to the condition $m_f^2\ll|q_{\|}^2|\ll m_{f,n}^2\sim n|q_f eB|$. Consequently,  in Eq.~({\ref{eq:I}}), $I(y_{\|})=0$,  indicating that quark excitations are suppressed in the vacuum. Therefore, the vacuum HLL quark-loop does not contribute to the retarded gluon self-energy, i.e.,
\begin{align}
\mathrm{Re} \Pi^{00}_{R,\mathrm{quark},T=0,n\geq 1}&=0,\\
\mathrm{Im} \Pi^{00}_{R,\mathrm{quark},T=0,n\geq 1}&=0.
\end{align}

\subsubsection*{A-3.~Medium LLL quark-loop contribution to retarded  gluon self-energy}
Next, we proceed with the computation of the terms related to temperature-dependent quark propagators in Eq.~(\ref{eq:general_tensor}). Using  $\delta^{-1}_{-1}=0$, we can express the temporal component of  $L^{\mu\nu}$ as $L^{00}=\alpha_{0n}(k_0p_0+k_zp_z+m_f^2+2n|q_feB|)\exp(\frac{-\bm{q}_{\perp}^2}{2|q_feB|})\frac{|q_fe B|}{\pi}$. Subsequently, the real part of $\Pi_{11,\rm quark}^{00}$ at finite temperature and magnetic field in equilibrium can be derived as
\begin{align}\label{eq:RePi}
\mathrm{Re}\Pi_{11,\mathrm{quark},T\neq 0}^{00}=&-
\sum_{f}\sum_{n=0}^{\infty}\frac{g_s^2}{2}\int\frac{dk_z}{2\pi}
\nonumber\\
&\times\bigg[\frac{f_{+}^0(k_0)L^{00}(k_0=E^f_{p,n}-q_0)}{2E^f_{p,n}[(-q_0+E^f_{p,n})^2-(E^f_{k,n})^2]}\nonumber\\
&+\frac{f_{-}^0(-k_0)L^{00}(k_0=-E^f_{p,n}-q_0)}{2E^f_{p,n}[(-q_0-E^f_{p,n})^2-(E^f_{k,n})^2]}\nonumber\\
&+\frac{f_{+}^0(k_0)L^{00}(k_0=E^f_{k,n})}{2E^f_{k,n}[(q_0+E^f_{k,n})^2-(E^f_{p,n})^2]}\nonumber\\
&+\frac{f^0_{-}(-k_0)L^{00}(k_0=-E^f_{k,n})}{2E^f_{k,n}[(q_0-E^f_{k,n})^2-(E^f_{p,n})^2]}\bigg].
\end{align}
Here,  $L^{00}(k_0=\pm E^f_{k,n})=[2(E^f_{k,n})^2\pm q_0 E^f_{k,n}+q_zk_z]\exp(\frac{-\bm{q}_{\perp}^2}{2|q_feB|})\frac{|q_fe B|}{\pi}$ and $L^{00}(k_0=-q_0\pm E^f_{p,n})=[(E^f_{p,n})^2\mp q_0 E^f_{p,n}+(E^f_{k,n})^2+k_zq_z]\exp(\frac{-\bm{q}_{\perp}^2}{2|q_feB|})\frac{|q_fe B|}{\pi}$.
For vanishing chemical potential, $f^0_-(k_0)=f^0_+(k_0)$, and putting the first two (last two) integrands in the right-hand side of Eq.~(\ref{eq:RePi}) over a common denominator, which is given by
\begin{eqnarray}
&&(q_{\|}^2-2q_0E^f_{k,n}-2k_zq_z)(q_{\|}^2+2q_0E_{k,n}^f-2k_zq_z)=\nonumber\\
&&(q_{0}^2-2q_0E^f_{p,n}+2k_zq_z+q_z^2)(q_{0}^2+2q_0E^f_{p,n}+2k_zq_z+q_z^2)\nonumber\\
&&=-4[q_{\|}^2(k_z+\frac{q_z}{2})^2-\frac{1}{4}q_0^2(q_{\|}^2-4m_{f,n}^2)].
\end{eqnarray}
The two numerators except the distribution functions in Eq.~(\ref{eq:RePi}) respectively read as
\begin{equation}\label{eq:numerator1}
-8k_zq_z[(k_z+\frac{q_z}{2})^2-\frac{q_0^2}{4}]-4m_{f,n}^2(q_z^2+2k_zq_z),
\end{equation} 
and 
\begin{equation}\label{eq:numerator2}
8p_zq_z[(k_z+\frac{q_z}{2})^2-\frac{q_0^2}{4}]+4m_{f,n}^2(q_z^2+2k_zq_z).
\end{equation}
In the massless limit, the integration over $k_z$ leads to a cancellation of the two terms (Eq.~(\ref{eq:numerator1}) and Eq.~(\ref{eq:numerator2})) in the LLL, as a result,  the medium correction to the real part of $\Pi_{11,\mathrm{quark}}^{00}$ vanishes.

For the LLL  massive light quarks,  Eq.~(\ref{eq:RePi}) within the static limit ($q_0\to $ 0) is proportional to $m_f^2$ and can be expressed as  
 \begin{align}
\mathrm{Re}\Pi^{00}_{11,\mathrm{quark},T\neq 0,n=0}&\propto\frac{m_{f}^2}{q_z^2}\int dk_z\frac{(q_z^2+2q_zk_z)}{(k_z+q_z/2)^2}\nonumber\\
&\times \bigg[\frac{f^0_+(E^f_{k,n})+f^0_-(E^f_{k,n})}{E_{k,n}^f}\nonumber\\
 &-\frac{f^0_+(E^f_{p,n})+f^0_-(E^f_{p,n})}{E^f_{p,n}}\bigg],
 \end{align}
 which is also consistent with the result presented in Refs.~\cite{Fukushima:2015wck,Hattori:2022uzp}. 
 In the present work, we focus on the massless limit, thus, the medium LLL quark-loop contribution to the real part of the retarded gluon self-energy  vanishes, i.e.,
\begin{eqnarray}
\mathrm{Re}\Pi^{00}_{R,\mathrm{quark},T\neq 0,n=0}=\mathrm{Re}\Pi^{00}_{11,\mathrm{quark},T\neq 0,n=0}=0.
\end{eqnarray}

For the medium quark-loop contribution to the imaginary part of the $\Pi_{11,\mathrm{quark}}^{00}$,  it in equilibrium can be given as 
\begin{widetext}
\begin{align}\label{eq:ImPi11}
\mathrm{Im}\Pi^{00}_{11,\mathrm{quark},T\neq0}=&\sum_{f}\sum_{n=0}^{\infty}\frac{g_s^2}{2}\int \frac{dk_z}{2\pi}\frac{2\pi}{2E^f_{k,n}4E^f_{p,n}}\biggl\{\left[f^0_+(E^f_{k,n})+f^0_{+}(q_0+E^f_{k,n})-2f^0_{+}(E^f_{k,n})f^0_{+}(q_0+E^f_{k,n})\right]\nonumber\\
	&\times L^{00}(k_0=E^f_{k,n})\left[\delta(q_0+E^f_{k,n}-E^f_{p,n})+\delta(q_0+E^f_{k,n}+E^f_{p,n})\right]\nonumber\\
	&+\left[f^0_-(E^f_{k,n})+f^0_{-}(-q_0+E^f_{k,n})-2f^0_{-}(E^f_{k,n})f^0_{-}(-q_0+E^f_{k,n})\right]\nonumber\\
	&\times L^{00}(k_0=-E^f_{k,n})\left[\delta(-q_0-E^f_{k,n}-E^f_{p,n})+\delta(-q_0-E^f_{k,n}+E^f_{p,n})\right]\biggr\}.
	\end{align}
\end{widetext}
In the above four delta functions, 
only $\delta (q_0+E^f_{k,n}-E^f_{p,n})$ and $\delta (q_0-E^f_{k,n}+E^f_{p,n})$ (decay processes) 
contribute to the final result of interest. 
In the static limit $q_0\to 0$, the delta function of our interest  can be worked out explicitly as 
\begin{eqnarray}
 \delta (E^f_{k,n}-E^f_{p,n})&=& \delta (E^f_{k,n}-E^f_{k',n})\nonumber\\
 &=&(E^f_{q_z/2,n}/|q_z|)\delta(k_z+\frac{q_z}{2}),
\end{eqnarray}
where $k_z'=k_z+q_z$ is defined and $E^f_{q_z/2,n}=(m_{f,n}^2+q_z^2/4)^{1/2}$.
This indicates that a backward scattering with $k_z=-k_z'=q_z/2$ is satisfied for the HLL quarks.  Such a backward scattering process is not allowed for LLL massless quarks. Consequently, in the static limit, 
 the medium quark-loop contribution to the imaginary part of the retarded self-energy, $\mathrm{Im}\Pi^{00}_{R,\mathrm{quark},T\neq0}$, can be computed as follows:
\begin{align}
\frac{\mathrm{Im }\Pi^{00}_{R,\mathrm{quark},T\neq 0}}{q_0}\bigg|_{q_0\to 0}=&\tanh\left|\frac{q_0}{2T}\right|\frac{\mathrm{Im }\Pi^{00}_{11,\mathrm{quark},T\neq 0}}{q_0}\bigg|_{q_0\to 0},\\
=&\sum_{f}\sum_{n=0}^{\infty}\frac{\alpha_{0n}\pi s_f m_{f,n}^2}{2TE_{q_z/2,n}^f|q_z|}H^0_{f,n}\label{eq:IMPi}
.
\end{align}
Here, we have defined the function: $H_{f,n}^0(T,eB,q_z)=f^0_{+}(E^f_{q_z/2,n})(1-f^0_{+}(E^f_{q_z/2,n}))+f^0_{-}(E^f_{q_z/2,n})(1-f^0_{-}(E^f_{q_z/2,n}))$.
For massless light quarks in the LLL, Eq.~(\ref{eq:IMPi})  vanishes strictly. For massive light quarks in the LLL, the presence of a small term $m_{f}^2/(|q_z|E^f_{q_z/2,n})$ in Eq.~(\ref{eq:IMPi}) makes the LLL massive quark-loop contribution to the imaginary part of the retarded gluon self-energy negligible compared to the HLL quark-loop contribution. Since our focus is on the massless limit ($m_f=0$), the medium LLL quark-loop contribution to the imaginary part of the retarded self-energy is zero, i.e.,
\begin{eqnarray}\label{eq:ImPin=0}
\mathrm{Im}\Pi^{00}_{R,\mathrm{quark},T\neq 0,n=0}=0.
\end{eqnarray}

\subsubsection*{A-4.~Medium HLL quark-loop contribution to retarded gluon self-energy}
When light quarks occupy higher Landau level states ($n\geq 1$), the current mass of the quark becomes insignificant because of $neB\gg m_f^2$. Taking $q_0= 0$ and $q_z\to 0$, Eq.~(\ref{eq:RePi}) can easily reduce as 
\begin{align}\label{eq:RePi_quark}
\mathrm{Re}\Pi_{11,\mathrm{quark},T\neq0,n\geq 1}^{00}
=&\sum_{f}\sum_{n=1}^{\infty}\frac{g_s^2|q_feB|}{4\pi T}e^{\frac{-\bm{q}_{\perp}^2}{2|q_feB|}}\int\frac{dk_z}{\pi}\nonumber\\
&\times[f_{+}^0(E^f_{k,n})(1-f^0_{+}(E^f_{k,n}))\nonumber\\
&+f_{-}^0(E^f_{k,n})(1-f_{-}^0(E^f_{k,n}))].
\end{align}
Therefore, the medium HLL quark-loop contribution to the  real part of retarded gluon self-energy  in the  magnetized QGP medium is given as 
\begin{eqnarray}
\mathrm{Re}\Pi_{R,\mathrm{quark},T\neq0,n\geq 1}^{00}
= \mathrm{Re}\Pi_{11,\mathrm{quark},T\neq0,n\geq 1}^{00}.
\end{eqnarray}
Using Eq.~(\ref{eq:IMPi}) and Eq.~(\ref{eq:ImPin=0}), the medium HLL quark-loop contribution to the imaginary part of the retarded gluon self-energy can be computed as follows: 
\begin{align}\label{eq:IMPi_n}
\frac{\mathrm{Im }\Pi^{00}_{R,\mathrm{quark},T\neq 0,n\geq 1} }{q_0}\bigg|_{q_0\to 0}=
\sum_{f}\sum_{n=1}^{\infty}\frac{\pi s_f m_{f,n}^2}{TE_{q_z/2,n}^f|q_z|}H^0_{f,n}
\end{align}
Finally, the total quark-loop contribution to the retarded gluon self-energy is expressed as:
\begin{eqnarray}
    \Pi_{R,\rm quark}^{00}= \Pi_{R,\mathrm{quark},T=0,n=0}^{00}+ \Pi_{R,\mathrm{quark},T\neq 0,n\geq1}^{00}.
\end{eqnarray}
Here, the first term presents the vacuum contribution from the LLL quark-loop, and the second term corresponds to the medium contribution from the HLL quark-loop.

\subsection{Gluon-loop contribution to retarded gluon self-energy}
Since gluons are not directly influenced by magnetic fields, following  Refs.~\cite{Singh:2017nfa, Blaizot:2021xqa, Kapusta:2006pm, Brambilla:2008cx, Weldon:1982aq}, the gluon-loop contribution to the retarded gluon self-energy in a magnetized QGP medium is computed as follows:
\begin{equation}\label{eq:Pi_gluon}
\Pi_{R,\rm gluon}^{00}=
m_{D,g}^2\bigg[1-\frac{q_0}{2|\bm{q}|}(\ln\bigg|\frac{q_0+|\bm{q}|}{q_0-|\bm{q}|}\bigg|-i\pi\Theta(-Q^2))\bigg],
\end{equation}
where $m_{D,g}^2=g_s^2T^2$. In Eq.~(\ref{eq:Pi_gluon}), the gluon self-energy from gluon-loop contribution also possesses  an imaginary part in the space-like region. In the static limit $q_0\to 0$,  one gets \begin{equation}\label{Imeq:Pi_gluon}
\frac{\mathrm{Im}\Pi_{R,\rm gluon}^{00}}{q_0}\bigg|_{q_0\to 0}=\frac{m_{D,g}^2\pi }{2|\bm{q}|}.
\end{equation}

\section{HTL resummed gluon propagator in magnetized QGP medium }\label{sec:propagator}

Utilizing the results from Section~\ref{sec:gluon_self-energy}, the HTL resummed effective gluon propagator  in the magnetized QGP medium can be determined by the Dyson-Schwinger equation,
\begin{eqnarray}\label{eq:DSeq}
G^{*00}_{R/A}=G^{00}_{R/A}+G^{00}_{R/A}\Pi_{R/A}^{00}G_{R/A}^{*00},
\end{eqnarray}
with $G^{00}_{R/A}$ being the temporal component of bare retarded/advanced gluon propagator. The superscript ``$*$" denotes the resummed effect gluon propagator. Consequently, the HTL resummed retarded/advanced gluon propagator in the Coulomb gauge is expressed as follows:
\begin{align}
G_{R/A}^{*00}=\frac{1}{\bm{q}^2+\mathrm{Re}\Pi^{00}_{R}\pm i\mathrm{Im}\Pi^{00}_{R}}.
\end{align}
The total real and imaginary parts of the retarded gluon self-energy are given as
\begin{eqnarray}
\mathrm{Re}\Pi_{R}^{00}
&=&\mathrm{Re}\Pi_{R,\mathrm{quark}}^{00}
+\mathrm{Re}\Pi_{R,\mathrm{gluon}}^{00},\\
&=&\sum\limits_{f}\sum\limits_{n=0}\limits^{\infty}\frac{g_s^2{\alpha_{0n}}|q_feB|}{4\pi T}e^{\frac{-\bm{q}_{\perp}^2}{2|q_feB|}}\int\frac{dk_z}{2\pi}\nonumber\\
&&\times[f_{+}^0(E^f_{k,n})(1-f^0_{+}(E^f_{k,n}))\nonumber\\
&&+f_{-}^0(E^f_{k,n})(1-f_{-}^0(E^f_{k,n}))]+m_{D,g}^2.\label{eq:Re}\\
\mathrm{Im}\Pi_{R}^{00}&=&\mathrm{Im}\Pi_{R,\mathrm{quark},T=0,n=0}^{00}+\mathrm{Im }\Pi_{R,\mathrm{quark},T\neq 0,n\geq 1}^{00}\nonumber\\
&&+\mathrm{Im}\Pi_{R,\mathrm{gluon}	}^{00}\label{eq:Im}.
\end{eqnarray}
When z$\bm{q}_{\perp}\to 0$, $\mathrm{Re}\Pi_{R}^{00}=m_{D}^2$ is nothing but the Debye screening mass.
Thus, the real and imaginary parts of the HTL resummed retarded gluon propagator in the static limit $q_0\to 0$ are expressed as follows:
\begin{eqnarray}
    \mathrm{Re} G_{R}^{*00}&=&\frac{1}{\bm{q}^2+\mathrm{Re}\Pi_{R}^{00}},\\
    \mathrm{Im} G_{R}^{*00}&=&-\frac{ \mathrm{Im\Pi}^{00}_{R}}{(\bm{q}^2+\mathrm{Re\Pi}^{00}_{R})^2},\\
&=&-\frac{q_0\sum_{f}\pi s_f}{2(\bm{q}^2+\mathrm{Re\Pi}^{00}_{R})^2}[\delta(q_z)+\delta(-q_z)]\nonumber\\
&&-
\sum_{f}\sum_{n=1}^{\infty}\frac{q_0\pi s_fm_{f,n}^2H_{f,n}^0}{T(\bm{q}^2+\mathrm{Re}\Pi_{R}^{00})^2|q_z|E^f_{q_z/2,n}}\nonumber\\
&&-\frac{q_0\pi  m_{D,g}^2}{2|\bm{q}|(\bm{q}^2+\mathrm{Re}\Pi_{R}^{00})^2}.\label{eq:IMGR00}
\end{eqnarray}
Notice that Eq.~(\ref{eq:IMGR00}) is the crucial formula of our work.

The resummed symmetric gluon propagator can be determined through the KMS condition in equilibrium medium~\cite{Brambilla:2008cx}
\begin{equation}\label{eq:GF*}
G^{*00}_F(Q)=(1+2f^0_{B}(q_0))\mathrm{sgn}(q_0)(G^{*00}_R(Q)-G^{*00}_A(Q)).
\end{equation} 
In the static limit $q_0\to 0$, the imaginary part of Eq.~(\ref{eq:GF*}) is computed as  
\begin{eqnarray}\label{eq:IMGF00}
\mathrm{Im }G_F^{*00}
&=& \frac{2T}{q_0}2\mathrm{Im} G_{R}^{*00}=-\frac{2T}{q_0}\frac{2\mathrm{Im\Pi}^{00}_{R}}{(\bm{q}^2+\mathrm{Re\Pi}^{00}_{R})^2}.
\end{eqnarray}

\section{Gluon self-energy and resummed gluon propagator in magnetized viscous QGP medium}\label{sec:gluon_self-energy_bulk}
\subsection{ Magnetized viscous correction on light quark distribution function}
To investigate the properties of non-equilibrium QGP medium, we take into account the viscous effect and study the responses of viscous QGP medium to the magnetic field. Subsequently, we explore how the magnetized viscous effect impacts both the  HQ potential and momentum diffusion coefficient. In the strong magnetic limit within the LLL approximation, Landau level quantization ensures that only the longitudinal component of bulk viscosities, arising from the contributions of  LLL (anti)-quarks, exists~\cite{Hattori:2017qih}.
The response of viscous quark matter to the magnetic field is manifested in the longitudinal bulk viscous modified distribution function of light (anti-)quarks, which can be obtained using the Boltzmann kinetic theory. 
In  the presence of a magnetic field, the dynamical evolution of the light (anti-)quark distribution function is described by the 1+1 dimensional relativistic Boltzmann equation~\cite{Hattori:2016lqx, Hattori:2017qih}
\begin{equation}\label{eq:Bolzmann_eq}
k_{\|}^{\mu}\partial_{\|\mu}f_{\pm}(x,k_{z})=
C[f_{\pm}],
\end{equation}
where the covariant derivative  $\partial_{\|\mu }$ is given by  $\partial_{\|\mu }=u_{\mu}D+\nabla_{\|\mu}$, with $D=u^\mu \partial_\mu$ and $\nabla_{\|}^\mu=\Delta_{\|}^{\mu\nu} \partial_\nu$ corresponding to the time derivative and spatial gradient operator in the local rest frame, respectively. $u^{\nu}$ is the fluid four-velocity, and is normalized to $u^{\nu}u_{\nu}=1$. $\Delta_{\|}^{\mu\nu}=g^{\mu\nu}_{\|}-u^\mu u^\nu$ is the longitudinal projection operator. 
The right-hand side of Eq.~(\ref{eq:Bolzmann_eq}) is the collision term,  which describes the change rate of the single-particle distribution induced by scatterings. Given that the system slightly deviates from the local thermal equilibrium due to external perturbation, the relaxation time approximation (RTA) can be reasonably employed to obtain an analytic solution for the distribution function. In the RTA, the collision term is given as  
\begin{equation}
C[f_{\pm}]=-(u\cdot k_{\parallel})\dfrac{\delta f_{\pm}}{\tau_{R}},
\end{equation}
where $\tau_R$ denotes the relaxation time which quantifies how fast the system reaches the equilibrium again. The perturbative term for the distribution functions $\delta f_{\pm}$ can be written as 
\begin{equation}
\delta f_{\pm}=f_{\pm}-f^0_{\pm}.  
\end{equation}
The local equilibrium distribution functions of light (anti-)quarks in a magnetic field have the following form:
\begin{align}\label{eq:flocal}
{f}^{0}_{\pm}=\dfrac{1}{1+ \exp[\beta(  k_{\parallel}\cdot u\pm\alpha)]},
\end{align}
where $\beta=1/T$ is inverse local temperature,  and $u\cdot k_{\|}\equiv
=E^f_{k,n}$.  The notation $\alpha=\mu\beta$ represents the ratio of quark chemical potential to temperature. By inserting Eq.~(\ref{eq:flocal}) into  the left-hand side of Eq.~(\ref{eq:Bolzmann_eq}), the solution for non-equilibrium correction to the phase-space distribution function can be expanded as
\begin{align}\label{eq:deltaf}
\delta f_{\pm}=& \frac{\tau_R}{u\cdot k_{\|}}\bigg[k_{\|}^\mu k_{\|}^\nu\beta\partial_{\|\mu }u_{\nu} +k_{\|}^\mu (u\cdot k_{\|})\partial_{\|\mu } \beta\pm k_{\|}^\mu\partial_{\|\mu }\alpha\bigg]\nonumber\\
&f^0_{\pm}(1-f^0_{\pm}).
\end{align}
Considering a small gradient expansion around ideal hydrodynamics, we can ignore the higher-order gradient corrections and employ leading-order equations of motion for thermodynamic parameters:
\begin{align}
&D\beta=\chi_{\beta}\theta_{\|},~ \nabla_{\|}^\mu\beta=-\beta Du^\mu+\frac{n_q}{\epsilon_{\|}+P_{\|}}\nabla_{\|}^\mu\alpha,\label{eq:1}\\ &D\alpha=\chi_{\alpha}\theta_{\|}.\label{eq:2}
\end{align} 
The dimensionless coefficients $\chi_{\beta}$ and $\chi_{\alpha}$ in Eqs.~(\ref{eq:1}-\ref{eq:2}) are respectively given as 
\begin{align}
\chi_{\beta}&=\bigg[\frac{L_{10}^{0+}(\epsilon_{\|}+P_{\|})-L_{20}^{0-}n_q}{L_{30}^{0+}L_{10}^{0+}-(L_{20}^{0-})^2}\bigg],\nonumber\\
~\chi_{\alpha}&=\bigg[\frac{L_{20}^{0-}(\epsilon_{\|}+P_{\|})-L_{30}^{0+}n_q}{L_{30}^{0+}L_{10}^{0+}-(L_{20}^{0-})^2}\bigg], 
\end{align}
where $\epsilon_{\|}$ and $P_{\|}$ are longitudinal energy density and longitudinal pressure, respectively, $n_{q}$ denotes net density. We also introduce the thermodynamic function in the magnetic field, which is defined as
\begin{align}
L_{jm}^{\gamma\pm}=&N_c\sum_f\frac{|q_feB|}{2\pi}\sum_{n=0}^{\infty}\alpha_{0n}\int\frac{dk_z}{\pi}(E_{k,n}^f)^{j-2m-\gamma-1}\nonumber\\ &k_z^{2m}(-1)^m[f^0_-(1-f^0_-)+f^0_{+}(1-f_{+}^0)].
\end{align}
For vanishing chemical potential, $\chi_{\beta}/\beta\equiv c_{\|s}^2
=\frac{\partial P_{\|}}{\partial \varepsilon_{\|}}$, which is the squared speed of sound in the longitudinal direction. In the description of kinetic theory, the longitudinal bulk viscous correction to momentum-energy tensor is given by~\cite{Hattori:2017qih}
\begin{eqnarray}
\delta T^{\mu\nu}=-\Pi_{\|} \Delta_{\|}^{\mu\nu}.
\end{eqnarray} 
The longitudinal bulk viscous pressure, $\Pi_{\|}$, in the magnetized QGP medium is computed as~\cite{Hattori:2022hyo}
\begin{equation}
 \Pi_{\|}=-N_c\sum_{f}\sum_{n=0}^{\infty}\frac{|q_feB|}{2\pi}\int\frac{dk_z\alpha_{0n}}{2\pi E_{k,n}^f} \Delta_{\|}^{\mu\nu} k_\mu k_\nu(\delta f_-+\delta f_{+})\label{eq:Pi_munu}.
\end{equation}
By substituting Eq.~(\ref{eq:deltaf}) into Eq.~(\ref{eq:Pi_munu}), the leading-order longitudinal bulk viscous pressure can be rewritten as
\begin{equation}\label{eq:Re_pi_munu}
\Pi_{\|}=-\tau_R\beta_{\Pi_{\|}}\theta_{\|},
\end{equation}
where $\theta_{\|}\equiv \partial_{\|\mu }u^\mu$  denotes the longitudinal fluid expansion rate. 
Here, we have introduced the longitudinal bulk viscous coefficient $\beta_{\Pi_{\|}}$, which is defined as  
\begin{eqnarray}
\beta_{\Pi_{\|}}=
\beta(\frac{\chi_\beta}{\beta}L_{31}^{0+}+L_{42}^{1+}-\frac{\chi_{\alpha}}{\beta}L_{21}^{0-}).
\end{eqnarray}
By comparing Eq.~(\ref{eq:Re_pi_munu}) with the Naiver-Stokes equation for the dissipative quantities in a magnetic field up to the first order in derivative expansion, namely, $\Pi_{\|}=-\zeta_{\|} \theta_{\|}$, 
with $\zeta_{\|}$  being the longitudinal bulk viscosity,  we can obtain the longitudinal bulk viscous modified distribution functions for light (anti-)quarks
\begin{align}\label{eq:deltafqbarq}
\delta f^{}_{\pm}=&\frac{\beta^{}\Pi_{\parallel}}{\beta_{\Pi_{\parallel}}(u\cdot k_{\parallel})^{}}f_{\pm}^0(1-f_{\pm}^0)\nonumber\\
&\times\bigg[-(u\cdot k_{\parallel})^2\frac{\chi_\beta}{\beta}+k_{z}^2\pm(u\cdot k_{\parallel})\frac{\chi_\alpha}{\beta}\bigg].
\end{align}
In the local rest frame of fluid, Eq.~(\ref{eq:deltafqbarq}) takes the following form:
\begin{align}\delta f_{\pm}=&\frac{s}{\beta_{\Pi_{\|}}\tau}\bigg(\frac{\zeta_{\parallel}}{\mathit{s}}\bigg)\frac{\beta^{}}{E^f_{k,n}}\bigg[(E^f_{k,n})^2\frac{\chi_{\beta}}{\beta}-k_z^2\mp E^f_{k,n}\frac{\chi_{\alpha}}{\beta}\bigg]\nonumber\\
&\times f^0_{\pm}(1-f^0_{\pm}),
\end{align}
where the expansion parameter $\theta_{\|}$ is given by $\theta_{\|}=1/\tau$ with $\tau $ being proper time parameter, and we adopt $\tau=0.3~$fm/c, as utilized in Ref.~\cite{Kurian:2020kct}. It is clear that $\delta f_{\pm}$ vanishes for  LLL massless quarks. The LLL massless quarks along the magnetic field direction cannot suffer any dissipative effect from the medium because the scatterings are strictly forbidden according to chirality conservation and the linear dispersion relation in the LLL~\cite{Sadofyev:2015tmb}.

\subsection{Magnetized viscous correction on gluon self-energy and HTL resummed gluon propagator}
To clarify the impact of longitudinal bulk viscous effects on both the heavy quark potential and momentum diffusion coefficients, it is imperative to generalize the HTL resummation technique to the non-equilibrium scenario related to this investigation.
Given that the non-equilibrium effects in the absence of a magnetic field violate the KMS condition expressed in Eq.~(\ref{eq:KMS}) or Eq.~(\ref{eq:GF*})~\cite{Carrington:1997sq, Thakur:2021vbo, Thakur:2020ifi}, it is necessary to examine the validity of the KMS condition in the magnetized viscous QGP medium.
Following Ref.~\cite{Thakur:2020ifi}, the temporal component of the resummed retarded/advanced gluon propagator in the non-equilibrium system can still be derived through the Dyson-Schwinger equation (\ref{eq:DSeq}).
The longitudinal bulk viscous correction is incorporated into the retarded/advanced gluon self-energy by just replacing the thermal equilibrium distribution functions in Eq.~(\ref{eq:RePi_quark}) with longitudinal bulk viscous modified ones.  Consequently, the real part of the longitudinal bulk viscous modified retarded gluon self-energy is computed as 
\begin{align}
\mathrm{Re}\label{eq:ModifyRePi}{\Pi}_{R}^{00}=&m_{D,g}^2+\sum\limits_{f}\sum\limits_{n=0}\limits^{\infty}\frac{g_s^2\alpha_{0n}|q_feB|}{4\pi T}e^{\frac{-\bm{q}_{\perp}^2}{2|q_feB|}}\nonumber\\
	&\times\int\frac{dk_z}{2\pi}
	[f_{+}^0(E^f_{k,n})(1-f^0_{+}(E^f_{k,n}))+F_{+}(E_{k,n}^f)\nonumber\\
	&+f_{-}^0(E^f_{k,n})(1-f_{-}^0(E^f_{k,n}))+F_{-}(E_{k,n}^f)],
\end{align}
where we have defined the notation: $F_{\pm}(E_{k,n}^f)=\delta f_{\pm}(E_{k,n}^f)(1-2{f}^{0}_{\pm}(E_{k,n}^f))$. 
By taking $\bm{q}_{\perp}\to 0$ in Eq.~(\ref{eq:ModifyRePi}), we can derive the longitudinal bulk viscous modified Debye mass,  which is denoted as $\widetilde{m}_{D}^2$. The resummed symmetric gluon propagator in the non-equilibrium scenario is also obtained through the Dyson-Schwinger equation
\begin{equation}
{G}_{F}^{*00}=G^{00}_F+G^{00}_R{\Pi}_R^{00}{G}^{*00}_F+G^{00}_F{\Pi}_A^{00}{G}^{*00}_A
+G^{00}_R{\Pi}_F^{00}{G}^{*00}_A.
\end{equation}
By utilizing $G_{F}^{00}=(1+2f_B)\mathrm{sgn}(q_0)(G^{00}_{R}-G^{00}_{A})$, which remains valid in non-equilibrium conditions as well~\cite{Carrington:1997sq}, the solution for the ${G}^{*00}_F$ can be formulated as 
\begin{eqnarray}\label{eq:GF}
{G}^{*00}_F&=&(1+2{f}_{B}(q_0)\mathrm{sgn}(q_0))({G}^{*00}_R-{G}^{*00}_A)+{G}^{*00}_R\big[{\Pi}^{00}_F\nonumber\\
&&-(1+2{f}_B(q_0)\mathrm{sgn}(q_0))({\Pi}^{00}_R-{\Pi}^{00}_A)\big]{G}^{*00}_A.
\end{eqnarray}
In the equilibrium state, the second term on the right-hand side of Eq.~(\ref{eq:GF}) vanishes. We only focus on the viscous effect from quark contribution, ${f}_B$ retains its equilibrium form.  By utilizing the relations in Eq.~(\ref{eq:Pi_{RAF}}), after tedious calculation, the longitudinal bulk viscous modified symmetric gluon self-energy from quark-loop within the static limit is computed as
\begin{equation}
{\Pi}^{00}_{F,\mathrm{quark},T\neq0}=i\sum_{f}\sum_{n=0}^{\infty}\frac{2m_{f,n}^2 s_f }
{E^f_{q_z/2,n}|q_z|}H_{f,n}.
\end{equation}
 Here,  $H_{f,n}(T,eB,q_z)=H_{f,n}^0+\delta H_{f,n}$ with $\delta H_{f,n}=F_{-}(E_{q_z/2,n}^f)+F_{+}(E_{q_z/2,n}^f)$. 
 Similarly, we can derive the longitudinal bulk viscous modified imaginary part of retarded gluon self-energy from medium quark-loop contribution in a magnetic field, which is given as  $\mathrm{Im}{\Pi}^{00}_{R,\mathrm{quark}, T\neq0}/q_0=\sum\limits_{f}\sum\limits_{n=1}\limits^{\infty}\frac{\pi m_{f,n}^2s_f}{TE^f_{q_z/2,n}|q_z|}H_{f,n}$.
It is evident that the square bracket term on the right-hand side of Eq.~(\ref{eq:GF}) cancels out precisely in the static limit $q_0\to 0$,  revealing that the  KMS condition is still valid in the magnetized viscous quark matter. Therefore, in our calculations of gluon self-energies and resummed gluon propagators, we substitute the equilibrium distribution functions of the light (anti-) quark with the longitudinal viscous modified ones to further explore the impact of the longitudinal bulk viscous effect on various quantities. 
 
\section{HQ potential and thermal decay width in magnetized viscous QGP medium}\label{sec:HQ_potential}

In this section, we employ the HTL resummed gluon propagators derived in Sections~\ref{sec:propagator} and~\ref{sec:gluon_self-energy_bulk} to compute the static heavy quark (HQ) potential in a magnetized viscous QGP medium. 
In the vacuum, the static HQ potential studied by lattice QCD can be well parameterized by the Cornell potential: $V(r)=-C_F\alpha_s/r+\sigma r+c$~\cite{Eichten:1974af, Matsui:1986dk},
with $r\equiv|\bm{r}|$ being the heavy quark-antiquark separation distance. Here,  $\sigma$ is string tension~\cite{Jacobs:1986gv}, and $c$ is an additive  calibration parameter. This potential captures both the short-range Coulomb interaction and long-range color confinement. In the thermal bath without a magnetic field, many phenomenological HQ potential descriptions have been proposed to investigate the in-medium properties of heavy quarkonium states~\cite{Hasan:2020iwa, Khan:2021syq, Singh:2017nfa, Guo:2018vwy, Lafferty:2019jpr}. Given the success of these potential models, we proceed to explore how a magnetic field influences in-medium heavy quark-antiquark interactions.
To achieve this, we adopt the approach reported in Ref.~\cite{Guo:2018vwy,Megias:2007pq}, where the in-medium HQ potential is defined via the Fourier transform of the physical ``11" (time-ordered) component of the temporal effective gluon propagator in the static limit (for brevity, the superscript ``00" will hereinafter be omitted): 
\begin{align}\label{eq:V1}
V(r)=&-C_F g_s^2
\int \frac{d^3\bm{q}}{(2\pi)^{3}}(e^{i\bm{q}\cdot\bm{r}}-1)\nonumber\\
&\times [{G}_{11}^{*\mathrm{}}(q_0\to 0,\bm{q})+{G}_{11}^{\mathrm{*String}}(q_0\to 0,\bm{q})].
\end{align}
Here, ${G}_{11}^{*\mathrm{}}$ is the temporal component of the ``11" part of the HTL resummed gluon propagator, it can be decomposed as ${G}_{11}^{*\mathrm{}}=\frac{1}{2}({G}_R^{*\mathrm{}}+{G}_A^{*\mathrm{}}+{G}_F^{*\mathrm{}})$.
The ${G}_{11}^{*\mathrm{String}}=\frac{1}{2}({G}_R^{*\mathrm{String}}+{G}_A^{*\mathrm{String}}+{G}_F^{*\mathrm{String}})$ is a phenomenological gluon propagator to account for the non-perturbative effects arising from dimension-two gluon condensates~\cite{Guo:2018vwy,  Megias:2005ve,Megias:2007pq}.
Since the rotational symmetry is broken by the magnetic field, without loss of generality, we take $\bm{r}=(\rho,0,z)$ and the three-momentum transfer is written as $\bm{q}=|\bm{q}|(\sin\theta\cos\phi,\sin\theta\sin\phi,\cos\theta)$. Then, the real part and imaginary part of HQ potential in Eq.~(\ref{eq:V1}) can be computed as 
\begin{align}
\mathrm{Re}V(\rho,z)=&-\frac{C_F\alpha_s}{2\pi}\int_0^{\infty}
dq^2_{\perp}\int 
dq_z[J_{0}(q_{\perp}\rho)e^{iq_zz}-1]\nonumber\\
&\times
[\mathrm{Re} {G}^{*\mathrm{}}_{R}+\mathrm{Re}{G}^{\mathrm{*String}}_{R}],\label{eq:V2}\\
\mathrm{Im}V(\rho,z)=&-\frac{C_F\alpha_s}{4\pi }\int_0^{\infty} dq^2_{\perp}\int dq_z[J_{0}(q_{\perp}\rho)e^{iq_zz}-1]\nonumber\\
&\times\frac{4T}{q_0}
[\mathrm{Im} {G}^{*}_{R}+\mathrm{Im}{G}^{\mathrm{*String}}_{R}],\label{eq:ImV}
\end{align}
where 
the azimuthal integration gives rise to a first-kind of Bessel function $J_n(x)$. Guided by the simple prescription outlined in Refs.~\cite{Guo:2018vwy, Megias:2007pq, Megias:2005ve}, a minimal extension of HTL resummed retarded/advanced gluon propagator,
 a non-perturbative gluon propagator induced by the dimension-two gluon condensates is defined as ~\cite{Guo:2018vwy, Megias:2007pq}
\begin{align}\label{eq:GRString}
{G}_{R/A}^{*\mathrm{String}}=\frac{m_{G}^2}{(\bm{q}^2+{\Pi}^{00}_{R/A})^2},
\end{align}
where $m_G^2=2\sigma/(C_F\alpha_{s})$ is a dimensionful constant related to the dimension-two gluon condensates. 

\subsection{Real part of HQ potential}
According to Eq.~(\ref{eq:GRString}), the real part of the non-perturbative effective retarded gluon propagator is given as 
\begin{align}\label{eq:GF00_v2}
\mathrm{Re}{G}_R^{*\mathrm{String}}
=\frac{m_{G}^2}{(\bm{q}^2+\mathrm{Re}{\mathrm\Pi}_{R}^{00})^2}.
\end{align}
We expect the longitudinal bulk viscous correction also impact the non-perturbative effective gluon propagator. Inserting Eq.~(\ref{eq:Re}) and Eq.~(\ref{eq:GF00_v2}) into Eq.~(\ref{eq:V2}), and removing the asymmetry factor $\exp\left(\frac{-\bm{q}_{\perp}^2}{2|q_feB|}\right)$ from the denominator of Eq.~(\ref{eq:GF00_v2}), the real part of in-medium HQ potential in the magnetized viscous QGP medium finally is computed as 
\begin{eqnarray}\label{eq:ReV}
\mathrm{Re}V(r)
=-C_F\alpha_s(\frac{e^{-\hat{r}}}{r}+\widetilde{m}_D)-\frac{\sigma}{\widetilde{m}_D}(e^{-\hat{r}}-1),
\end{eqnarray}
where $\hat{r}=\widetilde{m}_Dr$. This result has the same form as the result under zero magnetic field~\cite{Guo:2018vwy} except for the Debye mass. In the small distance limit, $r\to 0$, it approaches the Cornell potential. It is noteworthy that the lattice QCD studies in Refs.~\cite{Bonati:2016kxj,Bonati:2018uwh} have indicated  $\mathrm{Re}V$ in the magnetic field at zero temperature is 
anisotropic, and the authors proposed an effective phenomenological description of the $\mathrm{Re}V$ by using the Cornell potential that incorporate angular dependent strong coupling constant and string tension: $\alpha(\theta)=\bar{\alpha}_s(1-\sum_{n=1}^{}c_{2n}^{\alpha}\cos(2n\theta))$ and $\sigma(\theta)=\bar{\sigma}(1-\sum_{n=1}^{}c_{2n}^{\sigma}\cos(2n\theta))$. Here, $\theta$ denotes the angle between the quark-antiquark separation and the the magnetic field direction, and the parameters $c_{2n}^{\alpha}$ and $c_{2n}^{\sigma}$ are determined by lattice data at $T=0$. 
Consequently, by utilizing the angular-dependent strong coupling constant $\alpha_{s} (\theta)$ and string tension $\sigma(\theta)$ from Refs.~\cite{Bonati:2016kxj,Bonati:2018uwh}, the in-medium real part of the potential  is no longer isotropic,  which will be discussed later.
  
\subsection{Imaginary part of HQ potential}
In addition to the color screening effect preventing the formation of quarkonium states, the imaginary part of the HQ potential $\mathrm{Im }V$, can also trigger the dissociation and suppression of in-medium quarkonium.
The $\mathrm{Im }V$ at least comes from two distinct mechanisms~\cite{Brambilla:2011sg, Brambilla:2008cx, Blaizot:2021xqa}: 
gluodissociation (dissociation of quarkonium by absorbing a time-like gluon from thermal bath) or called the singlet to octet thermal breaking-up and inelastic parton scattering (dissociation of quarkonium by scattering with gluons and light quarks in the medium) or called Landau damping. 
In this paper, we focus on the Landau damping phenomenon which relates to the imaginary part of the gluon self-energy in a space-like region. Extracting the imaginary part of the potential in lattice QCD simulations is challenging, and as far as we know, there are currently no lattice results available regarding $\mathrm{Im}V$ in the presence of a magnetic field. Therefore, we employ the potential model for phenomenological research to gain a stable qualitative understanding. 
By mimicking the $q_0$-dependence of the HTL resummed gluon propagator,  the imaginary part of the effective retarded gluon propagator induced by dimension-two gluon condensates is specified as 
\begin{eqnarray}\label{eq:IMG}
\mathrm{Im}{G}_{R}^{*\mathrm{String}}=
-\frac{\mathrm{Im}{\Pi}_{R}^{00}}{({\bm{q}^2}+\mathrm{Re}{\Pi}_{R}^{00})^2}
\frac{2m_{G}^2}{{\bm{q}^2}+\mathrm{Re}{\Pi}_{R}^{00}}.
\end{eqnarray}
Since the imaginary parts of the retarded gluon self-energy from LLL quark-loop, HLL quark-loop, and gluon-loop presented in Eq.~(\ref{eq:IMGR00})  have strikingly different forms, we will separately discuss these corresponding imaginary parts of the potential to better clarify their respective features.

\subsubsection{Imaginary part of HQ potential related to gluon-loop contribution of  gluon self-energy}
Inserting Eq.~(\ref{eq:IMGR00}) into Eq.~(\ref{eq:IMG}) and utilizing Eq.~(\ref{eq:ImV}),
we can first get the imaginary part of the HQ potential related to the gluon-loop contribution of the gluon self-energy, denoted as $\mathrm{Im}V^{\rm gluon}$, which is computed as 
\begin{align}
\mathrm{Im}V^{\rm gluon}(r)=&C_F\alpha_s
\int \frac{d^3\bm{q}}{2\pi^{2}}(e^{i\bm{q}\cdot\bm{r}}-1)\frac{2T}{q_0}\nonumber\\
&\times\left[\frac{\mathrm{Im}\Pi_{R,\rm gluon}^{00}}{(\bm{q}^2+\mathrm{Re}\Pi_{R}^{00})^2}+\frac{2m_{G}^2\mathrm{Im}\Pi_{R,\rm gluon}^{00}}{(\bm{q}^2+\mathrm{Re}\Pi_{R}^{00})^3}\right],\\
=&C_F\alpha_s
\int \frac{d^3\bm{q}}{2\pi^{2}}(e^{i\bm{q}\cdot\bm{r}}-1)\nonumber\\
&\times\left[\frac{\pi T m_{D,g}^2}{|\bm{q}|(\bm{q}^2+\widetilde{m}_D^2)^2}+\frac{2\pi T m_G^2 m_{D,g}^2}{|\bm{q}|(\bm{q}^2+\widetilde{m}_D^2)^3}\right]\nonumber\\
=&-\frac{C_F\alpha_sTm_{D,g}^2 }{\widetilde{m}_D^2} \phi_2(\hat{r})-\frac{4\sigma Tm_{D,g}^2}{\widetilde{m}_D^4}\phi_3(\hat{r}),
\end{align}
where the longitudinal bulk viscous correction is encoded in the Debye mass, the functions $\phi_2$ and  $\phi_3$ are defined by the formula $\phi(x)_n=2\int_{0}^{\infty}\frac{zdz}{(z^2+1)^{n}}[1-\frac{\sin(zx)}{zx}]$.

\subsubsection{Imaginary part of HQ potential related to LLL quark-loop contribution of gluon self-energy}
Next, we shall study the imaginary part of the HQ potential related to the quark-loop contribution of gluon self-energy.
We first study the imaginary part of the potential related to the LLL quark-loop contribution of the gluon self-energy($\mathrm{Im }\Pi^{00}_{R,\mathrm{quark}, T= 0,n=0}$), denoted as  $\mathrm{Im}V^{\rm LLL~quark}$, which is usually overlooked in previous studies~\cite{Ghosh:2022sxi, Singh:2017nfa}. 
By incorporating the second term of  Eq.~(\ref{eq:IMGR00}) into Eq.~(\ref{eq:IMG}), and utilizing Eq.~(\ref{eq:ImV}), we can get 
  \begin{align}
  \mathrm{Im}V^{\rm LLL~quark}(\rho)=&C_F\alpha_s\int \frac{d^3 \bm{q}}{2\pi^2}(e^{i\bm{q}\cdot\bm{r}}-1)\frac{2T}{q_0}\nonumber\\
&\times\bigg[\frac{\mathrm{Im}\Pi_{R,\mathrm{quark}, T=0,n=0}^{00}}{(\bm{q}^2+\mathrm{Re}\Pi_{R}^{00})^2}\nonumber\\
&+\frac{2m_{G}^2\mathrm{Im}\Pi_{R,\mathrm{quark}, T=0,n=0}^{00}}{(\bm{q}^2+\mathrm{Re}\Pi_{R}^{00})^3}\bigg]\\
=&\frac{C_F\alpha_s}{2\pi}\int_0^{\infty} dx\int dq_z \left(J_{0}(\sqrt{x}\rho)e^{iq_z z}-1\right),\nonumber\\
  &\times\frac{\sum_{f}\pi 
  	s_fT}{(q_z^2+x+\widetilde{m}_{D}^2)^2}\left(1+\frac{2m_G^2}{q_z^2+x+\widetilde{m}_{D}^2}\right)\nonumber\\
  &\times\left[\delta(q_z)+\delta(-q_z)\right],\\
  =&\frac{C_F\alpha_s}{\pi}\int_0^{\infty} dx\left(J_{0}(\sqrt{x}\rho)-1\right)\nonumber\\
  &\times\frac{\sum_{f}\pi 
  	s_f(x)T}{(x+\widetilde{m}_{D}^2)^2}\left(1+\frac{2m_G^2}{x+\widetilde{m}_{D}^2}\right),
  \end{align}
with $x={q}_{\perp}^2$ for convenience. We note that when the light quarks occupy the LLL,  the presence of the delta function $\delta(\pm q_z)$ results in the heavy quark-antiquark dipole axis being aligned only in the plane perpendicular to the magnetic field direction.

\subsubsection{Imaginary part of HQ potential related to HLL  quark-loop contribution of gluon self-energy}
By inserting the first term of Eq.~(\ref{eq:IMGR00}) into Eq.~(\ref{eq:IMG}) and utilizing Eq.~(\ref{eq:ImV}), we can obtain the imaginary part of the HQ potential related to the HLL quark-loop contribution of the gluon self-energy ($\mathrm{Im}\Pi^{00}_{R,\mathrm{quark}, T\neq 0,n\geq 1}$), denoted as $\mathrm{Im} V^{\rm HLL~quark}$.
Due to the presence of factor $|q_z|E^f_{q_z/2,n}$ in the denominator of Eq.~(\ref{eq:IMGR00}), we need to carefully consider the longitudinal and transverse momentum components during the Fourier transform, the $\mathrm{Im} V^{\rm HLL~quark}$ can be computed as
\begin{align}\label{eq:V3}
\mathrm{Im}V^{\rm HLL~quark}(\rho,z)=&C_F\alpha_s\int \frac{d^3 \bm{q}}{2\pi^2}(e^{i\bm{q}\cdot\bm{r}}-1)\frac{2T}{q_0}\nonumber\\
&\times\bigg[\frac{\mathrm{Im}\Pi_{R,\mathrm{quark}, T\neq0,n\geq 1}^{00}}{(\bm{q}^2+\mathrm{Re}\Pi_{R}^{00})^2}\nonumber\\
&+\frac{2m_{G}^2\mathrm{Im}\Pi_{R,\mathrm{quark}, T\neq0,n\geq1}^{00}}{(\bm{q}^2+\mathrm{Re}\Pi_{R}^{00})^3}\bigg],\\
=&\frac{C_F\alpha_s}{2\pi}\int_0^{\infty} dx \int  dq_z\nonumber\\
&\times\left(J_{0}(\sqrt{x}\rho)e^{iq_zz}-1\right)
\nonumber\\
&\times\frac{\sum_f\sum_{n=1}^{\infty}4\pi s_{f}m_{f,n}^2H_{f,n}}{(q_z^2+x+\widetilde{m}_D^2)^2|q_z|E_{q_z/2,n}^f}\nonumber\\
&\times\left(1+\frac{2m_{G}^2}{q_z^2+x+\widetilde{m}_D^2}\right).
\end{align}
Finally, the imaginary part of the potential related to the total quark-loop contributions of the gluon self-energy is given as 
\begin{equation}
\mathrm{Im}V^{\rm quark}=\mathrm{Im}V^{\rm LLL~quark}+\mathrm{Im}V^{\rm HLL~quark}.
\end{equation} 
 The imaginary part of the total HQ potential can be written as:
\begin{equation}
\mathrm{Im}V=\mathrm{Im}V^{\rm quark}+\mathrm{Im}V^{\rm gluon}.
\end{equation}.
 
\subsection{Thermal decay widths of quarkonium states}
The presence of the imaginary part of the HQ potential indicates that the heavy quarkonium has a medium-induced decay width, and therefore a finite lifetime.  By applying the obtained HQ potential into the time-independent Schr$\ddot{\mathrm{o}}$dinger equation,  we can estimate the thermal decay widths of quarkonium states using the simple Coulomb wave functions $|\Psi' \rangle$~\cite{Srivastava:2018vxp}, 
\begin{eqnarray}
\Gamma_{\mathrm{decay}}&=&-\langle \Psi'|\mathrm{Im} V(r;T,B,\zeta_{\|}/s)|\Psi'\rangle.
\end{eqnarray}
After performing the integration in coordinate space by folding with probability density, the thermal decay widths of quarkonium states are given by 
\begin{eqnarray}
\Gamma_{\mathrm{decay}}=-\int d^3\bm{r}(\langle\bm{r}|\Psi'\rangle)^2\mathrm{Im} V(\bm{r};T,B,\zeta_{\|}/s).
\end{eqnarray}
We are only interested in the ground states of charmonium $J/\psi(1S)$ and bottomonium $\Upsilon (1S)$, the associated wave function is $\langle\bm{r}|1S\rangle=\frac{1}{\sqrt{\pi a_0^3}}e^{-\bm{r}/a_0}$, where $a_0=2/(C_F m_{HQ}\alpha_{s})$. In the numerical calculation, the charm and bottom masses are taken as $m_c=1.275~\mathrm{GeV}$ and $m_b=4.66~\mathrm{GeV}$, respectively.

\section{HQ momentum diffusion coefficient in magnetized viscous QGP medium}\label{sec:HQ_diffusion}
\subsection{formalism}
\begin{figure}
\subfloat{\includegraphics[scale=0.48]{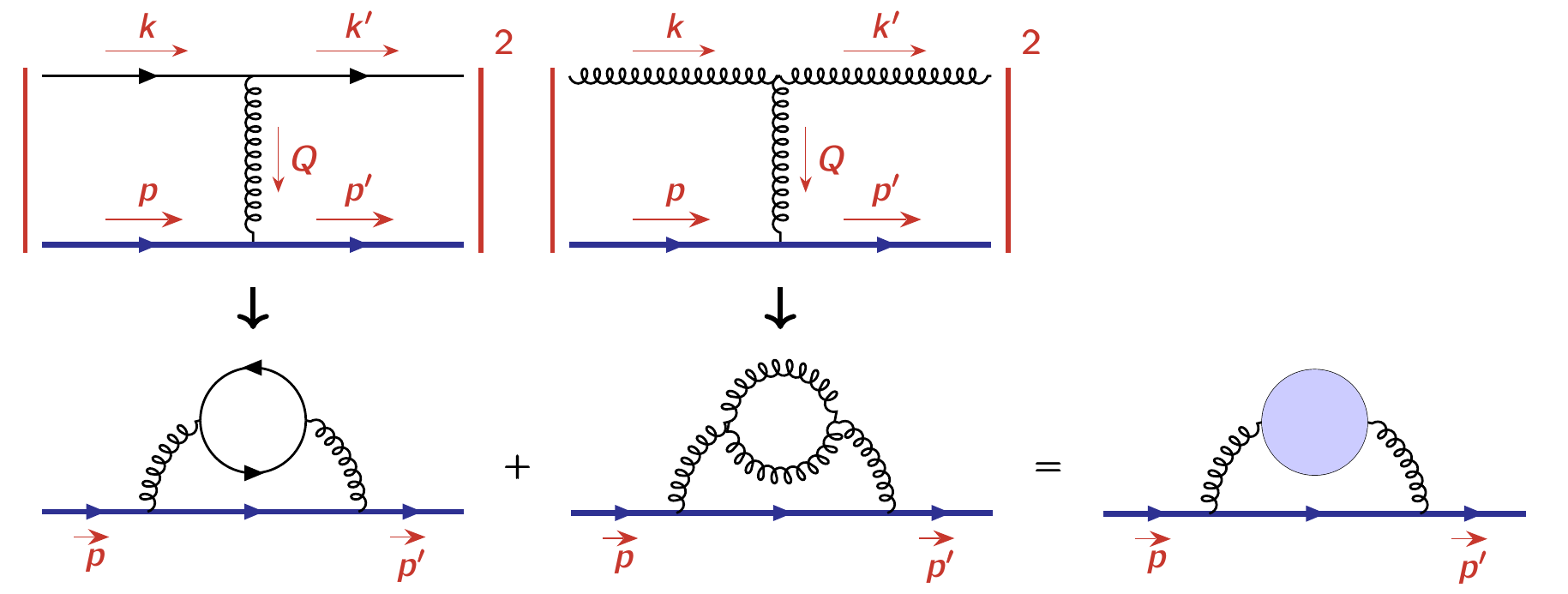}}\hspace{.1 cm}
	\caption{ (Upper panel) The $t$-channel scatterings between heavy quarks (blue lines) and thermal partons (light quarks and gluons) from QGP medium. (Lower panel) The corresponding squared scattering amplitude is directly related to the imaginary part of the HQ self-energy with the effective propagator of the exchanged gluon between a heavy quark and a thermal parton from the QGP medium.
 Both quark-loop and gluon-loop contributions to the gluon self-energy are considered in the computation of the effective gluon propagator. The blue blob denotes the HTL resummation.}
 \label{fig_HQ_diffusion}
\end{figure}

 As mentioned in the Introduction, a crucial input parameter in the Langevin transport  equation is the  HQ momentum diffusion coefficient $\kappa$, which encodes the interaction information between heavy quark and the QGP medium~\cite{Moore:2004tg}. Due to the breaking of the rotational symmetry induced by the magnetic field, it is necessary to separate longitudinal (denoted as $\|$) and transverse (denoted as$\perp$) HQ momentum diffusion coefficients. These can be expressed as follows:
\begin{eqnarray}\label{eq:kappa}
\kappa_{\|}=\int d^3\bm{q}\frac{d\Gamma(\bm{q})}{d^3\bm{q}}q_{z}^2,~\kappa_{\perp}=\frac{1}{2}\int d^3\bm{q}\frac{d\Gamma(\bm{q})}{d^3\bm{q}}\bm{q}_{\perp}^2,
\end{eqnarray}
where $d\Gamma/d^3\bm{q}$ represents the  HQ  scattering rate with thermal partons per unit volume of momentum transfer~$\bm{q}$. 
For the slowly moving heavy quarks, the dominant interactions in the magnetized QGP medium arise from  Coulomb scatterings, i.e., $HQ (p)+q/\bar{q}/g(k)\to HQ(p')+q/\bar{q}/g(k')$ ($t$-channel), as depicted in the upper panel of Fig.~\ref{fig_HQ_diffusion}. The general expression for the differential HQ scattering rate with thermal light quarks and gluons is given by
\begin{eqnarray}
\frac{d\Gamma}{d^3\bm{q}}=\frac{d\Gamma^{\rm quark}}{d^3\bm{q}}+\frac{d\Gamma^{\mathrm{gluon}}}{d^3\bm{q}}.
\end{eqnarray} 
In leading order perturbative QCD calculations, the  HQ scattering rate the QGP constituents via one-gluon exchange can be derived by evaluating the imaginary part of the HQ self-energy with the effective gluon propagator in the space-like region~\cite{Beraudo:2007ky, Fukushima:2015wck}, as illustrated in Fig.~\ref{fig_HQ_diffusion}.
As a consequence, we need to consider the imaginary part of the effective gluon propagator, i.e., Eq.~(\ref{eq:IMGR00}). Following Ref.~\cite{Fukushima:2015wck}, the contribution from perturbative QCD interactions to the HQ scattering rate with light quarks in the magnetic field is computed as
\begin{align}
\dfrac{d\Gamma_{\mathrm{HTL}}^{\rm quark}}{d^3{\bm q}}=&\dfrac{d\Gamma_{\mathrm{HTL}}^{\rm LLL~quark}}{d^3{\bm q}}+\dfrac{d\Gamma_{\mathrm{HTL}}^{\rm HLL~ quark}}{d^3{\bm q}},\label{eq:Gamma_quark_general}\\
=& \frac{C_R^{\mathrm{HQ}}g_s^2}{(2\pi)^3}\underset{q_0\rightarrow 0}
{\text{lim}}\frac{2T}{q_0}
\bigg[\frac{\mathrm{Im}\Pi_{R,\mathrm{ quark},T=0,n=0}^{00}}{({\bm{q}^2}+\mathrm{Re}\Pi_{R}^{00})^2}\nonumber\\
&+\frac{\mathrm{Im}\Pi_{R,\mathrm{quark},T\neq 0,n\geq 1}^{00}}{({\bm{q}^2}+\mathrm{Re}\Pi_{R}^{00})^2}\bigg]\label{eq:Gamma_quark_v0},
\end{align} 
where $C_R^{\mathrm{HQ}}=\frac{N_c^2-1}{2N_c}$ is the Casimir factor of heavy quark. The first (second) term in Eq.~(\ref{eq:Gamma_quark_general}) and Eq.~(\ref{eq:Gamma_quark_v0}) corresponds to the HQ scattering rate with the thermal LLL (HLL) quarks.
Utilizing the non-perturbative term of the effective gluon propagator in Eq.~(\ref{eq:IMG}), we can also obtain the contribution from the non-perturbative interactions to the HQ scattering rate with thermal LLL (HLL) quarks, 
\begin{eqnarray} 
\dfrac{d\Gamma^{\rm LLL~quark}_{\mathrm{String}}}{d^3{\bm q}}
&=& \dfrac{d\Gamma^{\rm LLL~quark}_{\mathrm{HTL}}}{d^3{\bm q}}\frac{2m_{G}^2}{\bm{q}^2+\mathrm{Re}\Pi_{R}^{00}},\label{eq:LLL_string}\\
\dfrac{d\Gamma^{\rm HLL~quark}_{\mathrm{String}}}{d^3{\bm q}}
&=& \dfrac{d\Gamma^{\rm HLL~quark}_{\mathrm{HTL}}}{d^3{\bm q}}\frac{2m_{G}^2}{\bm{q}^2+\mathrm{Re}\Pi_{R}^{00}}
.\label{eq:HLL_string}
\end{eqnarray} 
In the magnetic field, the gluon-loop contribution to the imaginary part of the retarded gluon self-energy ($\mathrm{Im} \Pi^{00}_{R,\rm gluon}$)  remains the same as in the absence of a magnetic field. The contribution from the perturbative QCD interactions to the HQ scattering rate with thermal gluons in the  magnetic field has the same form as that in the zero magnetic field~\cite{Moore:2004tg}, except for the screened Coulomb amplitude, and it is given as
\begin{align}\label{eq:G_gluon}
\frac{d\Gamma^{\rm gluon}_{\mathrm{HTL}}}{d^3\bm{q}}=&4\alpha_{s}^2N_c C_{R}^{\rm HQ}\int \frac{d^3 \bm{k}}{(2\pi)^3}\delta(|\bm{k}|-|\bm{k}-\bm{q}|)\nonumber\\ &\times\frac{1}{(\bm{q}^2+\mathrm{Re} \Pi^{00}_{R})^2}(1+\cos^2\theta_{\bm{k}\bm{k'}})\nonumber\\
&\times f_{B}(|\bm{k}|)(1+f_{B}(|\bm{k}'|)).
\end{align} 
 Accordingly, the contribution from the non-perturbative interactions to the HQ scattering rate with thermal gluons is given as
\begin{equation}\label{rate_gluon_NP}
\dfrac{d\Gamma^{\rm gluon}_{\mathrm{String}}}{d^3{\bm q}}
= \dfrac{d\Gamma^{\rm gluon}_{\mathrm{HTL}}}{d^3{\bm q}}\frac{2m_{G}^2}{\bm{q}^2+\mathrm{Re}\Pi_{R}^{00}}.
\end{equation} 

\subsection{Quark contribution to HQ momentum diffusion coefficient}
\subsubsection{LLL quark contribution to HQ momentum diffusion coefficient}
By incorporating the first term  from Eq.~(\ref{eq:Gamma_quark_v0}) into Eq.~(\ref{eq:kappa}), we can investigate the magnetic field effect and longitudinal bulk viscous correction on the LLL quark contribution to the HQ momentum diffusion coefficient. 
At the perturbative level, the LLL quark-loop contribution to the imaginary part of gluon self-energy arises only from the vacuum component ($\mathrm{Im}\Pi^{00}_{R,\mathrm{quark}, T=0,n=0}$), and the delta function in Eq.~(\ref{eq:ImPiT=0}) results in a nonzero transverse HQ momentum diffusion coefficient from the LLL quark contribution, denoted as $\kappa_{\perp,\mathrm{HTL}}^{\rm LLL~quark}$, which is computed as 
\begin{align}\label{eq:kappa_11}
\kappa_{\perp,\mathrm{HTL}}^{\rm LLL~quark}&=
\frac{1}{2}\int d^3\bm{q}\frac{d\Gamma^{\rm LLL~quark}_{\rm HTL}}{d^3\bm{q}}\bm{q}_{\perp}^2,\\
&=\frac{1}{2}\alpha_sC_R^{\rm HQ} T\int_0^{\infty} dx\frac{\sum_{f}xs_f(x)}{(x+\widetilde{m}_D^2)^2}.
\end{align} 
Inserting Eq.~(\ref{eq:LLL_string}) into Eq.~(\ref{eq:kappa}), the LLL quark contribution to the non-perturbative term of transverse HQ momentum diffusion coefficient, denoted as $\kappa_{\perp,\mathrm{String}}^{\rm LLL~quark}$, is computed  as 
\begin{align}
\kappa_{\perp,\mathrm{String}}^{\rm LLL~quark}&=
\frac{1}{2}\int d^3\bm{q}\frac{d\Gamma^{\rm LLL~quark}_{\rm String}}{d^3\bm{q}}\bm{q}_{\perp}^2,\\
&=\frac{4\sigma}{2} T\int_0^{\infty}dx\frac{\sum_{f}xs_f(x)}{(x+\widetilde{m}_D^2)^3}\label{eq:kappa_12}.
\end{align} 
The physical explanation behind Eqs.~(\ref{eq:kappa_11}-\ref{eq:kappa_12}) is that due to the chirality conservation, the scattering process between heavy quarks and the LLL quarks is forbidden along the magnetic field direction, as a result, the LLL quarks will pass through the heavy quarks without transferring momentum. If the heavy quarks are moving along the direction perpendicular to the magnetic field, the transverse HQ diffusion coefficient becomes non-zero.

\subsubsection{HLL quark contribution to HQ momentum diffusion coefficient}
Next, we consider the HLL quark contribution to the HQ momentum diffusion coefficient. By inserting the second term from Eq.~(\ref{eq:Gamma_quark_v0}) into Eq.~(\ref{eq:kappa}), the HLL quark contribution to the perturbative term of transverse HQ momentum diffusion coefficient, denoted as $\kappa_{\perp,\mathrm{HTL}}^{\rm HLL~quark}$, is computed as
\begin{align}
\kappa_{\perp,\mathrm{HTL}}^{\rm HLL~quark}=&
\frac{1}{2}\int d^3\bm{q}\frac{d\Gamma^{\rm HLL~quark}_{\rm HTL}}{d^3\bm{q}}\bm{q}_{\perp}^2,\\
=&\sum_{f}\sum_{n=1 }^{\infty}\frac{C_{R}^{\rm HQ}\alpha_{s}}{2}\int_0^{\infty} dx m_{f,n}^2 s_f(x)\nonumber\\
&\times\int dq_z\frac{2}{|q_z|E^f_{q_z/2,n}}\frac{xH_{f,n}(T,eB,q_z)}{[q_z^2+x+\widetilde{m}_D^2]^2}.
\end{align}
Similarly, by incorporating Eq.~(\ref{eq:HLL_string}) into Eq.~(\ref{eq:kappa}), the HLL quark contribution to the non-perturbative term of transverse HQ momentum diffusion coefficient,  denoted as $\kappa_{\perp,\mathrm{String}}^{\rm HLL~quark}$, is computed as
\begin{align}
\kappa_{\perp,\mathrm{String}}^{\rm HLL~quark}=&
\frac{1}{2}\int d^3\bm{q}\frac{d\Gamma^{\rm HLL~quark}_{\rm String}}{d^3\bm{q}}\bm{q}_{\perp}^2,\\
=&\sum_{f}\sum_{n=1}^{\infty}\frac{4\sigma}{2}\int_0^{\infty} dx m_{f,n}^2 s_f(x)\nonumber\\
&\times\int dq_z\frac{2}{|q_z|E^f_{q_z/2,n}}\frac{xH_{f,n}(T,eB,q_z)}{[q_z^2+x+\widetilde{m}_D^2]^3}\label{eq:kappa_perp_NP}.
\end{align}
Beyond the LLL approximation, the HLL quark contribution to the longitudinal HQ momentum diffusion coefficient is non-zero, and its perturbative term can be written as 
\begin{align}
\kappa_{\|,\mathrm{HTL}}^{\rm HLL~quark}=&
\int d^3\bm{q}\frac{d\Gamma^{\rm HLL~quark}_{\rm HTL}}{d^3\bm{q}}q_z^2,\\
=&\sum_{f}\sum_{n=1}^{\infty}C_{R}^{\rm HQ}\alpha_{s}\int_0^{\infty} dx  m_{f,n}^2 s_f(x)
\nonumber\\
&\times\int dq_z\frac{q_z^2}{|q_z|E^f_{q_z/2,n}} \frac{H_{f,n}(T,eB,q_z)}{[q_z^2+x+\widetilde{m}_D^2]^2}.
\end{align}
The HLL quark contribution to the non-perturbative term of longitudinal HQ momentum diffusion coefficient, denoted as $\kappa_{\|,\mathrm{String}}^{\rm HLL~quark}$, is also computed as 
\begin{align}
\kappa_{\|,\mathrm{String}}^{\rm HLL~quark}=&
\int d^3\bm{q}\frac{d\Gamma^{\rm HLL~quark}_{\rm String}}{d^3\bm{q}}q_z^2,\\
=&\sum_{f}\sum_{n=1}^{\infty}4\sigma\int_0^{\infty} dx  m_{f,n}^2 s_f(x)
\nonumber\\
&\times\int dq_z\frac{q_z^2}{|q_z|E^f_{q_z/2,n}} \frac{H_{f,n}(T,eB,q_z)}{[q_z^2+x+\widetilde{m}_D^2]^3}\label{eq:kappa_para_NP}.
\end{align}

\subsection{Gluon contribution to HQ momentum diffusion coefficient}
\begin{figure*}[htpb]
	\centering
	\subfloat{\includegraphics[scale=0.45]{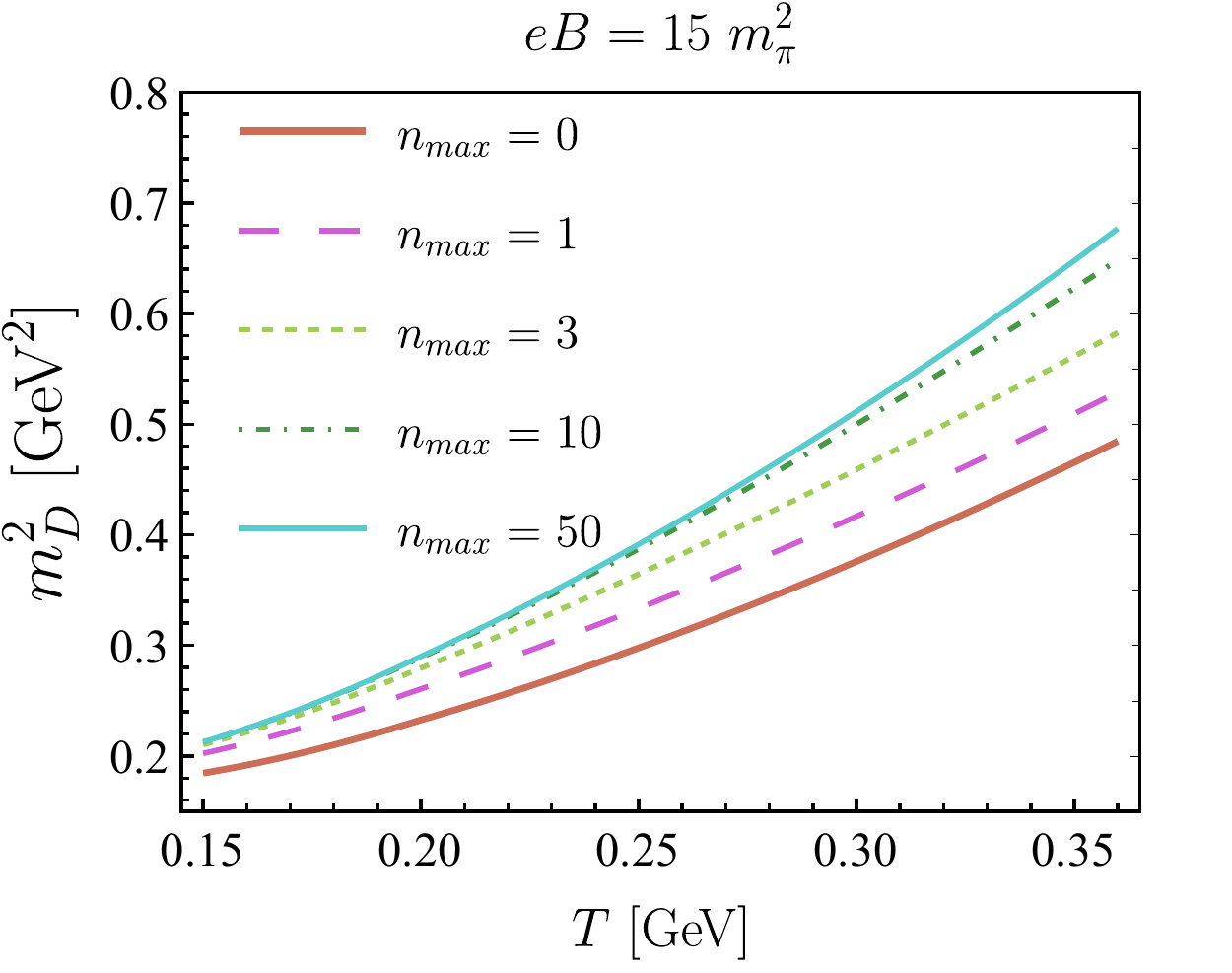}}\hspace{.01 cm}
	\subfloat{\includegraphics[scale=0.45]{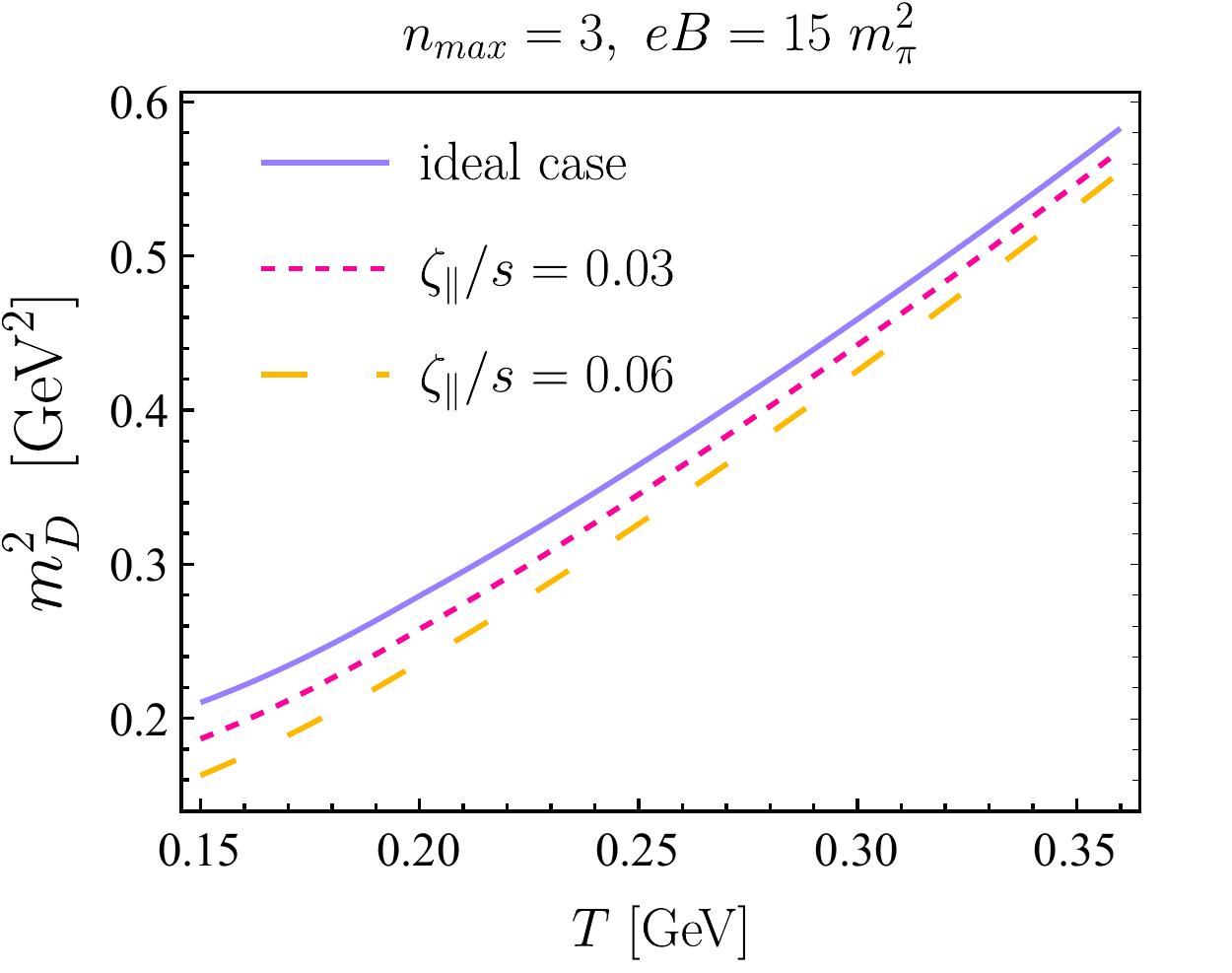}}\hspace{.01 cm}
	\subfloat{\includegraphics[scale=0.45]{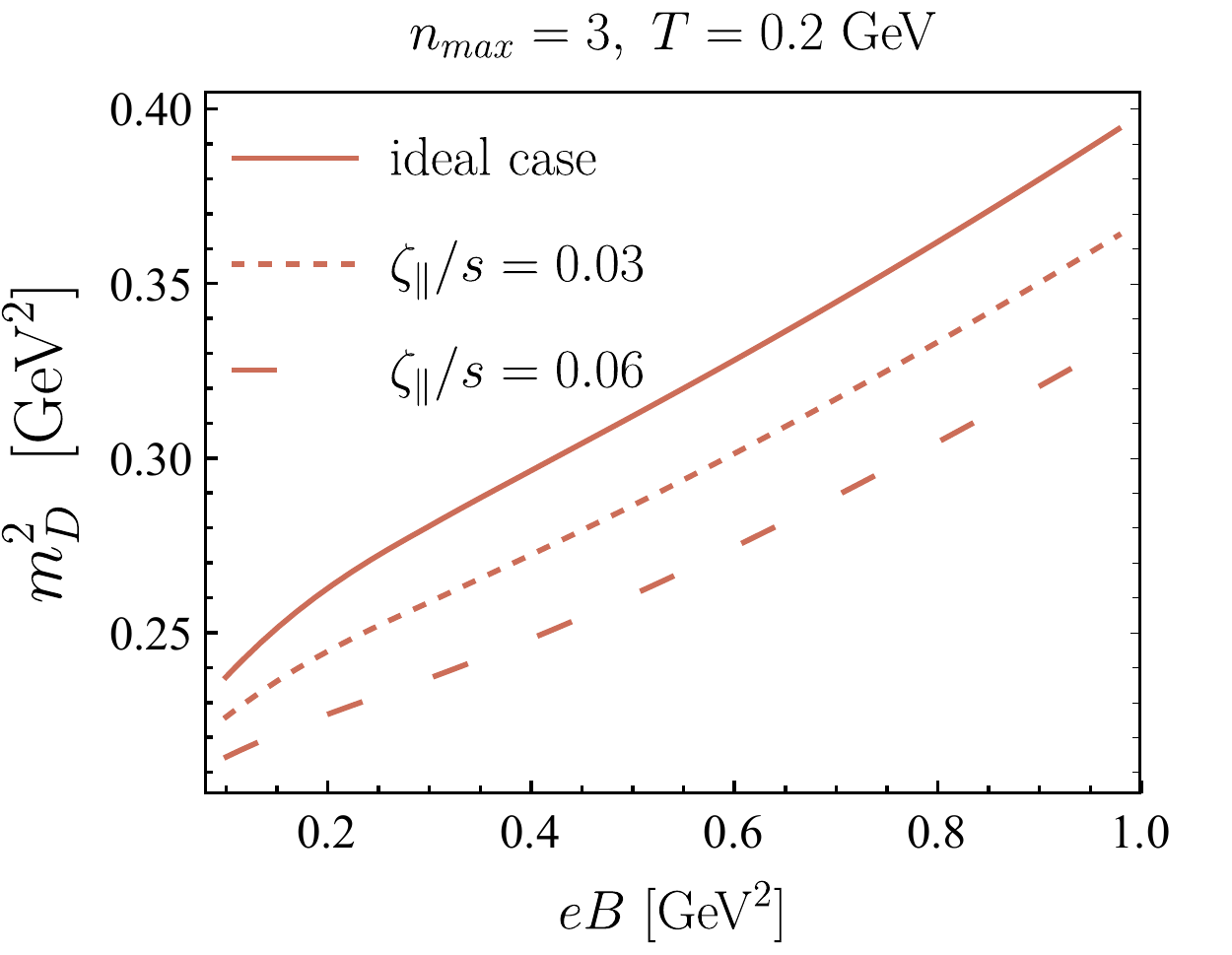}}\hspace{.01 cm}
	\caption{(Left panel) Temperature ($T$) dependence of squared Debye mass $m_D^2$ for different numbers of maximum Landau level ($n_{max}$) at $eB=15~m_\pi^2$. (Middle panel) Temperature dependence of  $m_D^2$ for different values of scaled longitudinal bulk viscosities ($\zeta_{\|}/s$) at $eB=15~m_\pi^2$ and $n_{max}=3$. (Right panel) 
	The variation of $m_D^2$ with magnetic field ($eB$) for different values of $\zeta_{\|}/s$ at fixed $T=0.2$~GeV and $n_{max}=3$.
	}
	\label{fig_Debyemass}
\end{figure*} 
Compared to the quark contribution, the estimation of the thermal gluon contribution to the HQ momentum diffusion coefficient is relatively straightforward, as the effect of magnetic field and the longitudinal bulk viscous correction are only incorporated through the screened Coulomb amplitude in Eq.~(\ref{eq:G_gluon}).
Therefore,  utilizing the relation $\delta (\bm{k}-|\bm{k}-\bm{q}|)=|\bm{q}|^{-1}\delta [\cos \theta_{\bm{k}\bm{q}}-|\bm{q}|/(2|\bm{k}|)]\Theta (|\bm{k}|-|\bm{q}|/2)$ and $\cos\theta_{\bm{k}\bm{k'}}=1-|\bm{q}|^2/(2|\bm{k}|^2)$ 
into Eq.~(\ref{eq:G_gluon}), the gluon contribution to the perturbative term of longitudinal or transverse HQ momentum diffusion coefficient,  denoted as $\kappa^{\rm gluon}_{\|/\perp,\mathrm{HTL}}$, can be written as follows:
\begin{align}
\kappa^{\rm gluon}_{\|/\perp,\mathrm{HTL}}=&
\frac{1}{3}\int d^3\bm{q}\frac{d\Gamma^{\rm gluon}_{\rm HTL}}{d^3\bm{q}}\bm{q}^2,
\\
=&\dfrac{4}{{3}\pi }\alpha_{s}^2N_cC_R^{\mathrm{HQ}}
\int_{0}^{\infty} |\bm{k}|^2d|\bm{k}|\int_{0}^{2|\bm{k}|}|\bm{q}|d|\bm{q}|f^0_B(|\bm{k}|)\nonumber\\
&\times(1+ f^0_B(|\bm{k}|))\dfrac{|\bm{q}|^2}{\big(\bm{q}^2+\widetilde{m}_D^2\big)^3}\nonumber\\
&\times\Big[2-\frac{|\bm{q}|^2}{|\bm{k}|^2}+\frac{|\bm{q}|^4}{4|\bm{k}|^4}\Big].
\end{align}

By inserting Eq.~(\ref{rate_gluon_NP}) into Eq.~(\ref{eq:kappa}), the gluon contribution to the non-perturbative term of longitudinal or transverse HQ momentum  diffusion coefficient,  denoted as $\kappa^{\rm gluon}_{\|/\perp,\mathrm{String}}$, can also be computed as
\begin{align}
\kappa^{\rm gluon}_{\|/\perp,\mathrm{String}}=&
\frac{1}{3}\int d^3\bm{q}\frac{d\Gamma^{\rm gluon}_{\rm String}}{d^3\bm{q}}\bm{q}^2,\\
=&\dfrac{16\sigma}{{3}}\pi \alpha_{s}N_c
\int_{0}^{\infty} |\bm{k}|^2d|\bm{k}|\int_{0}^{2|\bm{k}|}|\bm{q}|d|\bm{q}|f^0_B(|\bm{k}|)\nonumber\\
&\times(1+ f^0_B(|\bm{k}|))\dfrac{|\bm{q}|^2}{\big(\bm{q}^2+\widetilde{m}_D^2\big)^3}\nonumber\\
&\times\Big[2-\frac{|\bm{q}|^2}{|\bm{k}|^2}+\frac{|\bm{q}|^4}{4|\bm{k}|^4}\Big]\label{eq:kappa_gluon_NP}.
\end{align}
Here, we employ $q_z^2=|\bm{q}|^2/3$ due to the approximated rotational symmetry for thermal gluons. 
Later, we would see that the temperature-dependent form of string tension ($\sigma$) plays a pivotal role in determining the thermal evolution of the non-perturbative contribution to the HQ momentum diffusion coefficient.

\section{numerical Results and Discussions}\label{sec:discussions}
\begin{figure*}[htbp]
	\centering
	\subfloat{\includegraphics[scale=0.4]{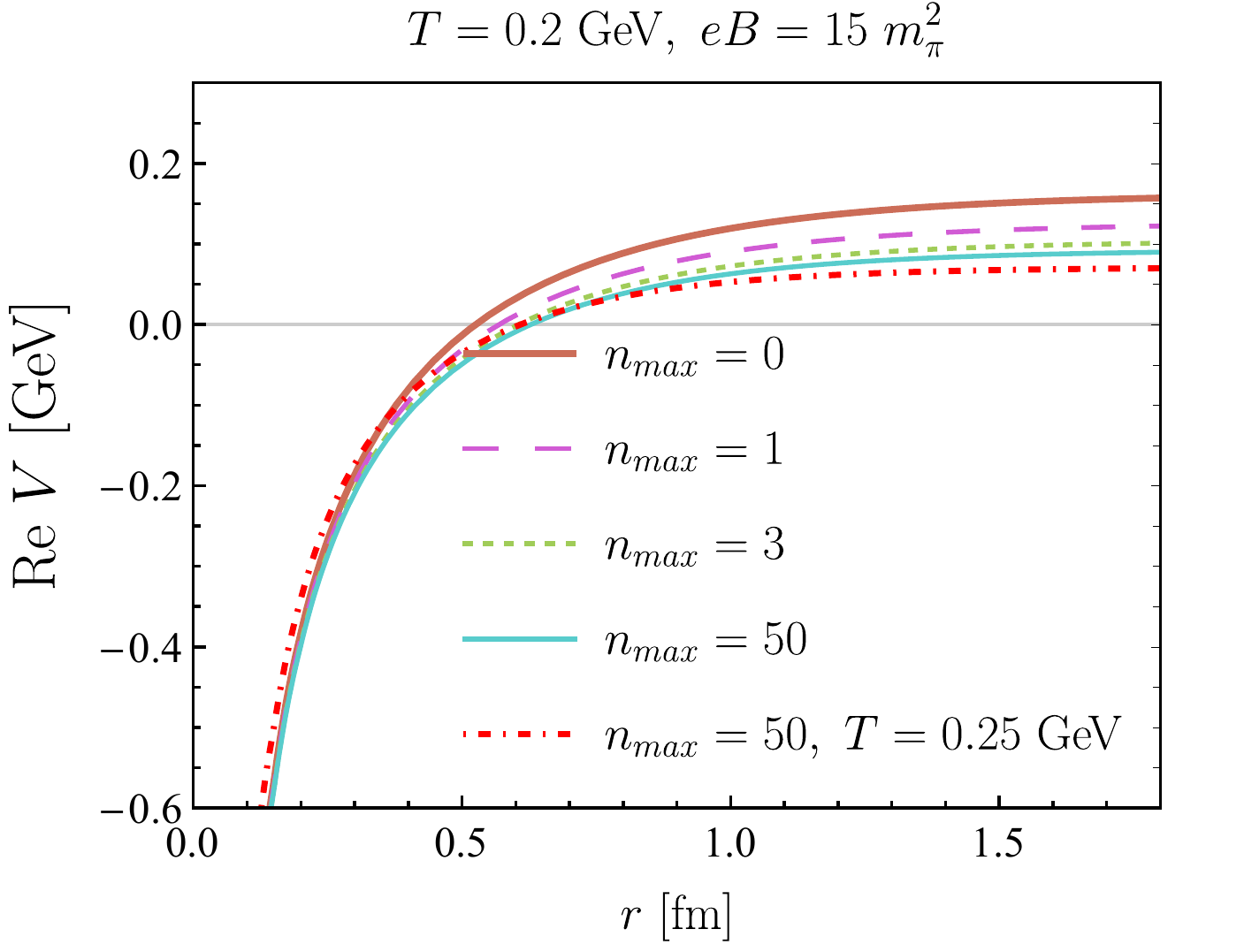}}\hspace{.05 cm}
	\subfloat{\includegraphics[scale=0.45]{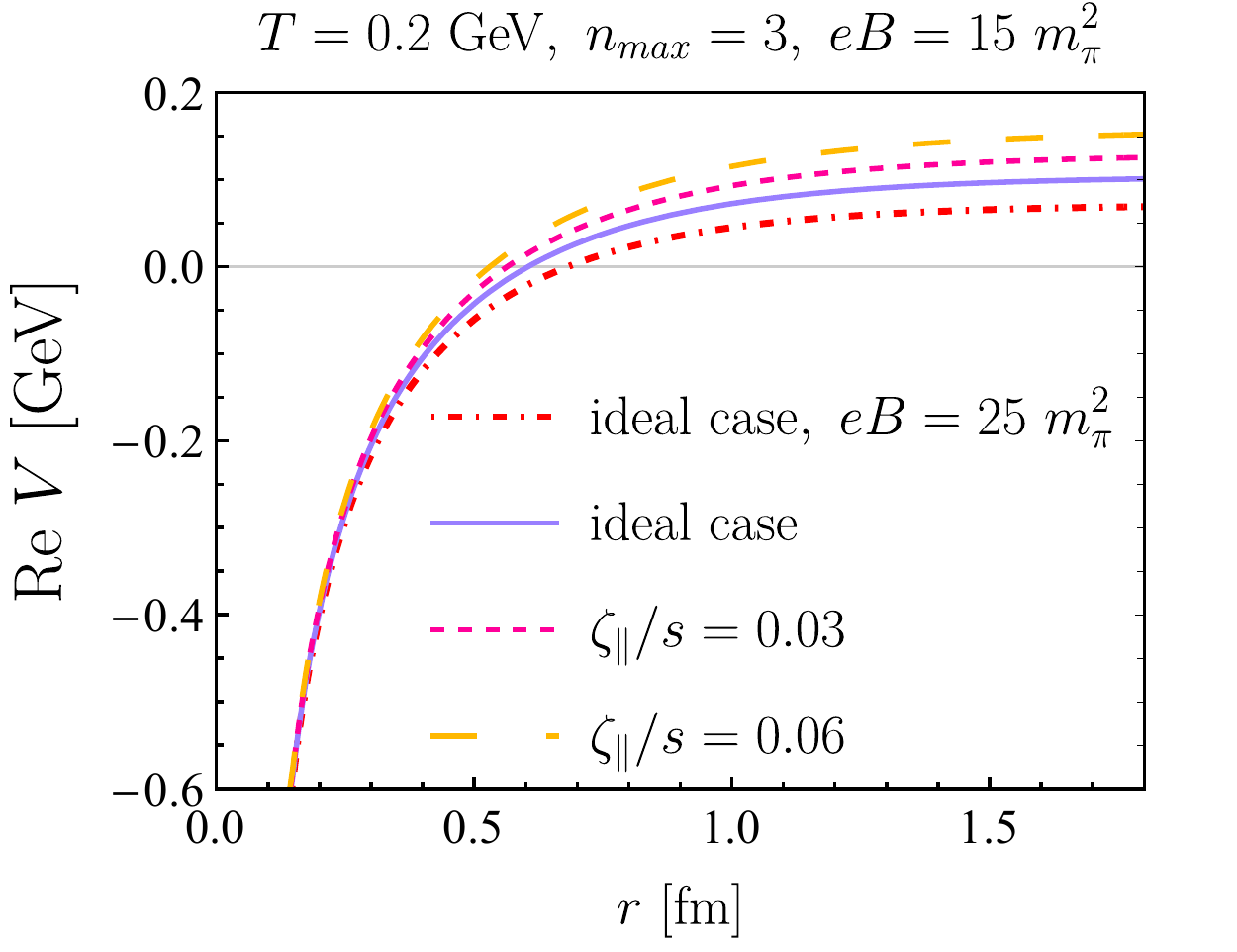}}\hspace{.05 cm}
	\subfloat{\includegraphics[scale=0.45]{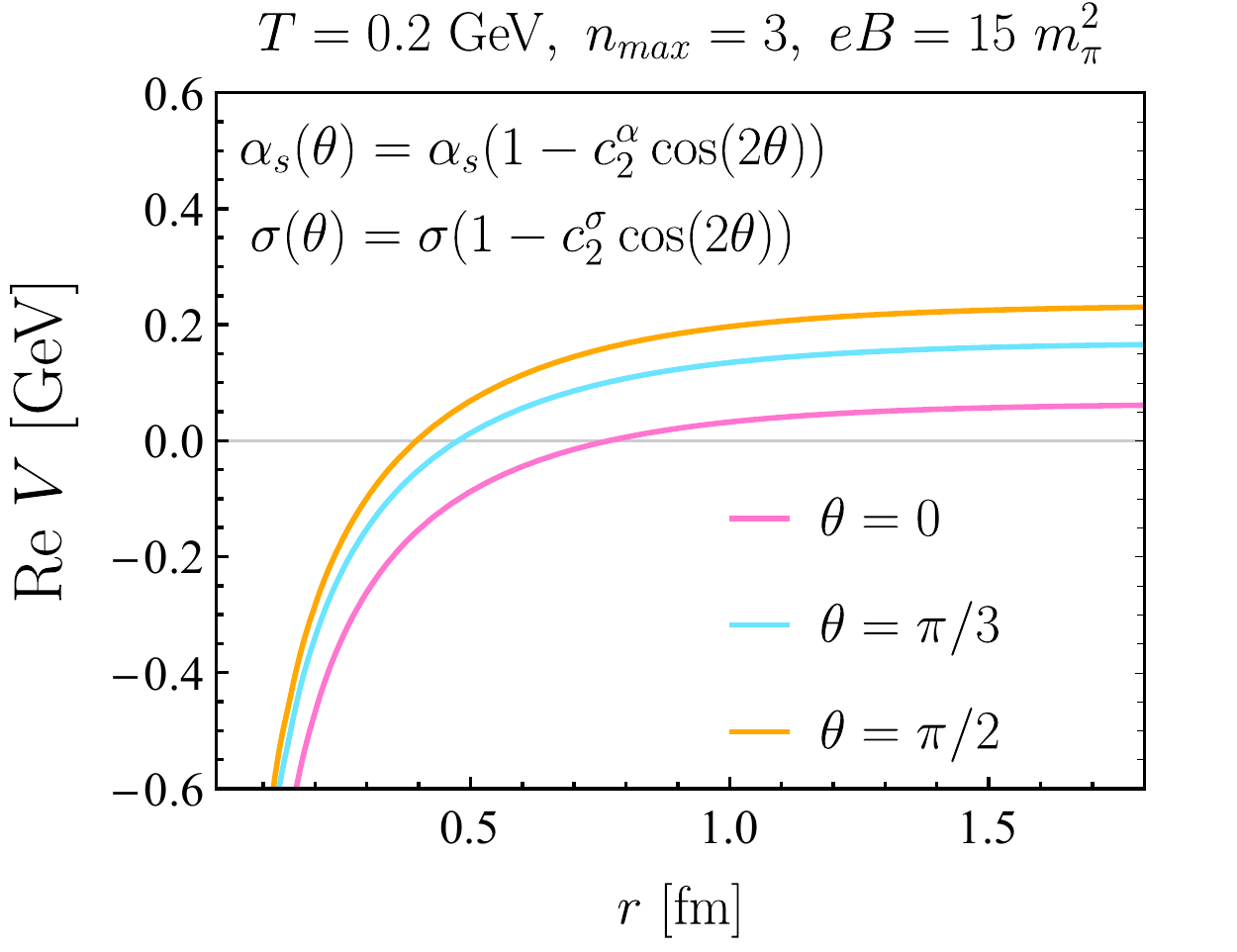}}
	\caption{(Left panel) The variation of the real part of the HQ potential $\mathrm{Re }V$ with quark-antiquark separation distance ($r$) for different numbers of maximum Landau level ($n_{max}$) and different temperatures ($T$) at  $eB=15~m_{\pi}^2$. (Middle panel) The variation of the $\mathrm{Re}V$ with $r$  for different scaled longitudinal bulk viscosities ($\zeta_{\|}/s$ )and different magnetic fields ($eB$) at fixed $T=0.2~\mathrm{GeV}$ and $n_{max}=3$.  (Right panel) The variation of the $\mathrm{Re }V$ with $r$ for several values of the angle ($\theta$) between the quark-antiquark separation and  the magnetic field direction, where the parameterized forms of the angular-dependent strong coupling constant and string tension are quoted from Ref.~\cite{Bonati:2016kxj}.}
	\label{fig_ReV}
\end{figure*} 
\begin{figure*}[htpb]
	\centering
	\subfloat{\includegraphics[scale=0.6]{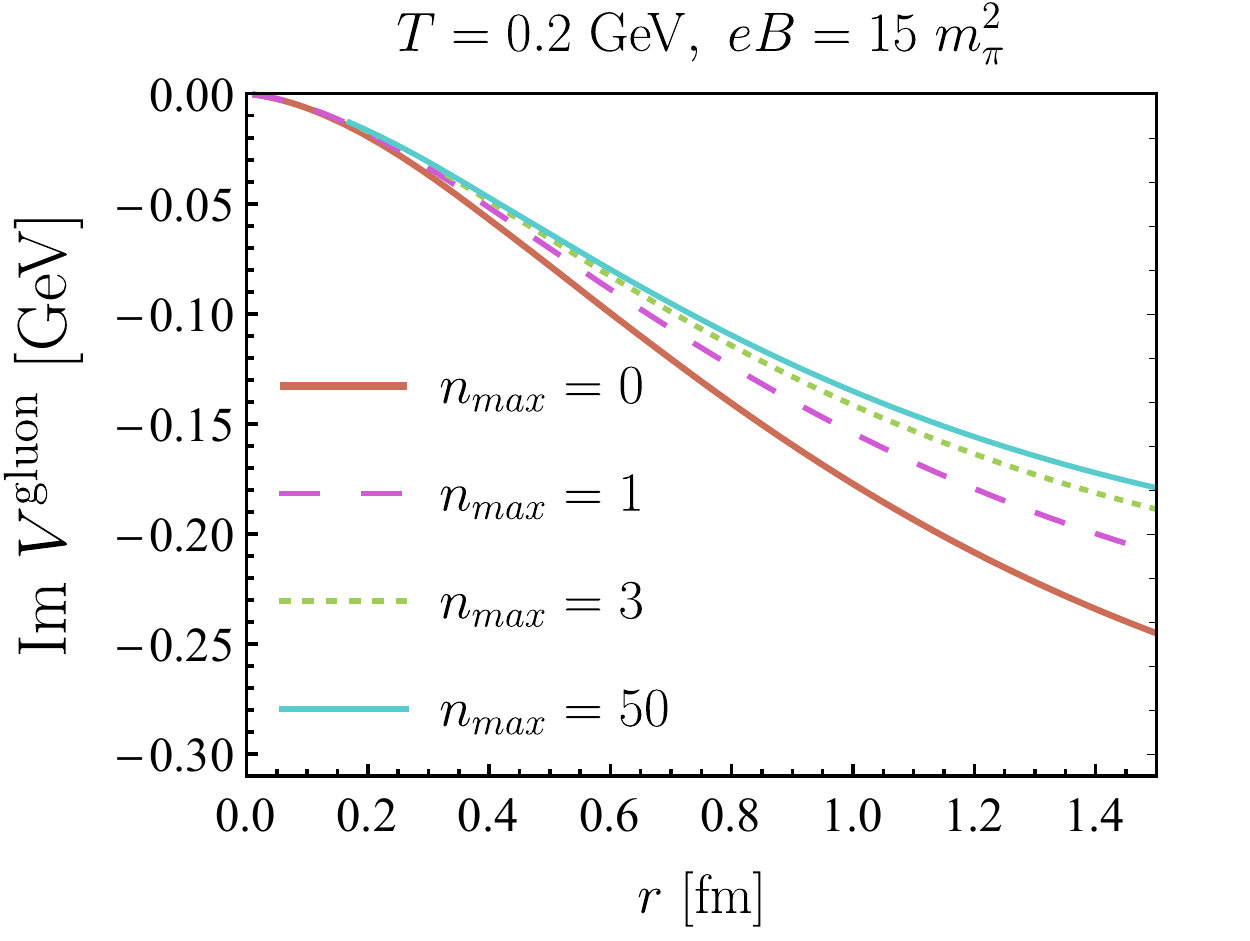}}\hspace{.1 cm}
	\subfloat{\includegraphics[scale=0.6]{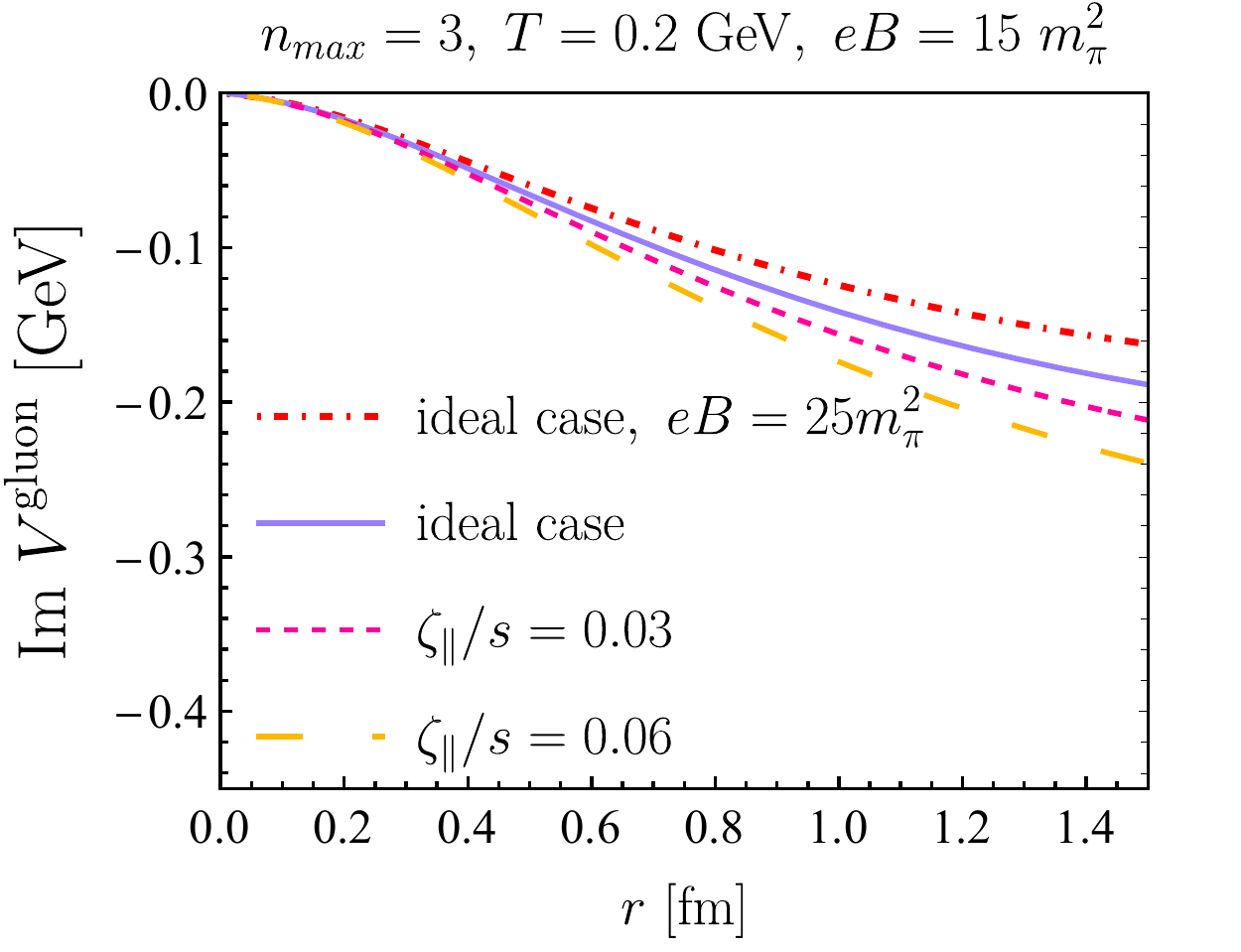}}\hspace{.1 cm}
	\caption{(Left panel) The variation of the imaginary part of the HQ potential related to gluon-loop contribution of the gluon self-energy $\mathrm{Im}V^{\rm gluon}$ with quark-antiquark separation distance ($r$) for different numbers of maximum Landau level ($n_{max}$) at fixed $T=0.2~\mathrm{GeV}$ and $eB=15m_{\pi}^2$. (Right panel) The variation of the $\mathrm{Im}V^{\rm gluon}$  with  $r$ for different scaled longitudinal bulk viscosities ($\zeta_{\|}/s$) and different magnetic fields at fixed $T=0.2~\mathrm{GeV}$. 
	}
	\label{fig_IMV_gluon}
\end{figure*} 

In this work, all numerical calculations are performed in the massless limit, i.e., $m_{u}=m_{d}=m_{s}=0$ at zero chemical potential. We first focus on the characteristics of the Debye mass, which can help us better understand the qualitative behaviors of both HQ  potential and HQ momentum diffusion coefficient.  In the left panel of Fig.~\ref{fig_Debyemass}, we illustrate the temperature dependence of squared Debye mass $m_{D}^2$ at a fixed magnetic field $eB=15~m_{\pi}^2$ for different numbers of the maximum Landau levels ($n_{max}$).  Consistent with most pioneering studies, we observe that the screening effect of the QGP medium increases with the temperature. 
The increase of $n_{max}$ can induce a decrease in screening length (or equivalently, an increase in the screening mass), which accelerates the dissociation of the quarkonia states. When the Landau levels exceed $50$, subsequent increases in $n_{max}$ have an invisible impact on $m_{D}^2$. To further investigate the longitudinal bulk viscous correction  to both HQ potential and HQ momentum diffusion coefficient, we consider several scaled longitudinal bulk viscosities ($\zeta_{\|}/s=0,~0.03,~0.06$). From the middle panel of Fig.~\ref{fig_Debyemass}, we observe that the longitudinal bulk viscous correction can suppress the color screening effect. Furthermore, as the magnetic field increases, the longitudinal bulk viscous correction  to the Debye mass becomes more pronounced, as demonstrated in the right panel of Fig.~\ref{fig_Debyemass}.


\subsection{Results of HQ potential}
\begin{figure*}[htpb]
	\centering
	\subfloat{\includegraphics[scale=0.51]{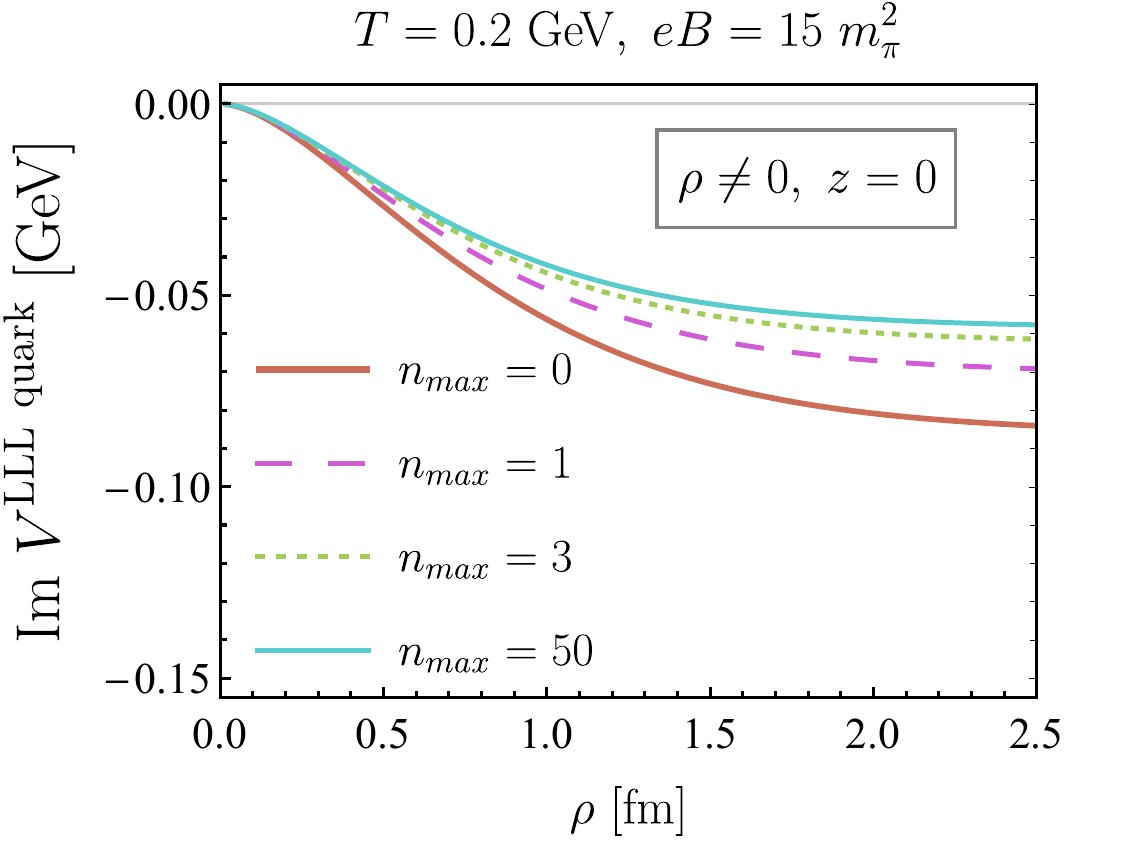}}\hspace{.01 cm}
	\subfloat{\includegraphics[scale=0.51]{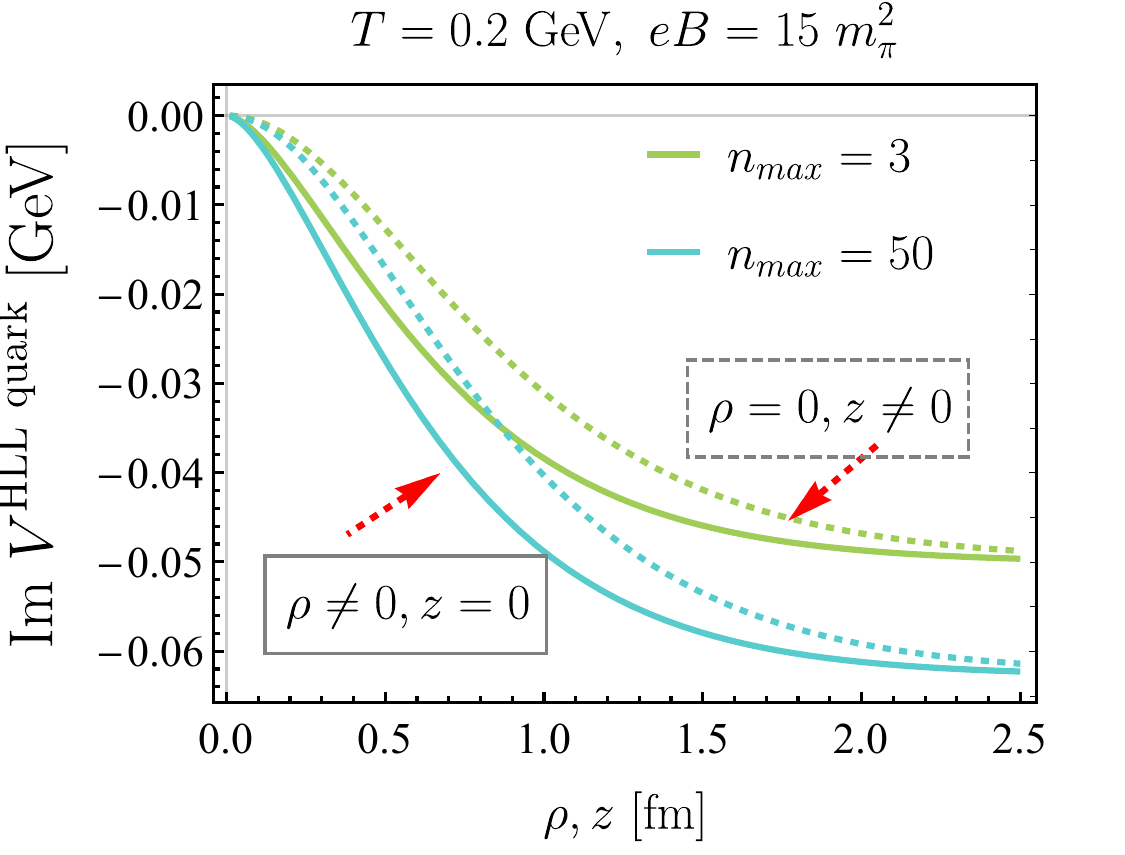}}\hspace{.01 cm}
	\subfloat{\includegraphics[scale=0.51]{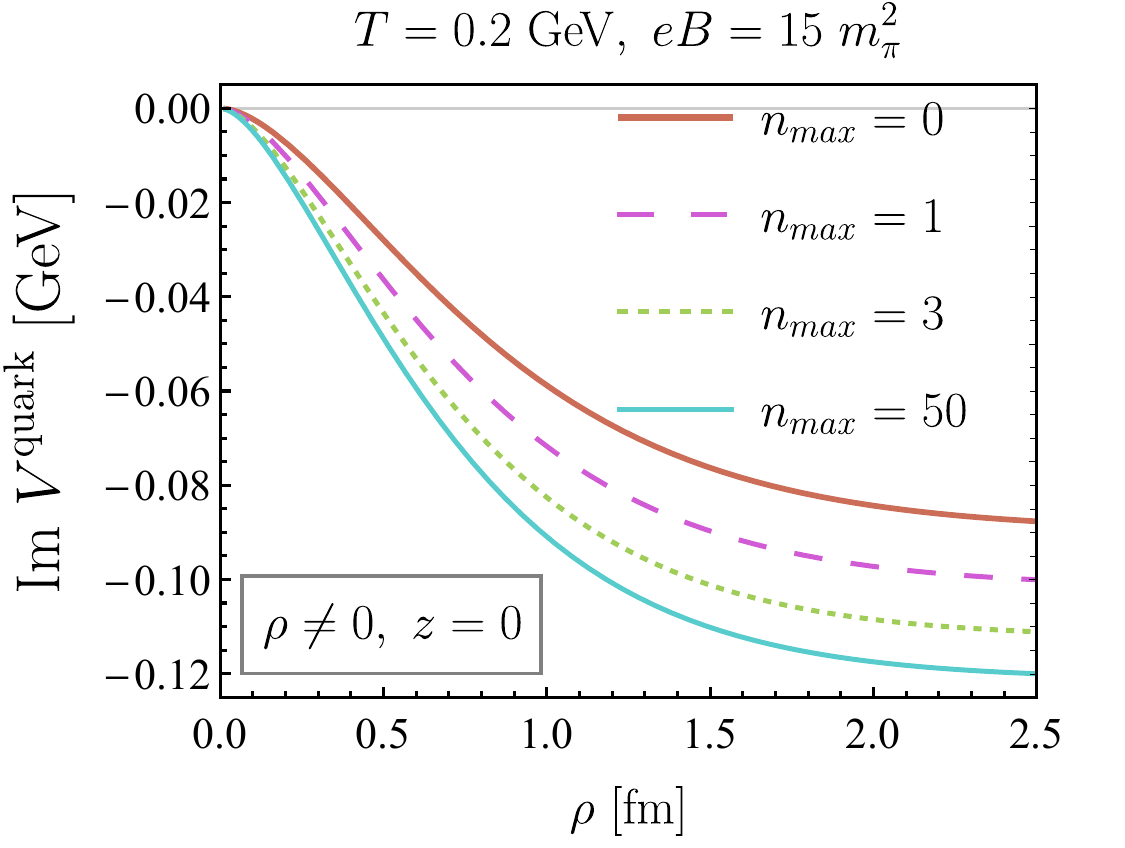}}
	\caption{(Right panel) The variation of the imaginary part of the HQ potential related to the LLL quark-loop contribution of the gluon self-energy $\mathrm{Im}V^{\rm LLL~quark}$ with respect to quark-antiquark separation distances for different numbers of maximum Landau level ($n_{max}$). (Middle panel) The variation of the imaginary part of the HQ  potential related to the HLL quark-loop contribution of the gluon self-energy $\mathrm{Im}V^{\rm HLL~quark}$ with separation distances for different $n_{max}$, where the solid lines and dotted lines represent the cases for the quark-antiquark separations perpendicular ($\rho\neq 0, z=0$) and parallel ($\rho= 0, z\neq 0$) to the  magnetic field direction, respectively. (Right panel) The variation of the imaginary part of the HQ potential related to total quark-loop contributions of the gluon self-energy $\mathrm{Im}V_{}^{\rm quark}=\mathrm{Im}V^{\rm LLL~quark}+\mathrm{Im}V^{\rm HLL~quark}$  with quark-antiquark separation distances for different $n_{max}$ in case of $\rho\neq 0,~z=0$. All the computations are performed at fixed $T=0.2~\mathrm{GeV}$ and $eB=15~m_{\pi}^2$. }
	\label{fig_IMV}
\end{figure*} 

By incorporating the HTL resummed gluon propagator and the gluon propagator induced by dimension-two gluon condensates, we investigate the characteristics of in-medium HQ potential in the magnetic field. 
Due to uncertainty in the exact temperature, magnetic field,  and angular-dependent form of string tension ($\sigma$), we first employ a constant $\sigma$, i.e., $\sigma=0.44~\mathrm{GeV}^2$~\cite{Satz:2005hx} for simplicity.
In the left panel of Fig.~\ref{fig_ReV}, we present the variation of the real part of the potential, $\mathrm{Re }V$, with respect to the quark-antiquark separation distance ($r$) for different numbers of maximum Landau level ($n_{max}$). The numerical calculation is performed at $T=0.2~\mathrm{GeV}$ and $eB=15~m_{\pi}^2$.  We see that the $\mathrm {Re }V $ flattens with increasing $r$. Neglecting the asymmetry factor in Eq.~(\ref{eq:GF00_v2}),  there is no additional source of anisotropy in the $\mathrm{Re}V$ apart from the string tension $\sigma$ and strong coupling constant $\alpha_s$. 
Furthermore, as both $n_{max}$ and $T$ increase, $\mathrm{Re}V$ decreases, as a consequence, the binding energy of heavy quark bound states decreases, thereby the formation of quarkonia states is suppressed. 
Since the longitudinal bulk viscous correction significantly depresses the screening effect, thereby enhancing the $\mathrm{Re}V$ as shown in the middle panel of Fig.~\ref{fig_ReV}. We also attempt to utilize the parameterization of angular-dependent strong coupling constant $\alpha_{s}(\theta)$ and string tension $\sigma (\theta)$ proposed in Ref.~\cite{Bonati:2016kxj}, to explore the anisotropic response of the $\mathrm{Re}V$ to the magnetic field. As depicted in the right panel of Fig.~\ref{fig_ReV}, the $\mathrm{Re}V$ exhibits anisotropic behavior. 
 In particular, at large separation distances,  the $\mathrm{Re}V$  when the quark-antiquark dipole axis is parallel to the magnetic field direction (i.e.,  $\theta=0$) is smaller than when it is perpendicular to the magnetic field direction, (i.e., $\theta=\pi/2$).

\begin{figure*}[htpb]
	\centering
	\subfloat{\includegraphics[scale=0.51]{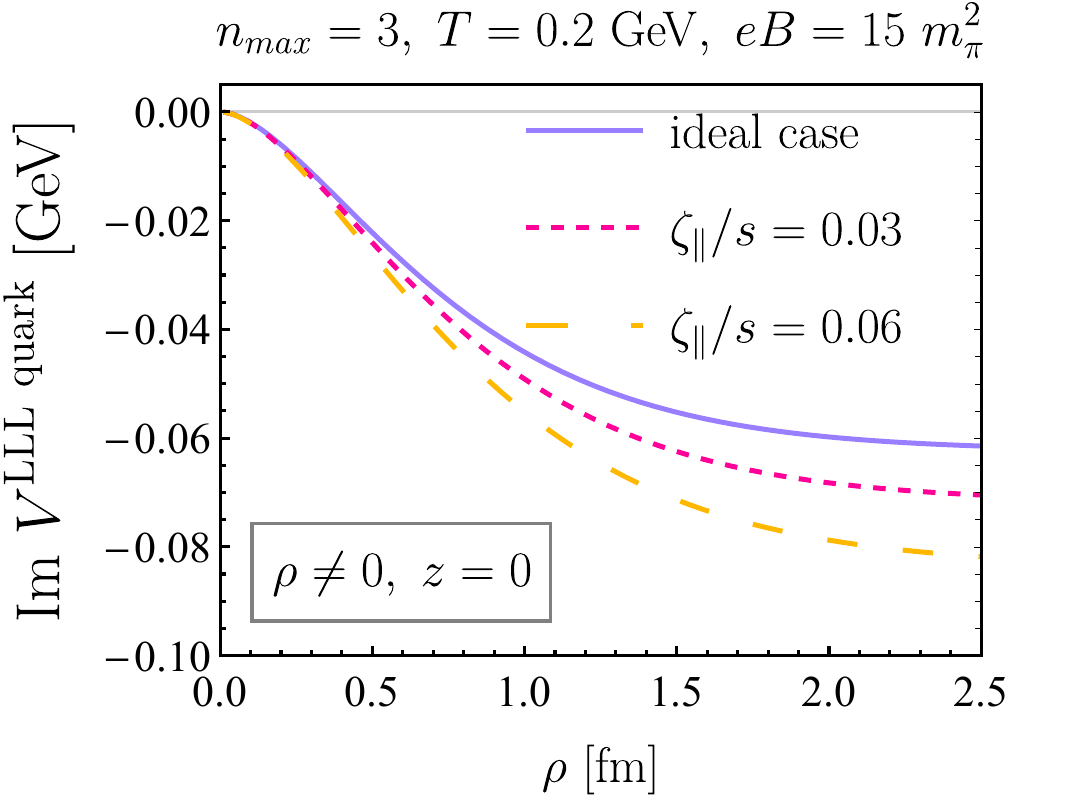}}
	\subfloat{\includegraphics[scale=0.51]{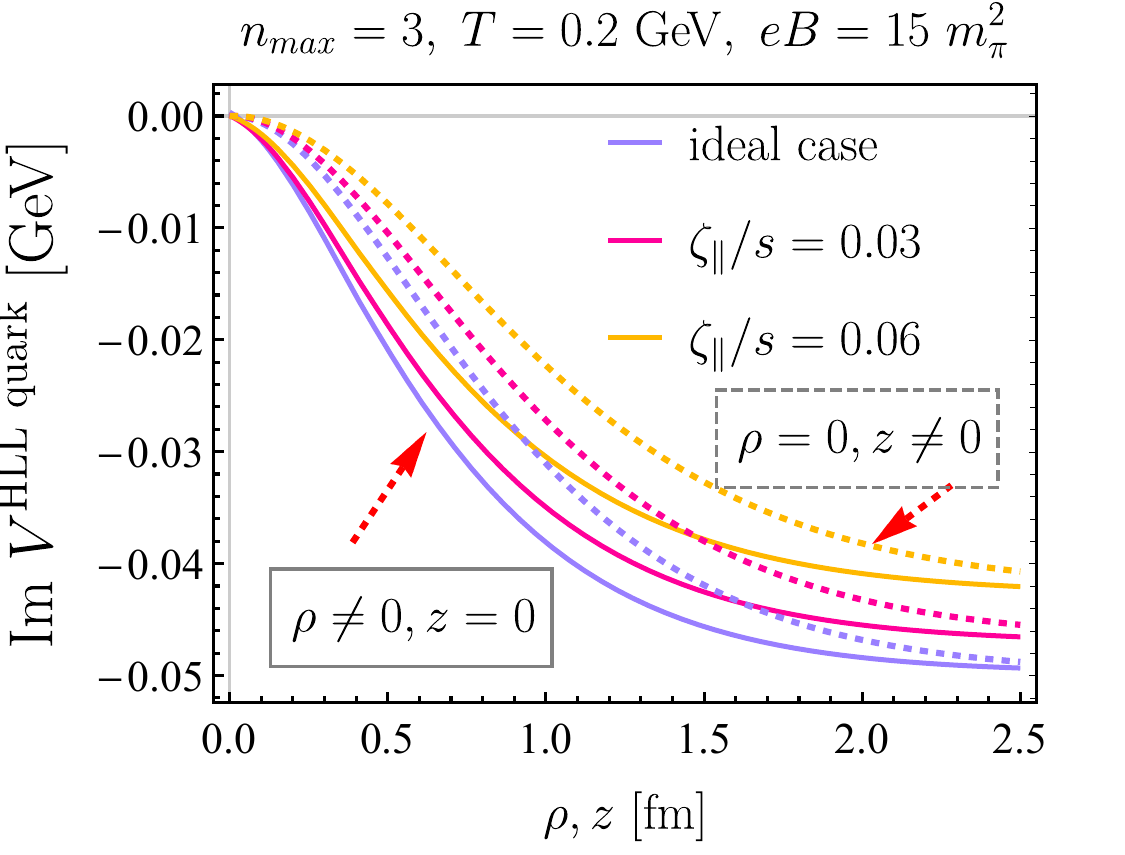}}
	\subfloat{\includegraphics[scale=0.51]{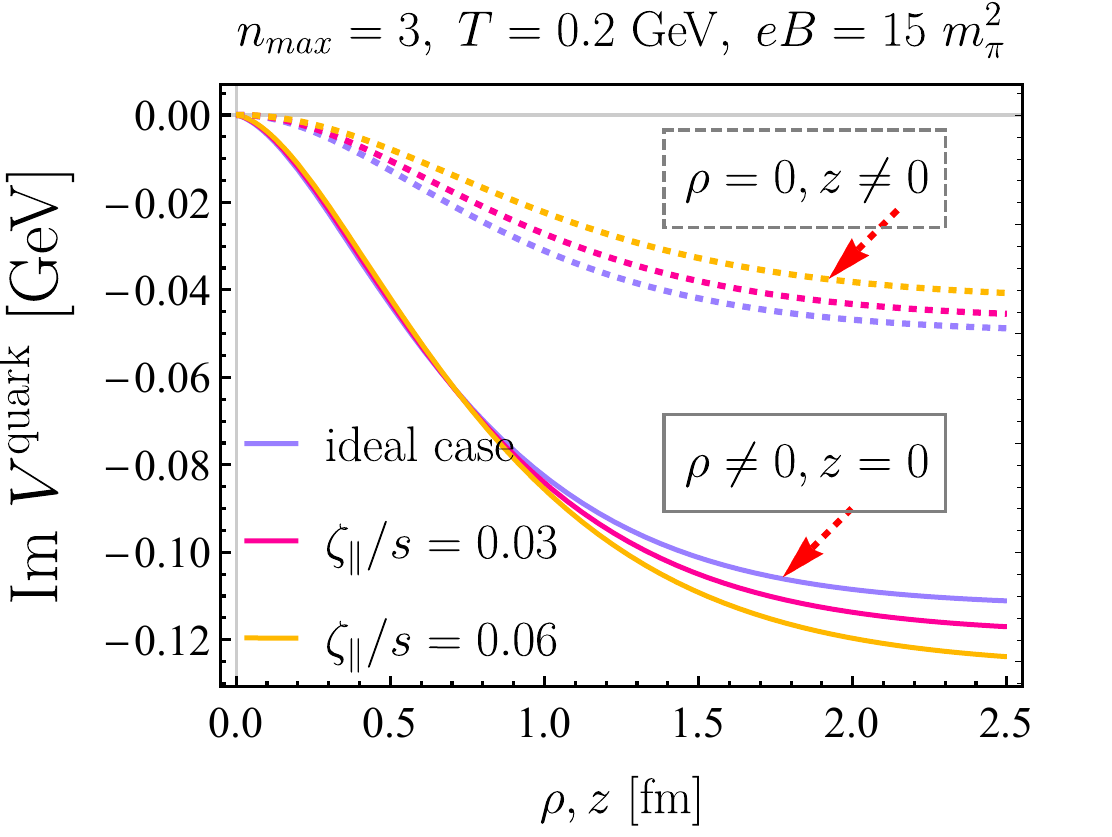}}
	\caption{Similar to Fig.~\ref{fig_IMV} but for different scaled longitudinal bulk viscosities ($\zeta_{\|}/s$) at fixed $n_{max}=3$.
	}
	\label{fig_IMV_bulk}
\end{figure*} 

In the left panel of Fig.~\ref{fig_IMV_gluon} we present the variation of the imaginary part of the HQ potential related to gluon-loop contribution of the gluon self-energy, $\mathrm{Im}~V^{\rm gluon}$, with respect to quark-antiquark separation distances ($r$) for different numbers of maximum Landau level ($n_{max}$) at $eB=15m_{\pi}^2$ and $T=0.2$ GeV. It is observed that the magnitude of $\mathrm{Im}V^{\rm gluon}$ decreases with increasing $n_{max}$, which qualitatively agrees with the studies of the HQ potential using dielectric permittivity~\cite{Ghosh:2022sxi}. \footnote{In the appendix of Ref.~\cite{Ghosh:2022sxi}, the result of $\mathrm{Im}V^{\rm gluon}$ should be corrected by multiplying an overall factor $m_{D,g}^2/m_{D}^2$ instead of just replacing the Debye mass in the expression of $\mathrm{Im}V^{\rm gluon}$ for the zero magnetic field with a magnetic field-dependent Debye mass.
Consequently, the expression of $\mathrm{Im}V^{\rm gluon}$ in a nonzero magnetic field should be rewritten as:
\begin{equation}
\mathrm{Im~V}^{\rm gluon}(r)=-C_F\alpha_s T\frac{m_{D,g}^2}{m_D^2}\phi_2(m_Dr)-\frac{2\sigma T m_{D,g}^2}{m_D^4}\chi(m_D r),
\end{equation} 
with $\chi(x)=2\int_{0}^{\infty} dz\frac{1}{z(z^2+1)^2}\left(1-\frac{\sin (zx)}{zx}\right)$.
}
Since the longitudinal bulk viscous correction to the $\mathrm{Im}V^{\rm gluon}$ only manifests through the Debye mass,  increasing  $\zeta_{\|}/s$  enhances the magnitude of $\mathrm{Im}V^{\rm gluon}$, as depicted in the right panel of Fig.~\ref{fig_IMV_gluon}.

\begin{figure*}[htpb]
	\centering
	\subfloat{\includegraphics[scale=0.5]{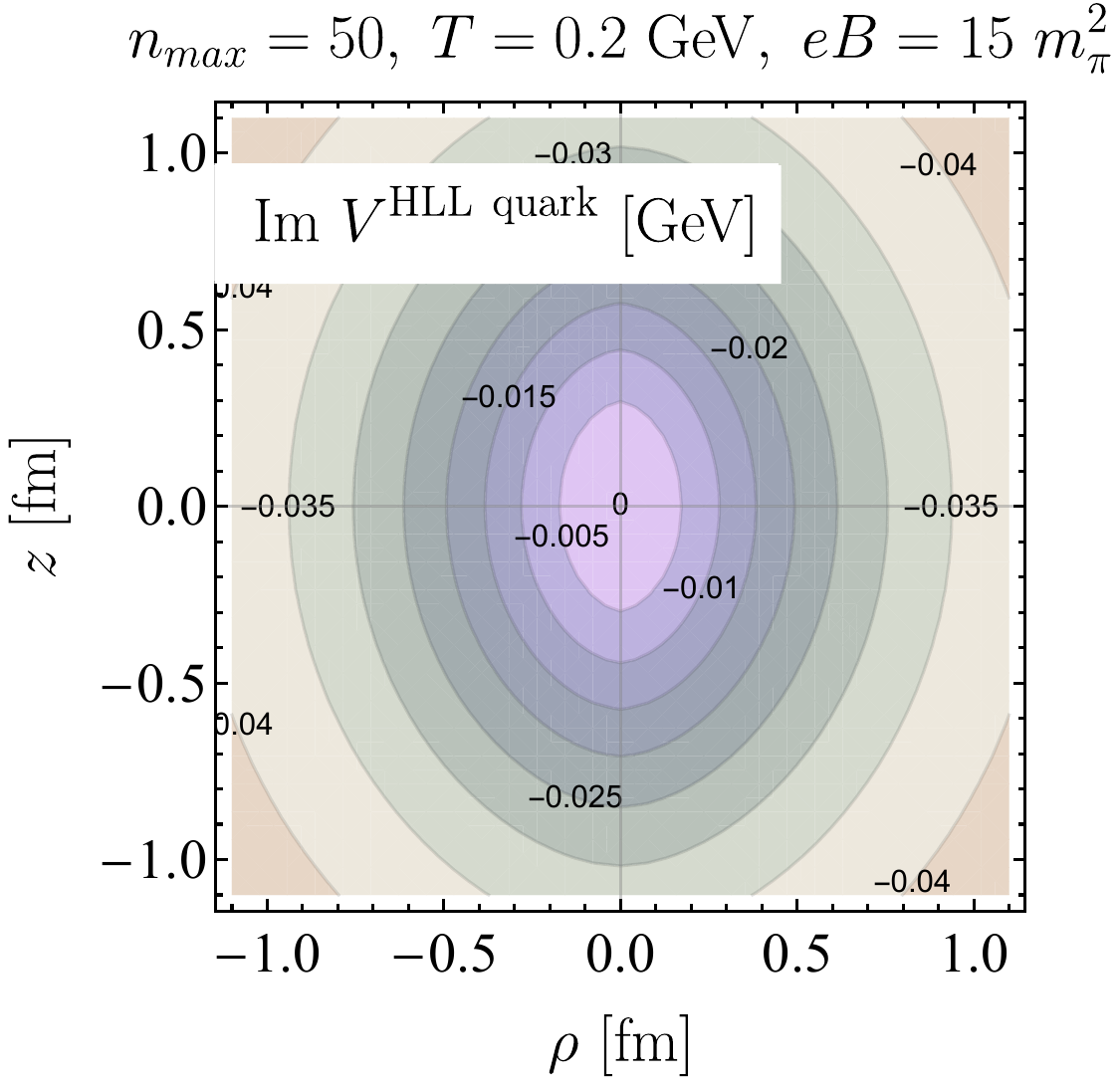}}
	\subfloat{\includegraphics[scale=0.5]{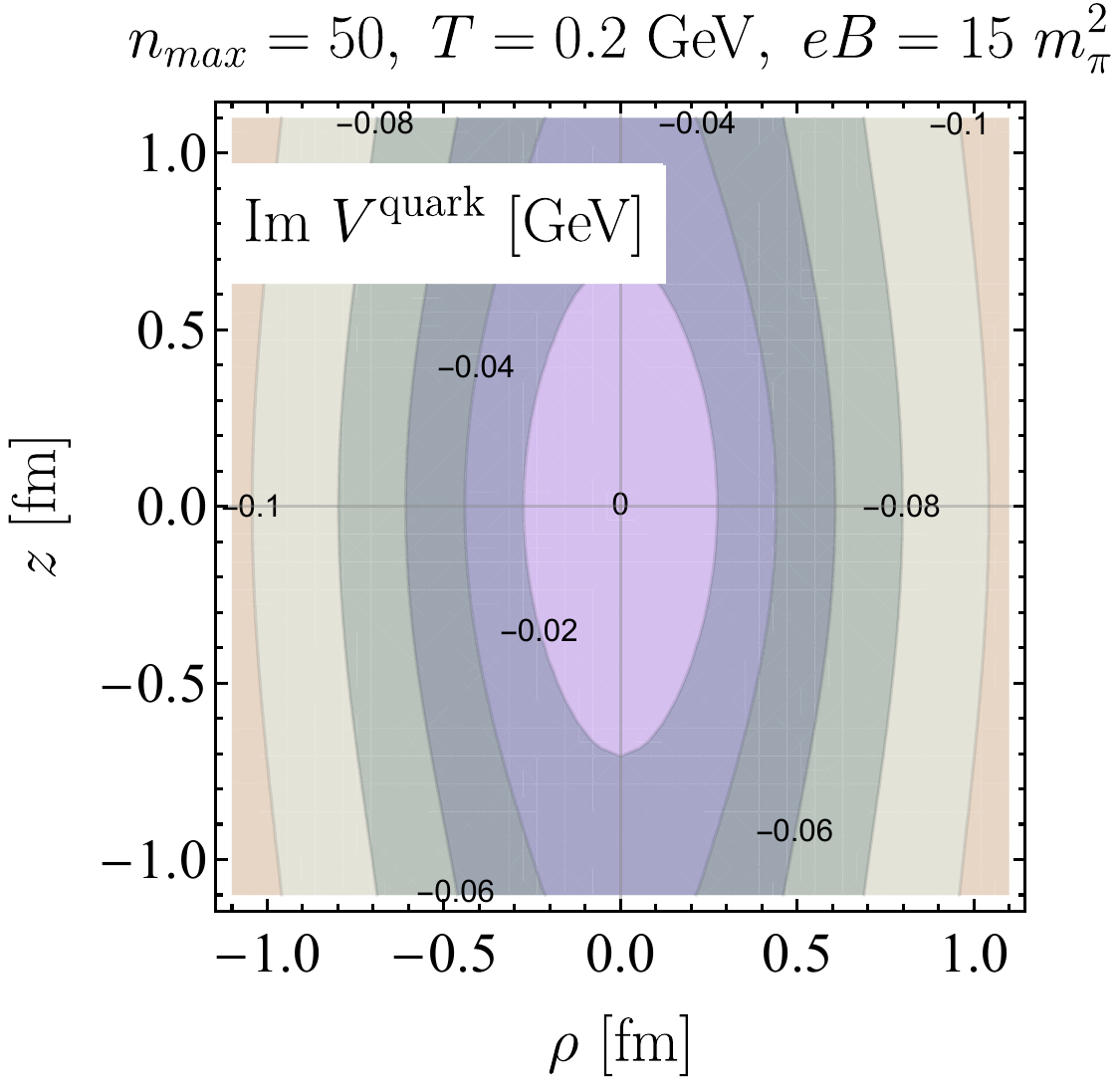}}
	\subfloat{\includegraphics[scale=0.5]{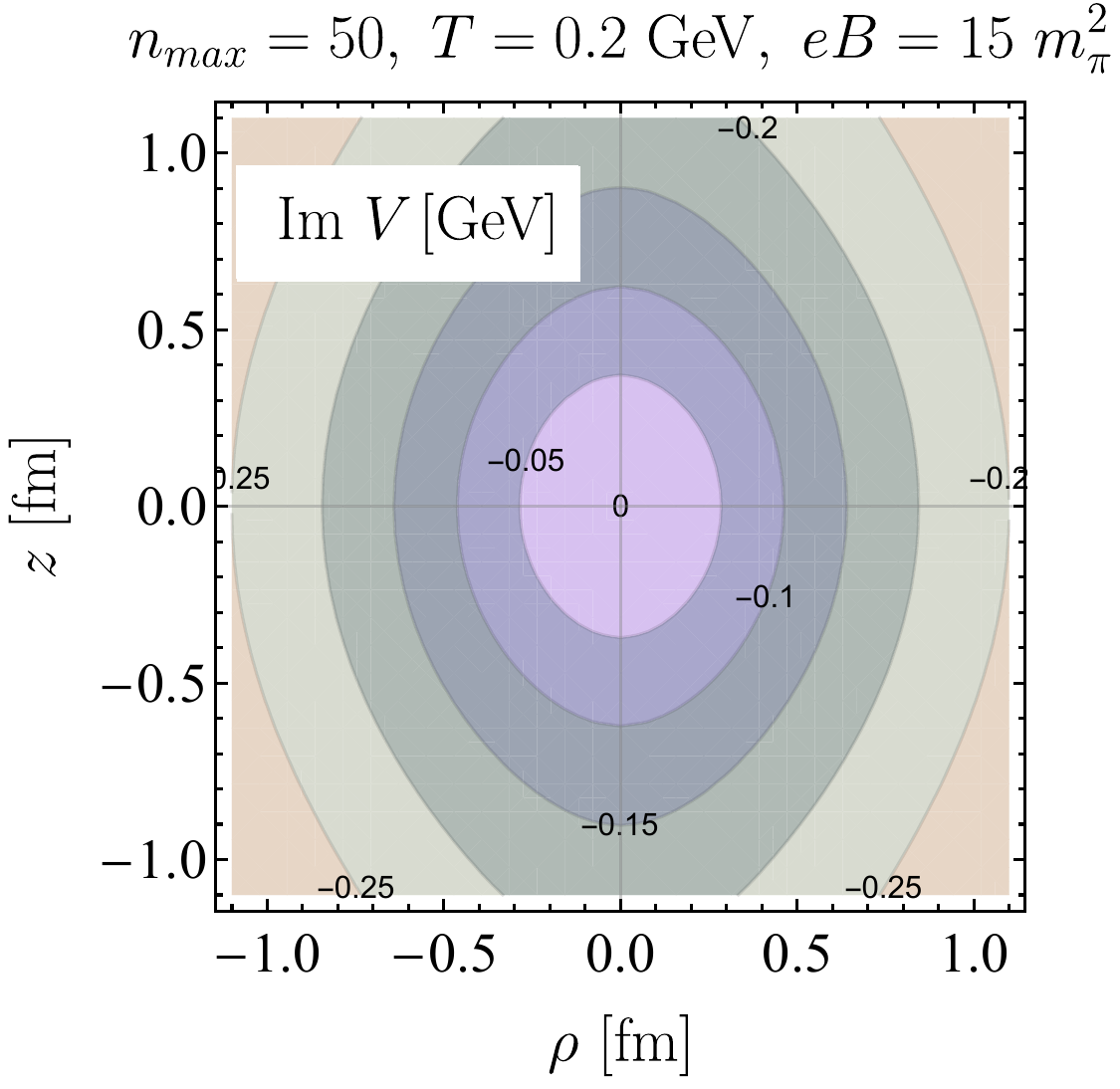}}
	\caption{The contour spatial maps for $\mathrm{Im}V^{\rm HLL~quark}$ (left panel), $\mathrm{Im}V^{\rm quark}=\mathrm{Im}V^{\rm LLL~quark}+\mathrm{Im}V^{\rm HLL~quark}$ (middle panel), and  $\mathrm{Im}V=\mathrm{Im}V^{\rm quark}+\mathrm{Im}V^{\rm gluon}$ (right panel) at $T=0.2$~GeV and $eB=15~m_{\pi}^2$ for $n_{max}=50$. 
	}\label{fig_IMV_total}
\end{figure*}

Different from the original argument in Ref.~\cite{Singh:2017nfa}, which states that the imaginary part of the HQ potential in the LLL approximation completely originates from the gluon-loop contribution of the gluon self-energy and exhibits  isotropy, we observe in the left panel of Fig.~\ref{fig_IMV},  a non-zero imaginary part of the potential related to the LLL quark-loop contribution of the gluon self-energy, $\mathrm{Im}V^{\rm LLL~quark}$, when the heavy quark-antiquark dipole is aligned orthogonal to magnetic field direction, i.e.,  $\bm{r}=(\rho\neq 0, z=0)$. As the numbers of maximum Landau level ($n_{max}$) increase, the magnitude  of $\mathrm{Im}V^{\rm LLL~quark}$ decreases. In the middle panel of Fig.~\ref{fig_IMV}, we present the variation of the imaginary part of the HQ potential related to HLL quark-loop contribution of the gluon self-energy, $\mathrm{Im}V^{\rm HLL~quark}$, with respect to the quark-antiquark separation distances for various $n_{max}$.  
We consider two special cases:  first, the heavy quark-antiquark separation is parallel to the magnetic field direction, i.e., $ \bm{r}=(\rho= 0, z\neq0)$,  and second, it is perpendicular to the magnetic field direction, i.e., $\bm{r}=(\rho\neq 0, z=0)$. As shown in middle panel of Fig.~\ref{fig_IMV}, the $\mathrm{Im}V^{\rm HLL~quark}$ for $\bm{r}=(\rho\neq 0,~z=0)$ at $T=0.2$~GeV is steeper than that for $\bm{r}=(\rho= 0,~z\neq0)$. 
As $n_{max}$ increases, there is an overall enhancement in the magnitude of $\mathrm{Im}V^{\rm HLL~quark}$. When we combine $\mathrm{Im}V^{\rm LLL~quark}$ and $\mathrm{Im}V^{\rm HLL~quark}$, as shown in the right panel of Fig.~\ref{fig_IMV}, we observe that when quark-antiquark dipole axis is perpendicular to magnetic field direction, i.e., $\bm{r}=(\rho\neq 0,~z=0)$, the imaginary part of HQ potential associated with the total quark-loop contributions to the gluon self-energy, denoted as  $\mathrm{Im} V^{\rm quark}$, increases in magnitude as $n_{max}$ increases.

From the left and middle panels of Fig.~\ref{fig_IMV_bulk}, we observe that the longitudinal bulk viscous correction enhances the magnitude of $\mathrm{Im}V^{\rm LLL~quark}$ whereas suppresses the magnitude of $\mathrm{Im}V^{\rm HLL~quark}$. In particular, when the heavy quark-antiquark dipole axis is perpendicular to the magnetic  field direction, i.e., $\bm{r}=(\rho\neq 0, z=0)$, the increasing trend of $|\mathrm{Im}V^{\rm LLL~quark}|$ with $\zeta_{\|}/s$ dominates over the decreasing trend of $|\mathrm{Im}V^{\rm HLL~quark}|$ with $\zeta_{\|}/s$. Consequently, at large separation distances, $|\mathrm{Im}V^{\rm quark}|$ becomes    a notably increasing function of $\zeta_{\|}$, as shown in the right panel of Fig.~\ref{fig_IMV_bulk}. 
	\begin{figure*}[htpb]
	\centering
	\subfloat{\includegraphics[scale=0.44]{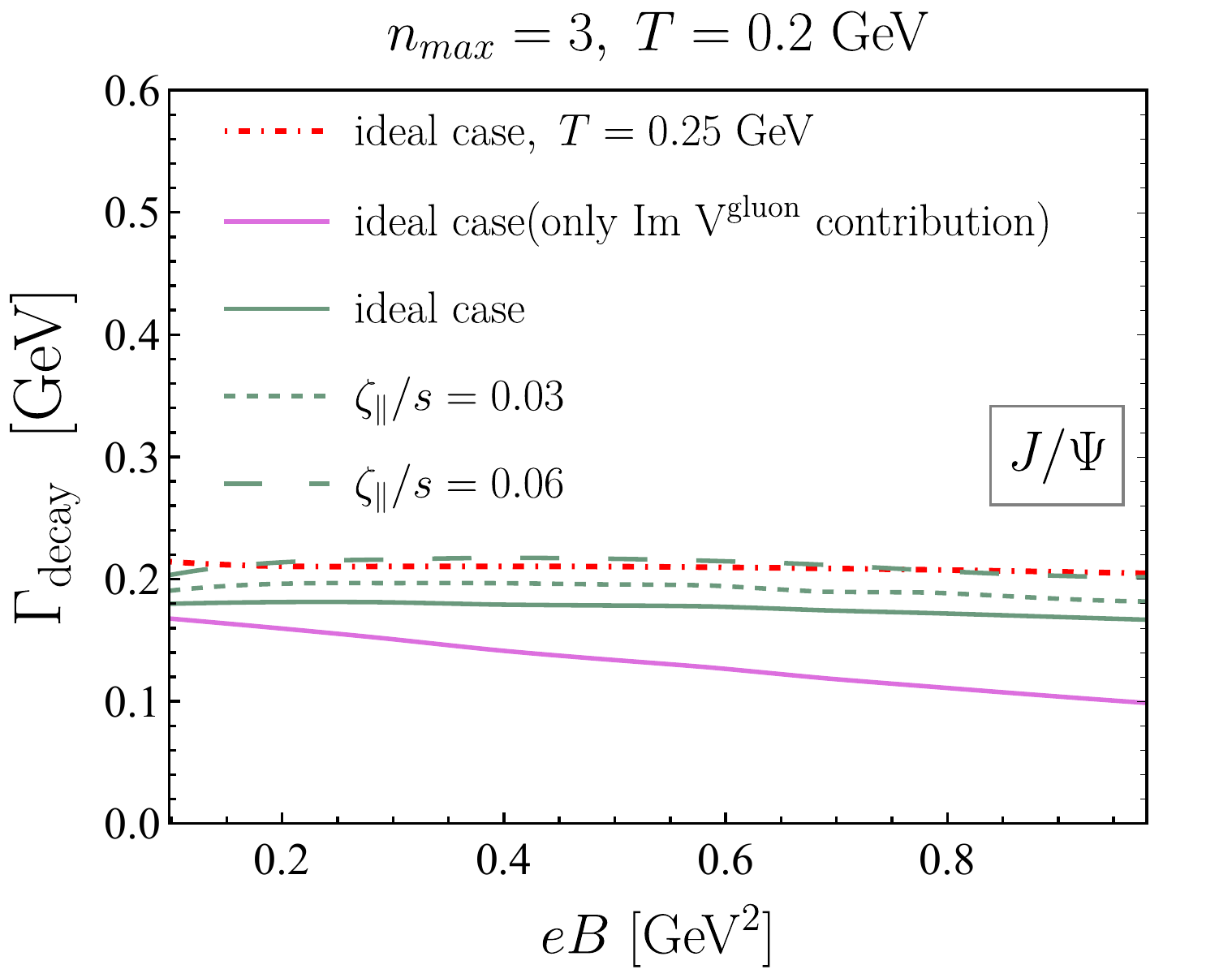}}\hspace{.1 cm}
	\subfloat{\includegraphics[scale=0.44]{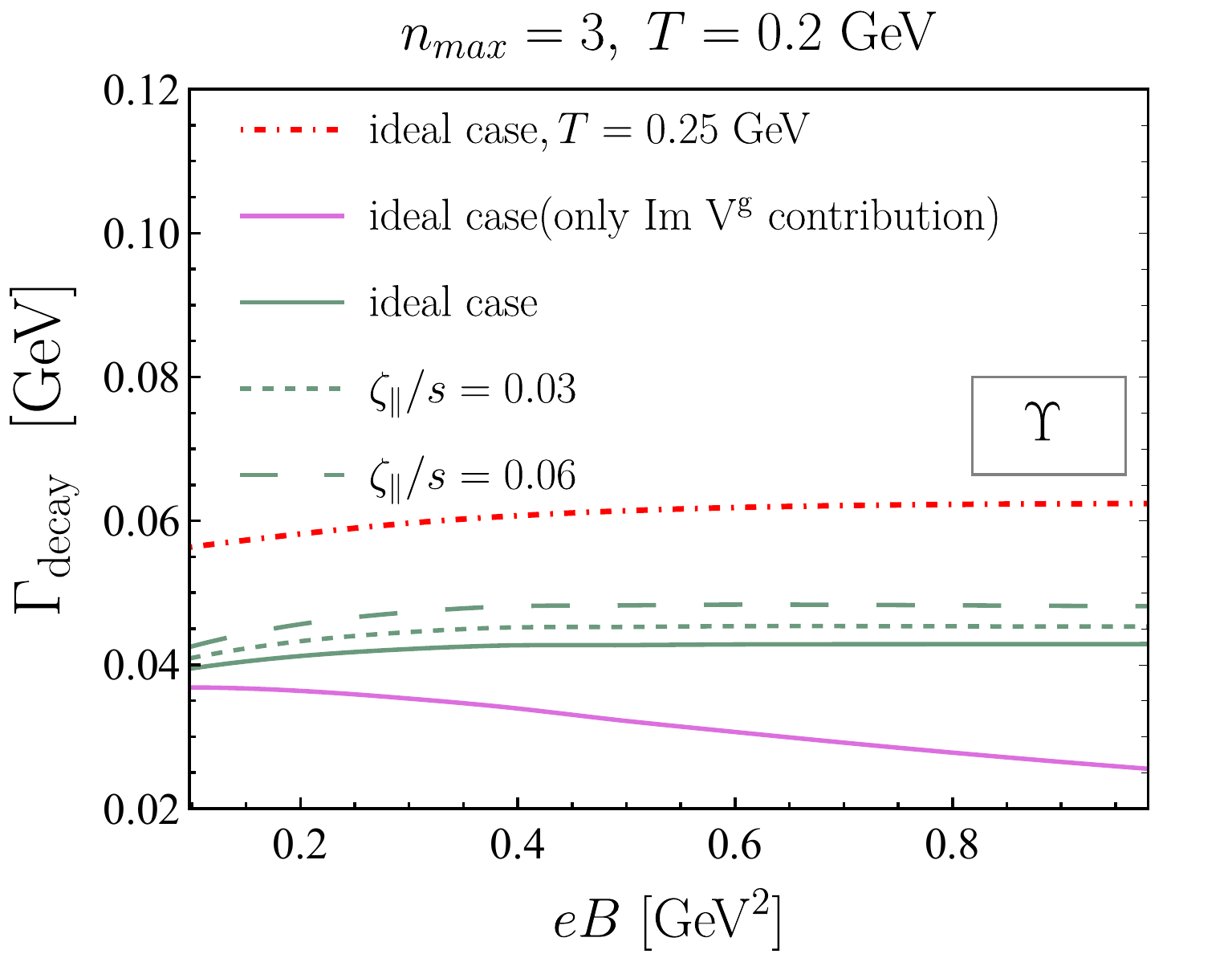}}\hspace{.1 cm}
	\caption{Thermal decay widths for charmonium $J/\Psi$ (left panel) and bottomonium $\Upsilon$ (right panel)  as a function of the magnetic field ($eB$) at different scaled longitudinal bulk viscosities ($\zeta_{\|}/s$) and different temperatures ($T$). The magenta lines are the thermal decay width results without considering the imaginary part of the HQ potential from the quark-loop contribution.}
	\label{fig_decay}
\end{figure*} 
\begin{figure*}[htpb]
	\centering
	\subfloat{\includegraphics[scale=0.42]{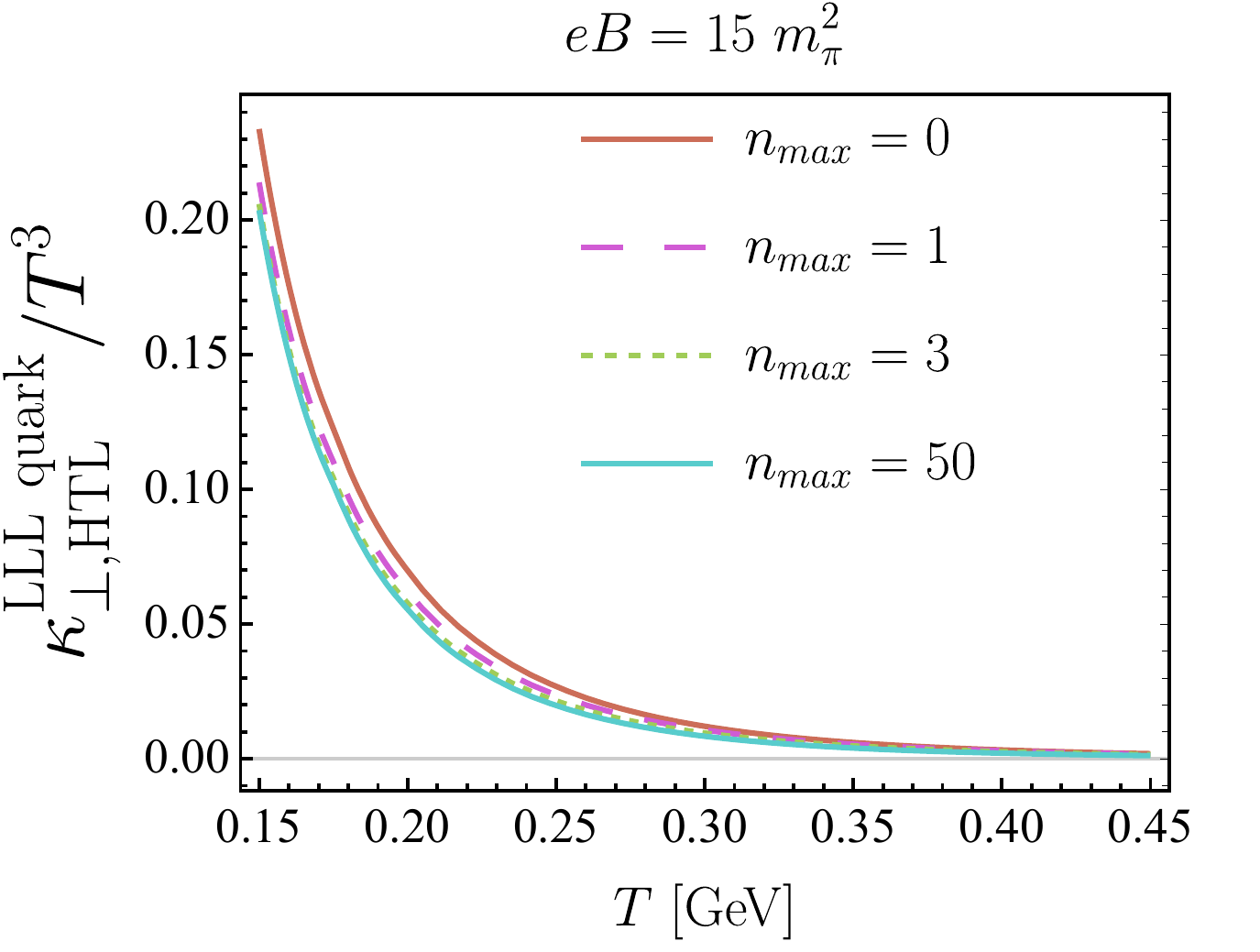}}\hspace{.01 cm}
	\subfloat{\includegraphics[scale=0.42]{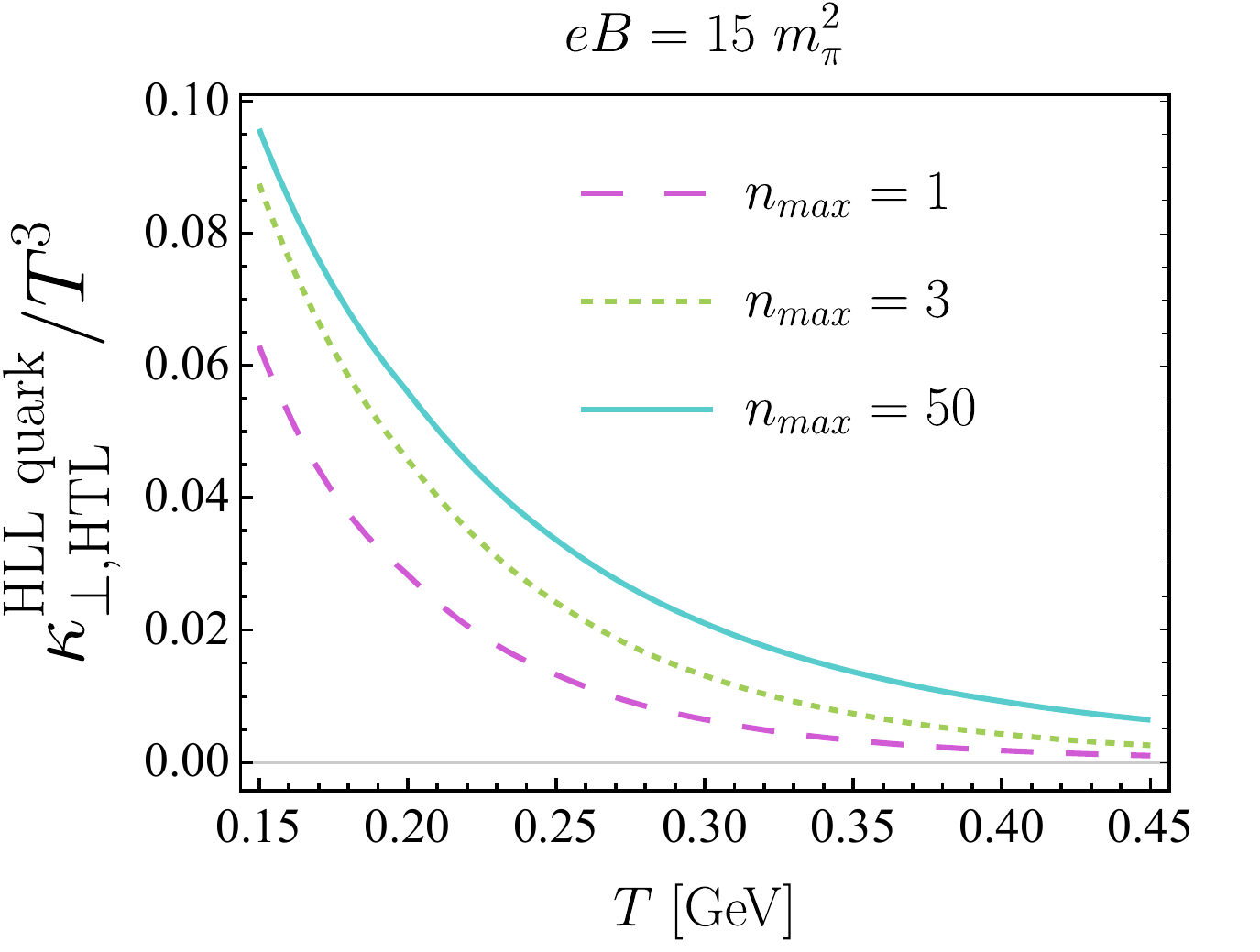}}\hspace{.01 cm}
	\subfloat{\includegraphics[scale=0.42]{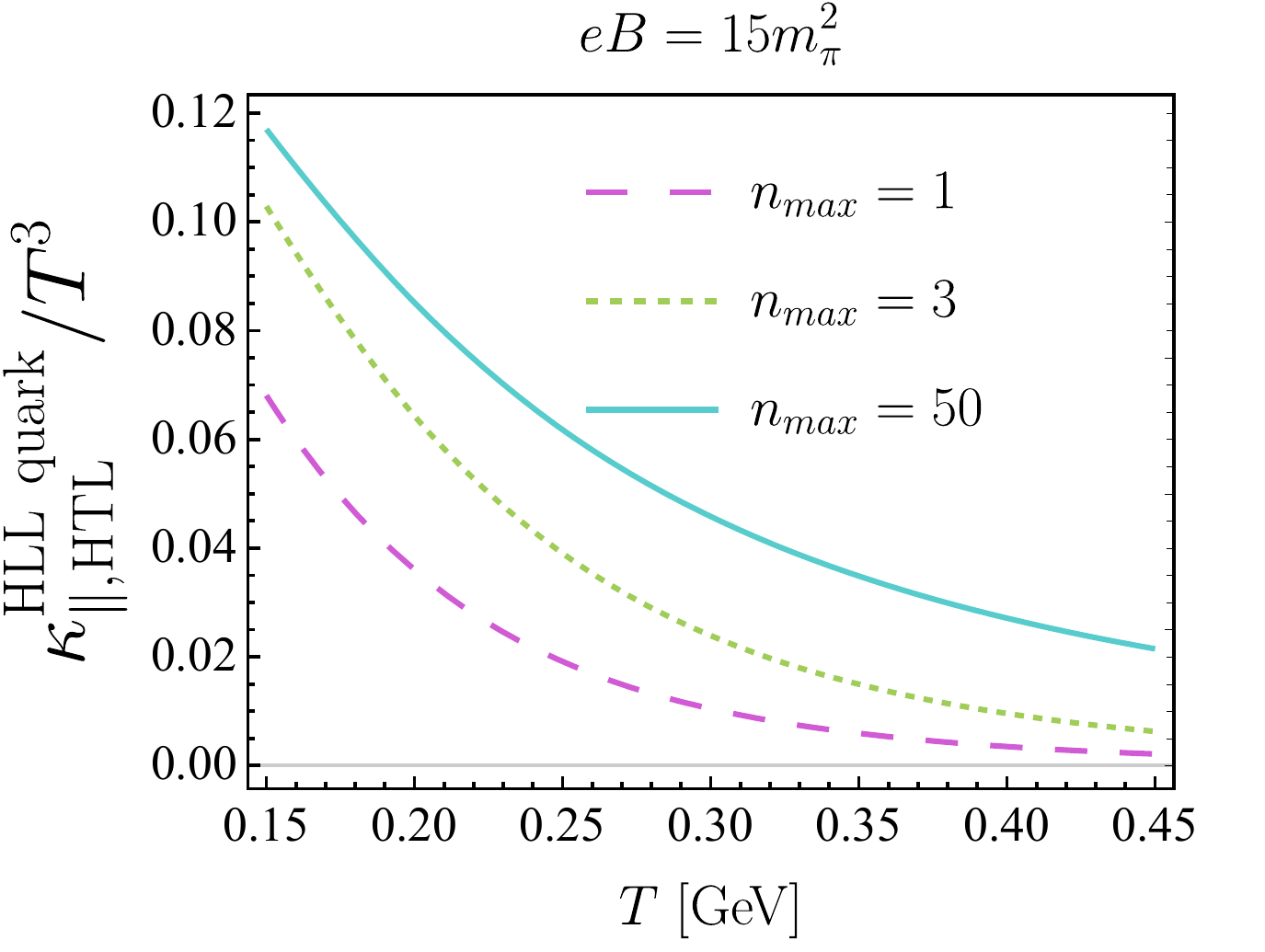}}\hspace{.01 cm}
	\caption{The LLL quark contribution to the perturbative term of scaled transverse HQ momentum diffusion coefficient  $\kappa^{\mathrm{LLL~quark}}_{\perp,\mathrm{HTL}}/T^3$ (left panel), the HLL quark contribution to the perturbative term of scaled transverse HQ momentum diffusion coefficient $\kappa^{\mathrm{HLL~quark}}_{\perp,\mathrm{HTL}}/T^3$ (middle panel), as well as the HLL quark contribution to the perturbative term of scaled longitudinal HQ momentum diffusion coefficient  $\kappa^{\mathrm{HLL~quark}}_{\|,\mathrm{HTL}}/T^3$ (right panel) as functions of $T$ for different $n_{max}$ at $eB=15 m_{\pi}^2$. 
	 } 
	\label{fig_kappa_quark}
\end{figure*} 

Except for the two special cases mentioned above, the left panel of  Fig.~\ref{fig_IMV_total} shows a spatial contour plot of the imaginary part of the HQ potential related to HLL quark-loop contribution of gluon self-energy, $\mathrm{Im}V^{\rm HLL~quark}$. We observe that although strong coupling constant and string tension are angular independent, the $\mathrm{Im}V^{\rm HLL~quark}$  displays significant anisotropy and elongates along the magnetic field direction. When the contribution from the $\mathrm{Im}V^{\rm LLL~quark}$ is taken into account, the imaginary part of HQ potential related to total quark-loop contributions of the gluon self-energy, denoted as  $\mathrm{Im}V^{\rm quark}=\mathrm{Im}V^{\rm LLL~quark}+\mathrm{Im}V^{\rm HLL~quark}$, becomes more anisotropic, as illustrated in the middle panel of Fig.~\ref{fig_IMV_total}. 
 Furthermore, we present a spatial contour plot of the total imaginary part of HQ potential, $\mathrm{Im}V=\mathrm{Im}V^{\rm quark}+\mathrm{Im}V^{\rm gluon}$, in Fig.~\ref{fig_IMV_total}. Compared to the shape of the $\mathrm{Im}V^{\rm quark}$, the anisotropy degree of the $\mathrm{Im} V$ can be weakened due to the significant magnitude of the isotropic $\mathrm{Im}V^{\rm gluon}$. Note that the anisotropic feature of the $\mathrm{Im}V$ has also been observed in the previous computations~\cite{Ghosh:2022sxi}. In Ref.~\cite{Ghosh:2022sxi}, the gluon self-energy was computed in imaginary time formalism of thermal field theory and the string-like term of the $\mathrm{Im}V$ was constructed by using the dielectric permittivity encoding the effects of the medium, although the contribution from $\mathrm{Im}V^{\rm LLL~quark}$ was overlooked in that computation.
 It is worth mentioning that for a fixed separation distance and temperature, the magnetic field dependence of $\mathrm{Im}V^{\rm LLL~quark}$  is opposite to that of $\mathrm{Im}V^{\rm HLL~quark}$. As the magnetic field increases, an increasing number of light quarks occupy the LLL. When the magnetic field is strong enough and the hierarchy of scale $T^2\ll eB$ is satisfied, the contribution from $\mathrm{Im}V^{\rm gluon}$ can be neglected. Consequently, the anisotropy of $\mathrm{Im}V$ arises solely from the LLL quark-loop contribution to the gluon self-energy.

\subsection{Results of thermal decay widths of quarkonium states}
\begin{figure*}[htbp]
	\centering
	\subfloat{\includegraphics[scale=0.44]{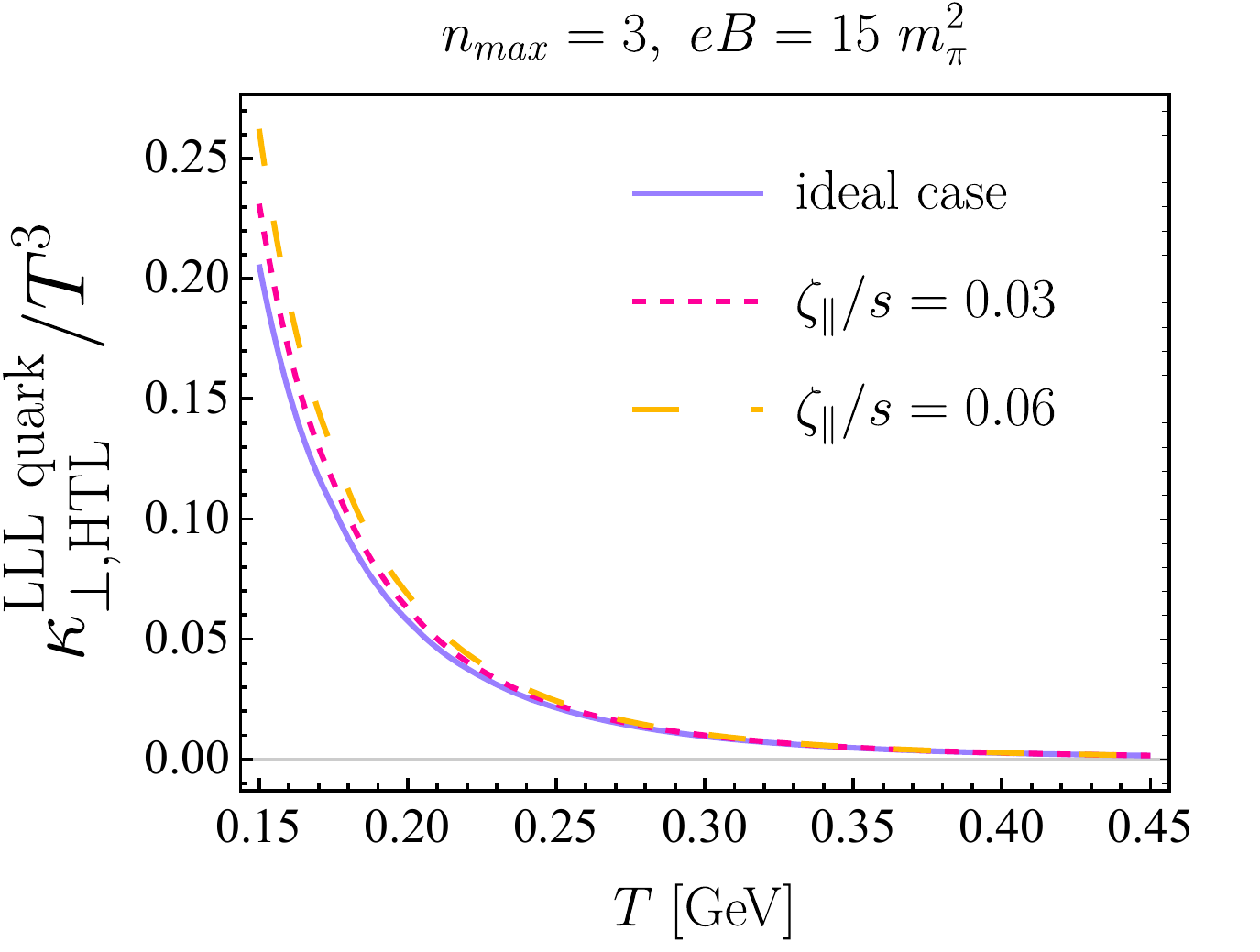}}
	\subfloat{\includegraphics[scale=0.44]{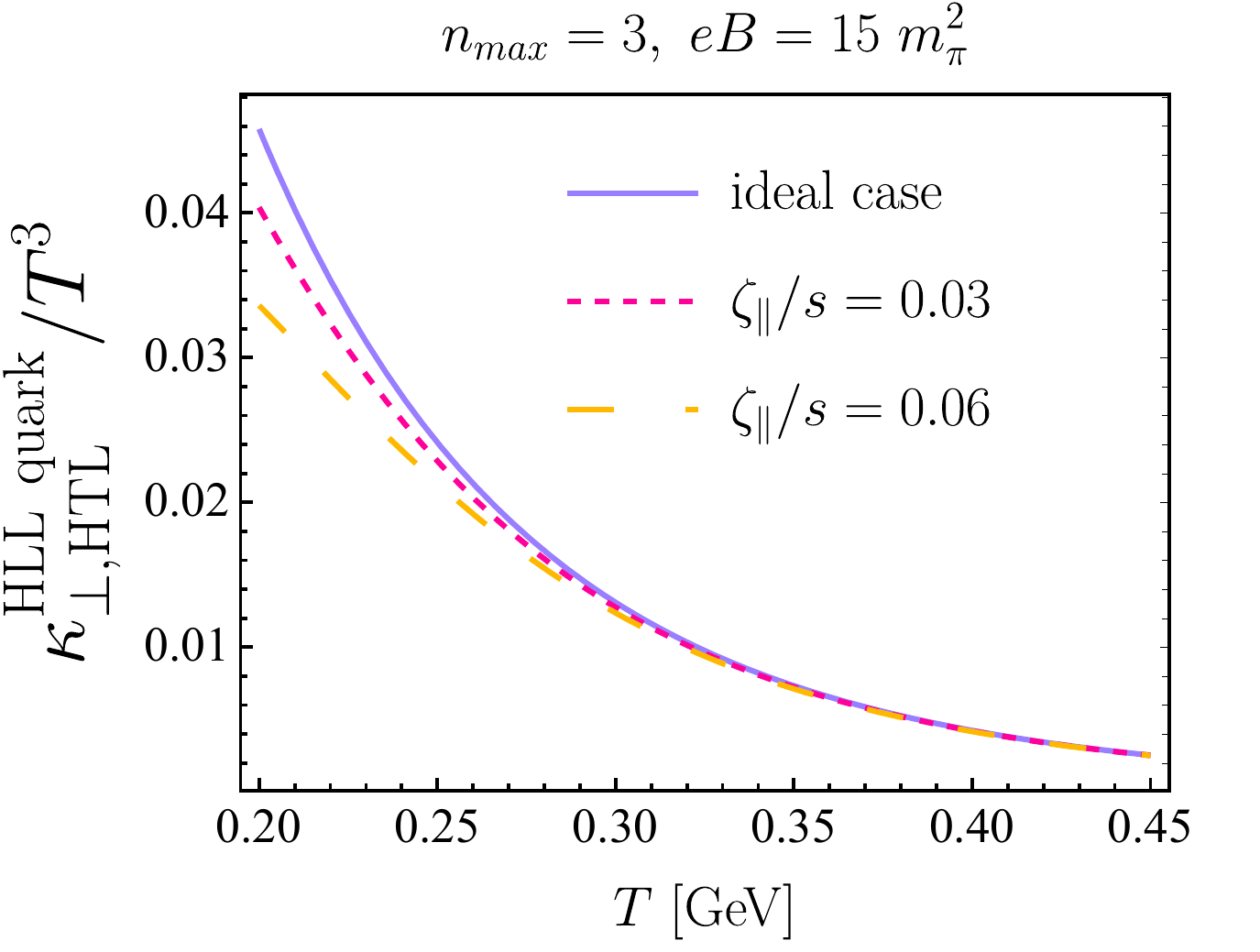}}
	\subfloat{\includegraphics[scale=0.44]{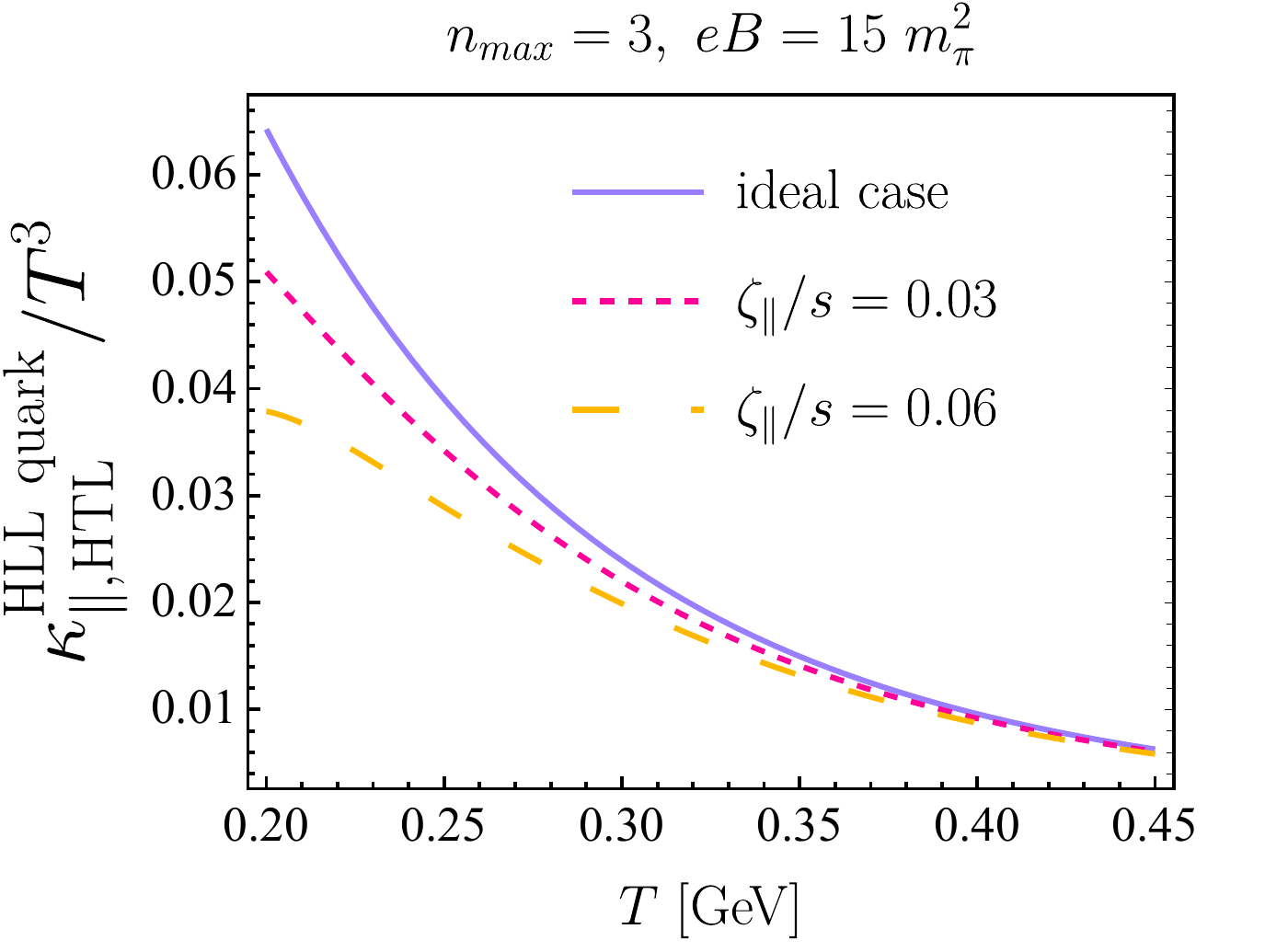}}
	\caption{Similar to Fig.~\ref{fig_kappa_quark} but for different scaled longitudinal bulk viscosities ($\zeta_{\|}/s$) at fixed $n_{max}=3$.}
	\label{fig_kappa_quark_bulk}
\end{figure*} 
\begin{figure}[htbp]
	\centering
	\subfloat{\includegraphics[scale=0.45]{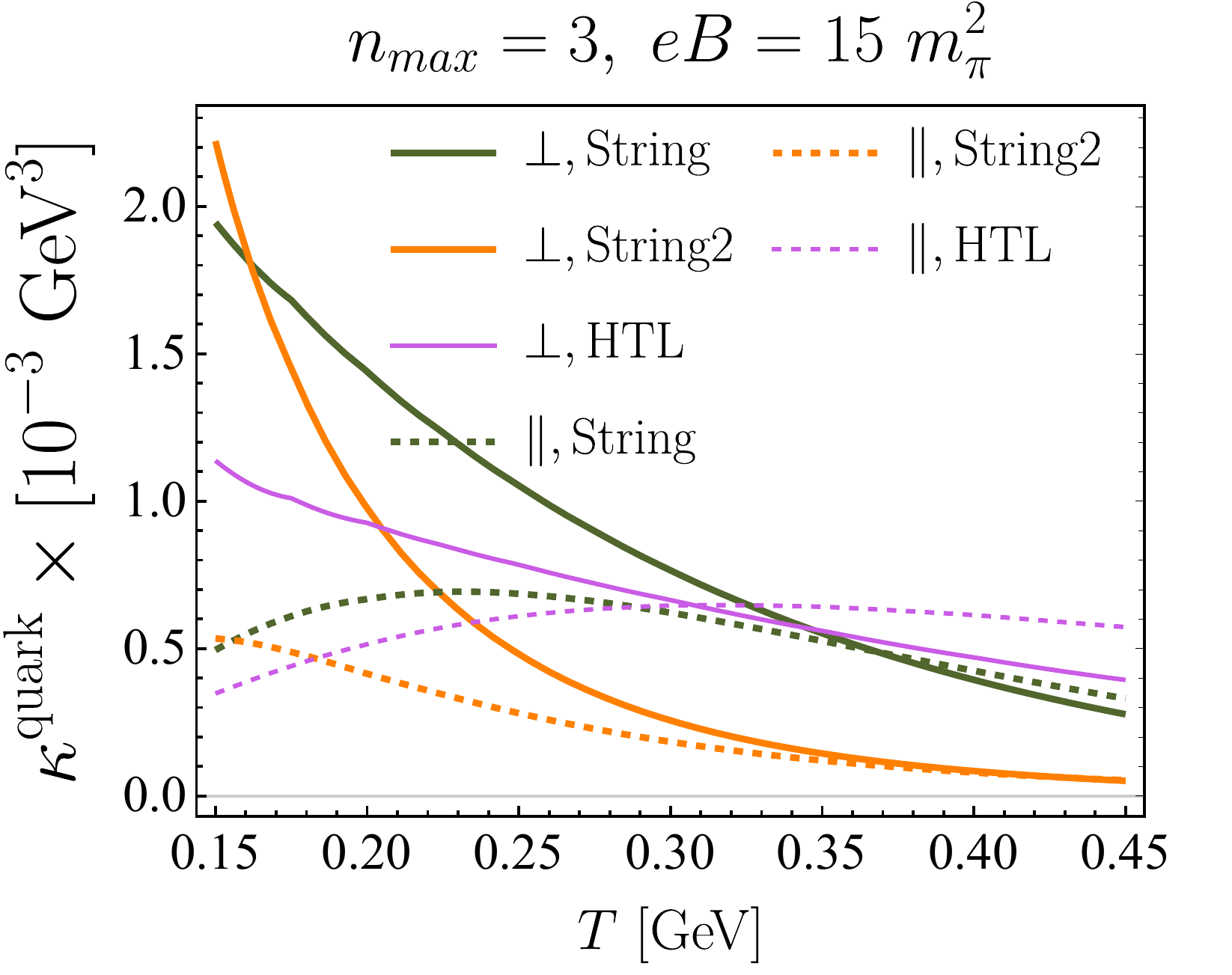}}\hspace{.1 cm}
	\caption{The comparison between the perturbative terms of $\kappa^{\rm quark}$  and the non-perturbative terms of $\kappa^{\rm quark}$  at $n_{max}=3$ and $eB=15~m_{\pi}^2$. The solid lines and dotted lines are the transverse and longitudinal parts of $\kappa^{\rm quark}$, respectively. "String" and "String2" represent the non-perturbative terms employing $\sigma(T)=\sigma_0\sqrt{1-\frac{\pi T^2}{3\sigma_0}}$~\cite{Mazumder:2022jjo,Caristo:2021tbk} and  employing $\sigma(T)=aT_1^2/T^2\sigma(T_1)$, respectively. }
	\label{fig_comparison_kappa_quark}
\end{figure} 
Utilizing the derived $\mathrm{Im}V$, we can further evaluate thermal decay widths for heavy quarkonia. Since the spatial anisotropy of $\mathrm{Im}V$ is integrated into the estimation of thermal decay widths, 
we can compare the thermal widths with and without taking into account the contribution from $\mathrm{Im}V^{\rm quark}$, to clarify how the  magnetic field-induced anisotropy effect affects thermal decay widths. In Fig.~\ref{fig_decay}, we show the thermal decay widths for charmonium $J/\Psi$ and bottomonium $\Upsilon$ as functions of the magnetic field at $T=0.2~\mathrm{GeV}$ and $T=0.25~\mathrm{GeV}$ for $n_{max}=3$. We observe that thermal decay widths increase with  increasing temperature, and the thermal decay width for $\Upsilon$ is quantitatively smaller than that for $J/\Psi$. 
From Fig.~\ref{fig_decay}, we observe that if excluding the contribution from the $\mathrm{Im}V^{\rm quark}$ in the estimation, the thermal decay widths for both $J/\Psi$ and $\Upsilon$ states decrease with increasing magnetic field. After incorporating the contribution from the $\mathrm{Im}V^{\rm quark}$, the anisotropy effect becomes  more pronounced as the magnetic field increases, and the decreasing trend of thermal decay widths of quarkonia is largely mitigated. As a result, the variation of thermal decay widths for $J/\Psi$ and $\Upsilon$ states with the magnetic field becomes inappreciable. Furthermore, the longitudinal bulk viscous correction significantly broadens the thermal widths. This broadening  arises from the dominant contribution of $\mathrm{Im} V^{\rm gluon}(r;\zeta_{\|}/s)$, which contrasts with the decreasing trend of $m_{D}^2(\zeta_{\|}/s)$ observed in Fig.~\ref{fig_Debyemass}. 
\begin{figure*}[htpb]
	\centering
	\subfloat{\includegraphics[scale=0.5]{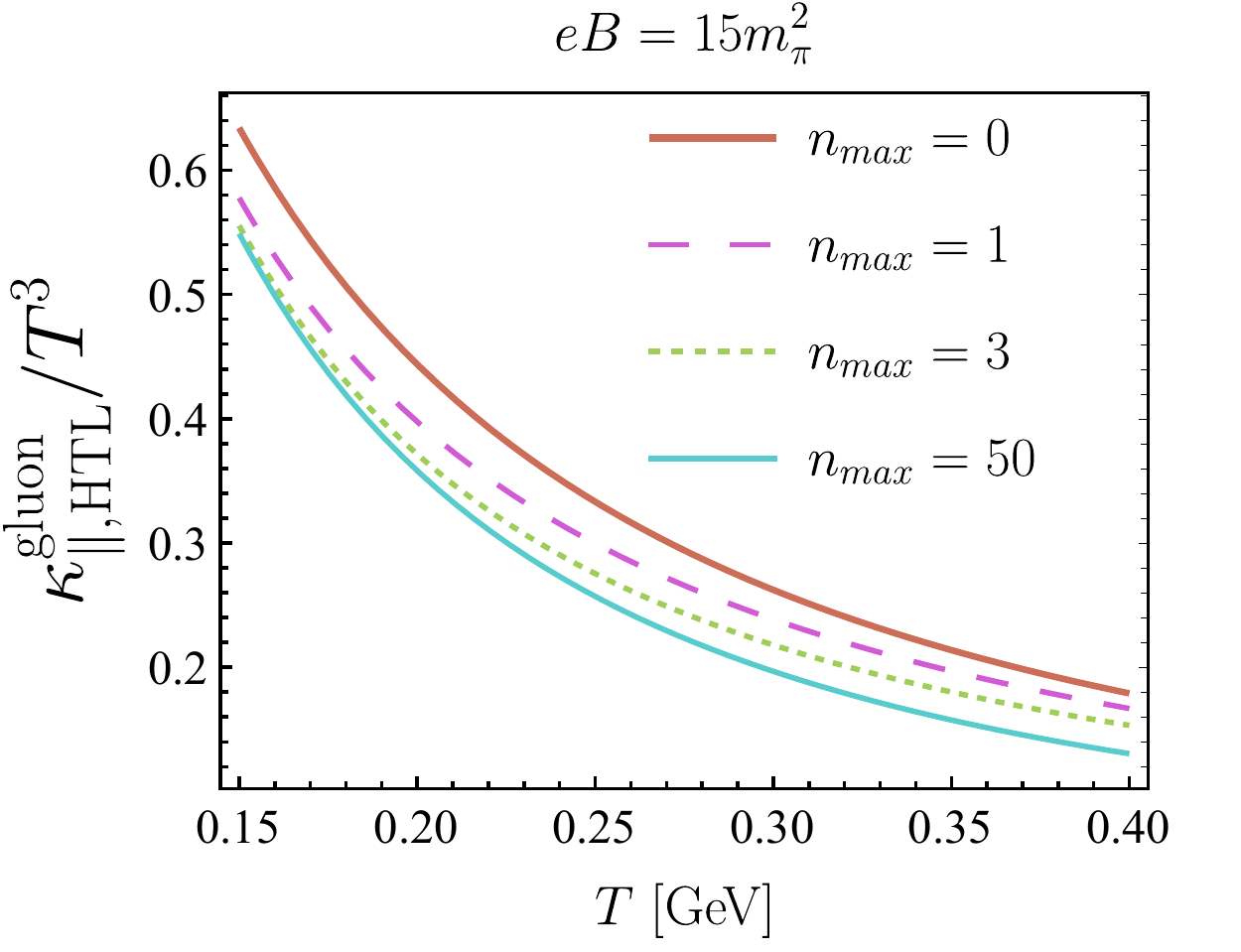}}\hspace{.1 cm}
	\subfloat{\includegraphics[scale=0.5]{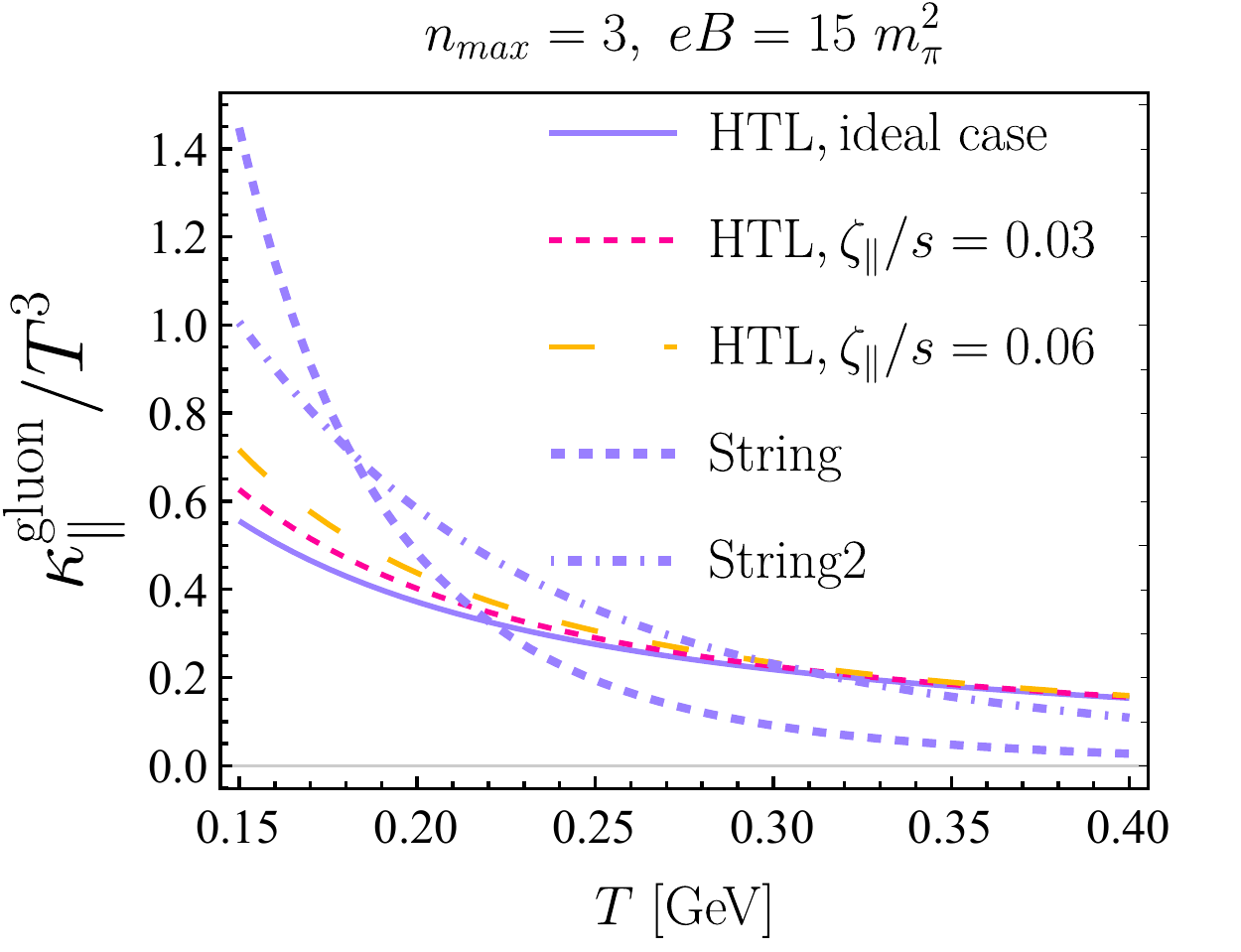}}\hspace{.1 cm}
	\caption{The gluon contribution to the  perturbative term of scaled longitudinal HQ momentum diffusion coefficient $\kappa_{\|,\rm HTL}^{\mathrm{gluon}}/T^3$ as a function of temperature for different numbers of maximum Landau level (left panel) and different scaled longitudinal bulk viscosities (right panel) at $eB=15m_{\pi}^2$. In the right panel, we also display the non-perturbative terms of the gluon contribution to the scaled longitudinal HQ momentum diffusion coefficient $\kappa_{\|,\rm String}^{\mathrm{gluon}}/T^3$, utilizing different $T$-dependent string-tensions.}
	\label{fig_kappa_gluon}
\end{figure*} 
\begin{figure}[htpb]
	\centering
	\subfloat{\includegraphics[scale=0.45]{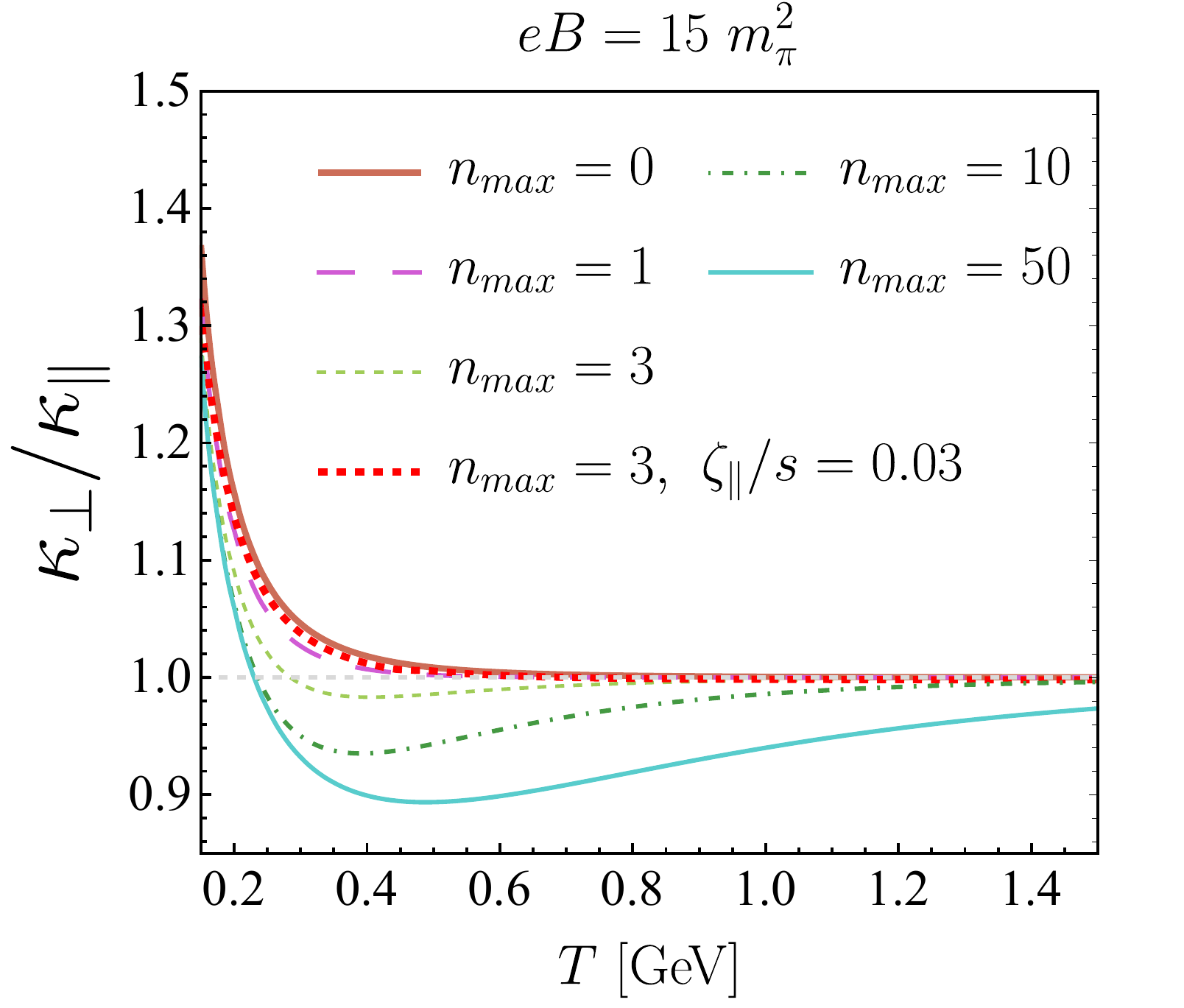}}
	\caption{Temperature dependence of the anisotropy ratio of the HQ momentum diffusion coefficient $\kappa_{\perp}/\kappa_{\|}$  for different numbers of maximum Landau level ($n_{max}$) at $eB=15m_{\pi}^2$. The red dotted line is the result incorporating finite longitudinal bulk viscous effect i.e., $\zeta_{\|}/s=0.03$ at $n_{max}=3$. }
	\label{fig_ratio}
\end{figure} 
\begin{figure}[htpb]
	\centering
	\subfloat{\includegraphics[scale=0.5]{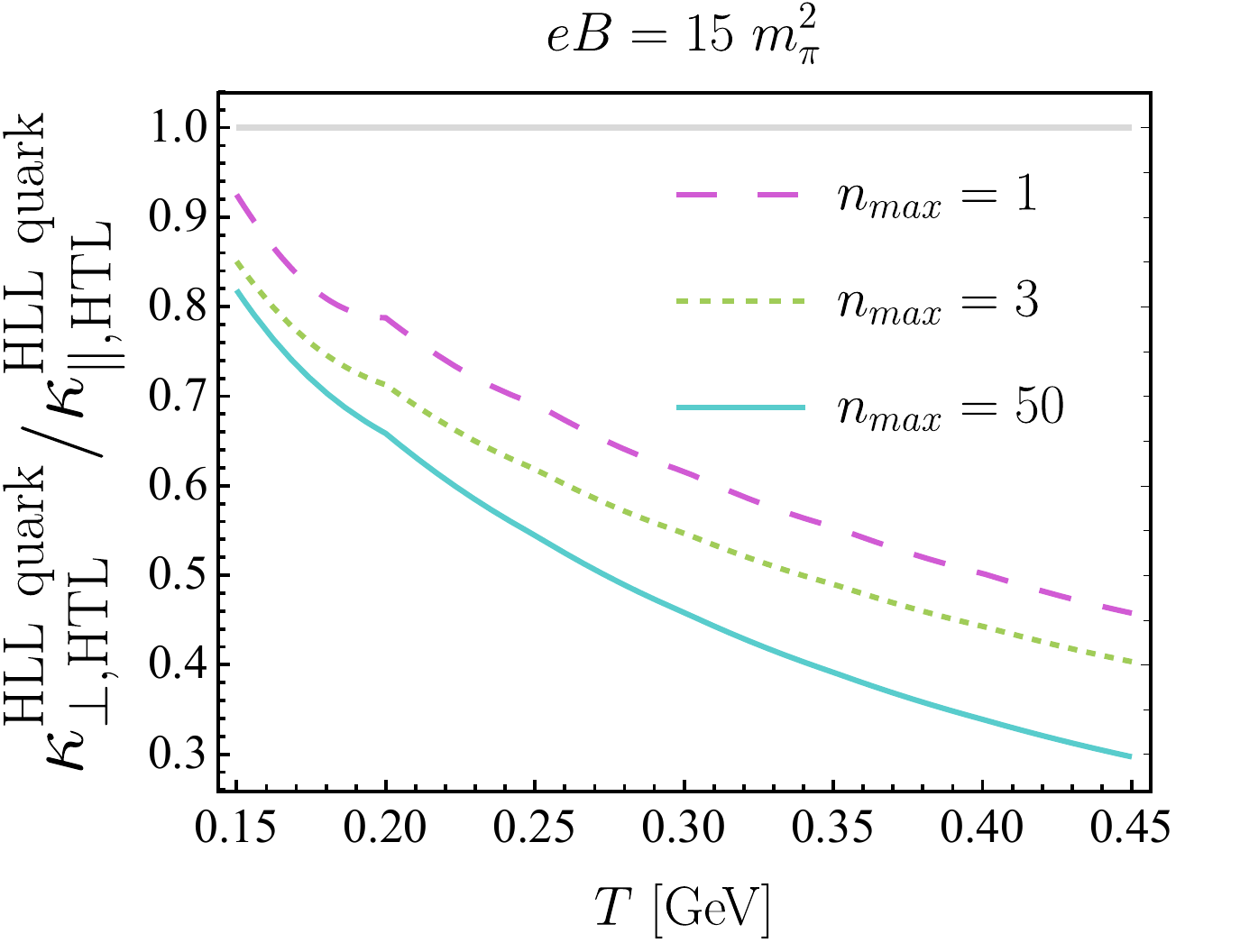}}
	\caption{The ratio of $\kappa^{\mathrm{HLL~quark}}_{\perp,\mathrm{HTL}}$ to  $\kappa^{\mathrm{HLL~quark}}_{\|,\mathrm{HTL}}$  as a function of temperature for different numbers of maximum Landau level ($n_{max}$) at $eB=15m_{\pi}^2$.}
	\label{fig_ratio_kappa_quark}
\end{figure}

\subsection{Results of HQ momentum diffusion coefficients }

We proceed to explore the impacts of temperature, magnetic field, HLL effect, and longitudinal bulk viscous correction on the HQ momentum diffusion coefficients. First, we display the temperature dependence of the quark contribution to the perturbative term of the scaled longitudinal and transverse HQ momentum diffusion coefficients at $eB=15m_\pi^2$ for different numbers of maximum Landau level. As shown in Fig.~\ref{fig_kappa_quark}, the LLL quark contribution to the perturbative term of scaled transverse HQ momentum diffusion coefficient, $\kappa_{\perp,\mathrm{HTL}}^{\rm LLL~quark}/T^3$, the HLL quark contribution to the perturbative term of scaled transverse HQ momentum diffusion coefficient, $\kappa_{\perp,\mathrm{HTL}}^{\rm HLL~quark}/T^3$ as well as the HLL quark contribution to the perturbative term of scaled longitudinal HQ momentum diffusion coefficient, $\kappa_{\|,\mathrm{HTL}}^{\rm HLL~quark}/T^3$, are decreasing functions of temperature. At low temperatures, the $\kappa_{\perp,\mathrm{HTL}}^{\rm LLL~quark}/T^3$ is larger than $\kappa_{\perp,\mathrm{HTL}}^{\rm HLL~quark}/T^3$. As $n_{max}$ increases, the $\kappa_{\perp,\mathrm{HTL}}^{\rm HLL~quark}/T^3$ has a pronounced enhancement, while the $\kappa_{\perp,\mathrm{HTL}}^{\rm LLL~quark}/T^3$ undergoes slight suppression. It is noted that the HLL effect on the HQ momentum diffusion coefficient has also been studied in Ref.~\cite{Kurian:2019nna}. The obtained results indicate that there is no quark contribution to the longitudinal HQ momentum diffusion coefficient beyond the LLL approximation, which is strikingly different from  our results. The discrepancy can be understood from the following two aspects.  Firstly, the expression for the HQ scattering rate with the HLL quarks in Ref.~\cite{ Kurian:2019nna} still utilized the form for the HQ scattering rate with the LLL quarks. In contrast, our Eq.~(\ref{eq:Gamma_quark_general}) and Eq.~(\ref{eq:Gamma_quark_v0}) clearly show that the HQ scattering rate with the HLL quarks is related to the imaginary part of retarded gluon self-energy from the HLL quark-loop contribution, $\mathrm{Im}\Pi^{00}_{R,\mathrm{quark}, T\neq0,n\geq 1}$, rather than the LLL quark-loop contribution, $\mathrm{Im}\Pi^{00}_{R,\mathrm{quark}, T=0,n=0}$.
Secondly, different from the estimation in Ref.~\cite{Kurian:2019nna}, where only the quark-loop contribution to the Debye mass is considered, our work incorporates both gluon-loop and quark-loop contributions to the Debye mass. 
From Fig.~\ref{fig_kappa_quark_bulk}, we can observe that as $\zeta_{\|}/s$ increases, both  $\kappa_{\perp,\mathrm{HTL}}^{\rm HLL~quark}/T^3$ and $\kappa_{\,\mathrm{HTL}}^{\rm HLL~quark}/T^3$ decrease at low temperatures, whereas the 
$\kappa_{\perp,\mathrm{HTL}}^{\rm LLL~quark}/T^3$ slightly increases.

In the above computation of the HQ momentum diffusion coefficient, only the contribution from perturbative QCD interactions is considered, we also attempt to explore the thermal behavior of the non-perturbative term of the heavy quark momentum diffusion coefficient. In Eqs.~(\ref{eq:kappa_12}),~(\ref{eq:kappa_perp_NP}),~(\ref{eq:kappa_para_NP}), and (\ref{eq:kappa_gluon_NP}), the string tension, which in principle decreases with increasing temperature, plays a crucial role in determining the temperature dependence of non-perturbative term of the HQ momentum diffusion coefficient. Inspired by Ref.~\cite{Lafferty:2019jpr}, we parameterize the temperature dependent $\sigma$  in this work as $\sigma(T)/\sigma(T_1)=aT^2_1/T^2$, where the $a$ is temporarily set to 1 and $\sigma(T_1)=0.215$~$\mathrm{GeV}^2$ for $T_1=0.113$~GeV~\cite{Burnier:2016mxc}. For comparison, we also employ the $T$-dependent form of $\sigma(T)=\sigma_0\sqrt{1-\frac{\pi T^2}{3\sigma_0}}$ with $\sigma_0=0.45$~$\mathrm{GeV}^2$ as reported in Ref.~\cite{Mazumder:2022jjo,Caristo:2021tbk}. In Fig.~\ref{fig_comparison_kappa_quark}, we present the temperature dependence of the quark contribution to the non-perturbative term of HQ momentum diffusion coefficient $\kappa^{\rm quark}$, utilizing different string tensions, and compare it with the perturbative term. 
We note that, in the low temperature region, the non-perturbative term of $\kappa^{\rm quark}$ is non-ignorable compared to its perturbative counterpart.
The inclusion of the contribution from non-perturbative interactions to $\kappa^{\rm quark}$ at low temperatures  can help heavy quark obtain larger elliptic flow $v_2$  from the later stage of the QGP~\cite{Xing:2021xwc}. 

In Fig.~\ref{fig_kappa_gluon}, we display the gluon contribution to the perturbative term of scaled longitudinal HQ momentum diffusion coefficient $\kappa_{\|,\mathrm{HTL}}^{\rm gluon}/T^3$ as a function of temperature for the different numbers of maximum Landau level ($n_{max}$) and different longitudinal scaled bulk viscosities ($\zeta_{\|}/s$). Similar to the $\kappa_{\|,\mathrm{HTL} }^{\rm gluon}/T^3$, the HLL effect (longitudinal bulk viscous correction) influences $\kappa_{\|,\mathrm{HTL} }^{\rm gluon}/T^3$ solely through the Debye mass in the Coulomb amplitude,  resulting in  a decrease (increase) in its value. Additionally, in the right panel of Fig.~\ref{fig_kappa_gluon}, we observe that when accounting for temperature-dependent string tensions, the non-perturbative contribution is indispensable for the $\kappa_{\|,\mathrm{HTL} }^{\rm gluon}/T^3$ in the low temperatures region.

To better demonstrate the anisotropic response of the HQ momentum diffusion coefficient to the magnetic field, an anisotropy ratio of the HQ momentum diffusion coefficient is defined as
 \begin{equation}
\frac{\kappa_{\perp}}{\kappa_{\parallel}}=\frac{\kappa_{\perp,\mathrm{HTL}}^{\mathrm{LLL~quark}}+\kappa_{\perp,\mathrm{HTL}}^{\mathrm{HLL~quark}}+\kappa_{\perp,\mathrm{HTL}}^{\rm gluon}}{\kappa_{\|,\mathrm{HTL}}^{\mathrm{HLL~quark}}+\kappa_{\|,\mathrm{HTL}}^{\rm gluon}},
 \end{equation}
 where only the perturbative terms are considered. As shown in Fig.~\ref{fig_ratio}, the anisotropy ratio in the LLL exceeds 1 at low temperature due to the dominant LLL quark contribution, i.e., $\kappa^{\mathrm{LLL~quark}}_{\perp,\mathrm{HTL}}$, and with increasing temperature, the rapid decrease in $\kappa^{\mathrm{LLL~quark}}_{\perp,\mathrm{HTL}}$ leads to a decline in the anisotropy ratio, which ultimately approaches 1.
As the number of maximum Landau level increases, the HLL quark contribution becomes increasingly significant at moderate temperatures, surpassing the LLL quark contribution. 
Due to the significant anisotropy feature of $\kappa^{\mathrm{HLL~quark}}_{\|,\mathrm{HTL}}$ being greater than $\kappa^{\mathrm{HLL~quark}}_{\perp,\mathrm{HTL}}$ depicted in Fig.~\ref{fig_ratio_kappa_quark}, consequently, the anisotropy ratio first exceeds 1 at low temperatures but then gradually falls below 1,  ultimately approaching $1$ at  sufficiently high temperatures.  We also note that the longitudinal bulk viscous correction can increase the anisotropy ratio.

\section{Summary}\label{sec:summary}
 In this work, we systematically investigate how magnetic fields and the longitudinal bulk viscous effect influence both the static HQ potential and HQ momentum diffusion coefficient in the QGP, beyond the LLL approximation. 
 To account for possible non-perturbative effects, a phenomenological gluon propagator induced by the dimension-two gluon condensates is introduced, based on the HTL resummed effective gluon propagator.
 The response of the viscous quark matter to the magnetic field, manifesting in the longitudinal bulk viscous modified distribution function of light quarks, is obtained by solving the relativistic 1+1 dimensional Boltzmann equation under the RTA. Subsequently, the magnetic field and longitudinal bulk viscous correction enter into the Debye screening and Landau damping, further influencing HQ potential and HQ momentum diffusion coefficient. Our main findings can be summarized as follows.

\begin{itemize}
    \item  The increase in temperature, magnetic field, and number of Landau levels can enhance the screening effect of the QGP medium, which accelerates the dissociation of quarkonium states. However, the longitudinal bulk viscous correction can mitigate the screening effect.

    \item 
 The anisotropic feature of the real part of the HQ potential in the magnetic field is  primarily encoded in the strong coupling constant and string tension. In contrast, significant anisotropy can be observed in the imaginary part of the potential even without taking into account angular-dependent strong coupling constant and string tension. In particular, the imaginary part of the potential gets steeper in the direction perpendicular to the magnetic field direction and flatter in the parallel one.

    \item 
Thermal decay widths of quarkonium states broaden significantly with increases in both temperature and longitudinal bulk viscous correction, whereas they are almost insensitive to the variation of the magnetic field.

  \item  At low temperatures,  the contribution of LLL quarks to the anisotropy of HQ momentum diffusion coefficient is dominant, resulting in the anisotropy ratio greater than 1. 
  With the increase of $T$, the significance of the LLL quark contribution diminishes, while the contribution from HLL quarks increases, causing the anisotropy ratio to decrease and even fall below 1. 
   As $T$ increases further, the influence of the magnetic field gradually weakens, ultimately leading to the anisotropy ratio approaching 1.

 \item  While heavy quark momentum diffusion coefficient at high temperatures is dominated by the perturbative QCD interactions, non-perturbative interactions are indispensable for understanding heavy quark dynamics in the low-temperature region.

\end{itemize}
Studying the anisotropic response of the HQ potential and HQ momentum diffusion coefficients to the magnetic field is crucial for understanding static heavy quark-antiquark interactions and heavy quark dynamics in the magnetized QGP medium. Additionally, this research is essential for the theoretical description of heavy quarkonia spectra and the collective flow coefficients of heavy flavor hadrons in heavy-ion collision experiments.

\section{Acknowledgments}
H. X. Zhang is grateful for the helpful discussions with Yuxin Xiao, Koichi Hattori, Ben-Wei Zhang, Hongxi Xing, and Xingyu Guo. This research is supported by Guangdong Major Project of Basic and Applied Basic Research with No. 22020B0301030008, 2022A1515010683, Natural Science Foundation of China with Project No. 12247132, and China Postdoctoral Science Foundation No. 2023M731159.


\appendix
\section*{Appendix}
\begin{widetext}
		\begin{align}
			&\int\frac{d^2\bm{k}_{\perp}}{(2\pi)^2}\exp\bigg(\frac{-\bm{k}_{\perp}^2}{|q_f eB|}\bigg)\exp\bigg(\frac{-\bm{p}_{\perp}^2}{|q_f eB|}\bigg)=\frac{|q_feB|}{8\pi}\exp\left(\frac{-\bm{q}_{\perp}^2}{2|q_feB|}\right),\\
	&\int\frac{d^2\bm{k}_{\perp}}{(2\pi)^2}\exp\bigg(\frac{-\bm{k}_{\perp}^2}{|q_f eB|}\bigg)\exp\bigg(\frac{-\bm{p}_{\perp}^2}{|q_f eB|}\bigg)L_{n}\bigg(\frac{2\bm{k}_{\perp}^2}{|q_feB|}\bigg)L_{l}\bigg(\frac{2\bm{p}_{\perp}^2}{|q_feB|}\bigg)=\frac{|q_feB|}{8\pi}\exp\bigg(\frac{-\bm{q}_{\perp}^2}{2|q_feB|}\bigg)\delta^n_l,\\
	&\int\frac{d^2\bm{k}_{\perp}}{(2\pi)^2}\bm{k}_{\perp}^2\exp\bigg(\frac{-\bm{k}_{\perp}^2}{|q_f eB|}\bigg)\exp\bigg(\frac{-\bm{p}_{\perp}^2}{|q_f eB|}\bigg)L_{n-1}^{1}\bigg(\frac{2\bm{k}_{\perp}^2}{|q_feB|}\bigg)L_{l-1}^{1}\bigg(\frac{2\bm{p}_{\perp}^2}{|q_feB|}\bigg)=-\frac{|q_feB|^2n}{16\pi}\exp\bigg(\frac{-\bm{q}_{\perp}^2}{2|q_feB|}\bigg)\delta^{l-1}_{n-1}.
	\end{align}
	\end{widetext}


{}


\begin{thebibliography}{99}
	
	
	\bibitem{STAR:2005gfr}
	J.~Adams \textit{et al.} [STAR],
	Nucl. Phys. A \textbf{757}, 102-183 (2005).
	
	
	\bibitem{PHENIX:2004vcz}
	K.~Adcox \textit{et al.} [PHENIX],
	Nucl. Phys. A \textbf{757}, 184-283 (2005).

 
	
	
	\bibitem{ALICE:2008ngc}
	K.~Aamodt \textit{et al.} [ALICE],
	JINST \textbf{3}, S08002 (2008).
	

\bibitem{magnetic}
K.~Tuchin,
Adv.\ High Energy Phys.\  {\bf 2013}, 490495 (2013).


\bibitem{Skokov:2009qp} 
V.~Skokov, A.~Y.~Illarionov and V.~Toneev,
Int.\ J.\ Mod.\ Phys.\ A {\bf 24}, 5925 (2009).





\bibitem{Bzdak:2011yy} 
A.~Bzdak and V.~Skokov,
Phys.\ Lett.\ B {\bf 710}, 171 (2012).



\bibitem{Deng:2012pc} 
W.~T.~Deng and X.~G.~Huang,
Phys.\ Rev.\ C {\bf 85}, 044907 (2012).




\bibitem{Kharzeev:2007jp} 
D.~E.~Kharzeev, L.~D.~McLerran and H.~J.~Warringa,
Nucl.\ Phys.\ A {\bf 803}, 227 (2008).


\bibitem{Voronyuk:2011jd}
V.~Voronyuk, V.~D.~Toneev, W.~Cassing, E.~L.~Bratkovskaya, V.~P.~Konchakovski and S.~A.~Voloshin,
Phys. Rev. C \textbf{83}, 054911 (2011).


\bibitem{Fukushima:2008xe}
K.~Fukushima, D.~E.~Kharzeev and H.~J.~Warringa,
Phys. Rev. D \textbf{78}, 074033 (2008)



\bibitem{Alver:2010gr}
B.~Alver and G.~Roland,
Phys. Rev. C \textbf{81}, 054905 (2010)
[erratum: Phys. Rev. C \textbf{82}, 039903 (2010)].



\bibitem{Moore:2004tg}
G.~D.~Moore and D.~Teaney,
Phys. Rev. C \textbf{71}, 064904 (2005).

\bibitem{Rapp:2009my}
R.~Rapp and H.~van Hees,
[arXiv:0903.1096 [hep-ph]].

\bibitem{Das:2016cwd}
S.~K.~Das, S.~Plumari, S.~Chatterjee, J.~Alam, F.~Scardina and V.~Greco,
Phys. Lett. B \textbf{768}, 260-264 (2017).


\bibitem{Chatterjee:2018lsx}
S.~Chatterjee and P.~Bozek,
Phys. Lett. B \textbf{798}, 134955 (2019).


\bibitem{STAR:2019clv}
J.~Adam \textit{et al.} [STAR],
Phys. Rev. Lett. \textbf{123}, no.16, 162301 (2019).



\bibitem{Ding:2010ga}
H.~T.~Ding, A.~Francis, O.~Kaczmarek, F.~Karsch, E.~Laermann and W.~Soeldner,
Phys. Rev. D \textbf{83}, 034504 (2011).

\bibitem{Amato:2013naa}
A.~Amato, G.~Aarts, C.~Allton, P.~Giudice, S.~Hands and J.~I.~Skullerud,
Phys. Rev. Lett. \textbf{111}, no.17, 172001 (2013).


\bibitem{ALICE:2019sgg}
S.~Acharya \textit{et al.} [ALICE],
Phys. Rev. Lett. \textbf{125}, no.2, 022301 (2020).


\bibitem{Jiang:2022uoe}
Z.~F.~Jiang, S.~Cao, W.~J.~Xing, X.~Y.~Wu, C.~B.~Yang and B.~W.~Zhang,
Phys. Rev. C \textbf{105}, no.5, 054907 (2022).



\bibitem{Sun:2020wkg}
Y.~Sun, S.~Plumari and V.~Greco,
Phys. Lett. B \textbf{816}, 136271 (2021).

\bibitem{RHMD1}
Andre Lichnerowicz, ``{\it Magnetohydrodynamics: waves and shock waves in curved space-time}"
(Kluwer Academic, 1994).

\bibitem{RHMD2}
A. M. Anile, ``{\it Relativistic Fluids and Magneto-fluids: With Applications in Astrophysics and
Plasma Physics}" (Cambridge University Press, 2005).

\bibitem{RHMD3}
 Hans Goedbloed, Rony Keppens, and Stefaan Poedts, ``{\it Magnetohydrodynamics of laboratory and astrophysical plasmas}" (Cambridge University Press, 2019).

\bibitem{RHMD4}
Shoji Kato and Jun Fukue, ``{\it Fundamentals of Astrophysical Fluid Dynamics: Hydrodynamics, Magnetohydrodynamics, and Radiation Hydrodynamics}" (Springer, 2020).

\bibitem{Hattori:2022hyo}
K.~Hattori, M.~Hongo and X.~G.~Huang,
Symmetry \textbf{14}, no.9, 1851 (2022).

\bibitem{Gusynin:1995nb}
V.~P.~Gusynin, V.~A.~Miransky and I.~A.~Shovkovy,
Nucl. Phys. B \textbf{462}, 249-290 (1996).

\bibitem{Akhiezer}
A.I. Akhiezer and V.B. Berestetsky, ``{\it Quantum Electrodynamics}" (Interscience, NY, 1965).

\bibitem{Fukushima:2015wck}
K.~Fukushima, K.~Hattori, H.~U.~Yee and Y.~Yin,
Phys. Rev. D \textbf{93}, no.7, 074028 (2016).


\bibitem{Bandyopadhyay:2021zlm}
A.~Bandyopadhyay, J.~Liao and H.~Xing,
Phys. Rev. D \textbf{105}, no.11, 11 (2022).





\bibitem{Kurian:2020kct}
M.~Kurian, V.~Chandra and S.~K.~Das,
Phys. Rev. D \textbf{101}, no.9, 094024 (2020).

\bibitem{Kurian:2019nna}
M.~Kurian, S.~K.~Das and V.~Chandra,
Phys. Rev. D \textbf{100}, no.7, 074003 (2019).

\bibitem{Bandyopadhyay:2023hiv}
A.~Bandyopadhyay,
Phys. Rev. D \textbf{109}, no.3, 034013 (2024).

\bibitem{Bonati:2016kxj}
C.~Bonati, M.~D'Elia, M.~Mariti, M.~Mesiti, F.~Negro, A.~Rucci and F.~Sanfilippo,
Phys. Rev. D \textbf{94}, no.9, 094007 (2016).
\bibitem{Bonati:2018uwh}
C.~Bonati, S.~Cal\`\i{}, M.~D'Elia, M.~Mesiti, F.~Negro, A.~Rucci and F.~Sanfilippo,
Phys. Rev. D \textbf{98}, no.5, 054501 (2018).

\bibitem{Hasan:2020iwa}
M.~Hasan and B.~K.~Patra,
Phys. Rev. D \textbf{102}, no.3, 036020 (2020).

\bibitem{Khan:2021syq}
S.~A.~Khan, M.~Hasan and B.~K.~Patra,
Nucl. Phys. A \textbf{1034}, 122643 (2023).


\bibitem{Singh:2017nfa}
B.~Singh, L.~Thakur and H.~Mishra,
Phys. Rev. D \textbf{97}, no.9, 096011 (2018).

\bibitem{Ghosh:2022sxi}
R.~Ghosh, A.~Bandyopadhyay, I.~Nilima and S.~Ghosh,
Phys. Rev. D \textbf{106}, no.5, 5 (2022).

\bibitem{Hattori:2017qih}
K.~Hattori, X.~G.~Huang, D.~H.~Rischke and D.~Satow,
Phys. Rev. D \textbf{96}, no.9, 094009 (2017).

\bibitem{Rath:2020beo}
S.~Rath and B.~K.~Patra,
Eur. Phys. J. C \textbf{81}, no.2, 139 (2021).

\bibitem{Kurian:2018qwb}
M.~Kurian, S.~Mitra, S.~Ghosh and V.~Chandra,
Eur. Phys. J. C \textbf{79}, no.2, 134 (2019).

\bibitem{Ghosh:2020wqx}
S.~Ghosh and S.~Ghosh,
Phys. Rev. D \textbf{103}, 096015 (2021).

\bibitem{Fukushima:2017lvb}
K.~Fukushima and Y.~Hidaka,
Phys. Rev. Lett. \textbf{120}, no.16, 162301 (2018).

\bibitem{Hattori:2017xoo}
K.~Hattori and D.~Satow,
Phys. Rev. D \textbf{97}, no.1, 014023 (2018).

\bibitem{Guo:2018vwy}
Y.~Guo, L.~Dong, J.~Pan and M.~R.~Moldes,
Phys. Rev. D \textbf{100}, no.3, 036011 (2019).

\bibitem{Megias:2007pq}
E.~Megias, E.~Ruiz Arriola and L.~L.~Salcedo,
Phys. Rev. D \textbf{75}, 105019 (2007).



\bibitem{Megias:2005ve}
E.~Megias, E.~Ruiz Arriola and L.~L.~Salcedo,
JHEP \textbf{01}, 073 (2006).


\bibitem{Mallik:2009pj}
S.~Mallik and S.~Sarkar,
Eur. Phys. J. C \textbf{61}, 489-494 (2009).

\bibitem{Miransky:2015ava}
V.~A.~Miransky and I.~A.~Shovkovy,
Phys. Rept. \textbf{576}, 1-209 (2015).

\bibitem{Kobes:1984vb}
R.~L.~Kobes, G.~W.~Semenoff and N.~Weiss,
Z. Phys. C \textbf{29}, 371 (1985).


\bibitem{Rath:2017fdv}
S.~Rath and B.~K.~Patra,
JHEP \textbf{12}, 098 (2017).

\bibitem{Keldysh:1964ud}
L.~V.~Keldysh,
Zh. Eksp. Teor. Fiz. \textbf{47}, 1515-1527 (1964).


\bibitem{Chou:1984es}
K.~c.~Chou, Z.~b.~Su, B.~l.~Hao and L.~Yu,
Phys. Rept. \textbf{118}, 1-131 (1985).

\bibitem{KMS1}
R.~Kubo,
J. Phys. Soc. Jap. \textbf{12}, 570-586 (1957).


\bibitem{KMS2}
P.~C.~Martin and J.~S.~Schwinger,
Phys. Rev. \textbf{115}, 1342-1373 (1959).


\bibitem{Landsman:1986uw}
N.~P.~Landsman and C.~G.~van Weert,
Phys. Rept. \textbf{145}, 141 (1987).


\bibitem{Ayala:2018wux}
A.~Ayala, C.~A.~Dominguez, S.~Hernandez-Ortiz, L.~A.~Hernandez, M.~Loewe, D.~Manreza Paret and R.~Zamora,
Phys. Rev. D \textbf{98}, no.3, 031501 (2018).


\bibitem{Bazavov:2012ka}
A.~Bazavov, N.~Brambilla, X.~Garcia Tormo, i, P.~Petreczky, J.~Soto and A.~Vairo,
Phys. Rev. D \textbf{86}, 114031 (2012).


\bibitem{Ayala:2018ina}
A.~Ayala, J.~D.~Casta\~no-Yepes, C.~A.~Dominguez, S.~Hern\'andez-Ortiz, L.~A.~Hern\'andez, M.~Loewe, D.~Manreza Paret and R.~Zamora,
Rev. Mex. Fis. \textbf{66}, no.4, 446-461 (2020).




\bibitem{Hattori:2022uzp}
K.~Hattori and K.~Itakura,
Annals Phys. \textbf{446}, 169114 (2022).




\bibitem{Fukushima:2011nu}
K.~Fukushima,
Phys. Rev. D \textbf{83}, 111501 (2011).


\bibitem{Schwinger:1962tn}
J.~S.~Schwinger,
Phys. Rev. \textbf{125}, 397-398 (1962).


\bibitem{Baier:1991gg}
R.~Baier and E.~Pilon,
Z. Phys. C \textbf{52}, 339-342 (1991).


\bibitem{Kobes:1985kc}
R.~L.~Kobes and G.~W.~Semenoff,
Nucl. Phys. B \textbf{260}, 714-746 (1985).



\bibitem{Fujimoto:1985me}
Y.~Fujimoto, M.~Morikawa and M.~Sasaki,
Phys. Rev. D \textbf{33}, 590 (1986).




\bibitem{Blaizot:2021xqa}
J.~P.~Blaizot and M.~\'A.~Escobedo,
Phys. Rev. D \textbf{104}, no.5, 054034 (2021).


\bibitem{Kapusta:2006pm}
J.~I.~Kapusta and C.~Gale,
``{\it Finite-temperature field theory: Principles and applications}''
(Cambridge University Press, 2011).

\bibitem{Brambilla:2008cx}
N.~Brambilla, J.~Ghiglieri, A.~Vairo and P.~Petreczky,
Phys. Rev. D \textbf{78}, 014017 (2008).


\bibitem{Weldon:1982aq}
H.~A.~Weldon,
Phys. Rev. D \textbf{26}, 1394 (1982).

\bibitem{Hattori:2016lqx}
K.~Hattori, S.~Li, D.~Satow and H.~U.~Yee,
Phys. Rev. D \textbf{95}, no.7, 076008 (2017)


\bibitem{Sadofyev:2015tmb}
A.~V.~Sadofyev and Y.~Yin,
Phys. Rev. D \textbf{93}, no.12, 125026 (2016).


\bibitem{Carrington:1997sq}
M.~E.~Carrington, D.~f.~Hou and M.~H.~Thoma,
Eur. Phys. J. C \textbf{7}, 347-354 (1999).




\bibitem{Thakur:2020ifi}
L.~Thakur, N.~Haque and Y.~Hirono,
JHEP \textbf{06}, 071 (2020).



\bibitem{Thakur:2021vbo}
L.~Thakur and Y.~Hirono,
JHEP \textbf{02}, 207 (2022).

\bibitem{Eichten:1974af}
E.~Eichten, K.~Gottfried, T.~Kinoshita, J.~B.~Kogut, K.~D.~Lane and T.~M.~Yan,
Phys. Rev. Lett. \textbf{34}, 369-372 (1975).

\bibitem{Matsui:1986dk}
T.~Matsui and H.~Satz,
Phys. Lett. B \textbf{178}, 416-422 (1986).

\bibitem{Jacobs:1986gv}
S.~Jacobs, M.~G.~Olsson and C.~Suchyta, III,
Phys. Rev. D \textbf{33}, 3338 (1986).



\bibitem{Lafferty:2019jpr}
D.~Lafferty and A.~Rothkopf,
Phys. Rev. D \textbf{101}, no.5, 056010 (2020).




\bibitem{Brambilla:2011sg}
N.~Brambilla, M.~A.~Escobedo, J.~Ghiglieri and A.~Vairo,
JHEP \textbf{12}, 116 (2011).


\bibitem{Srivastava:2018vxp}
P.~K.~Srivastava, O.~S.~K.~Chaturvedi and L.~Thakur,
Eur. Phys. J. C \textbf{78}, no.6, 440 (2018).

\bibitem{Beraudo:2007ky}
A.~Beraudo, J.~P.~Blaizot and C.~Ratti,
Nucl. Phys. A \textbf{806}, 312-338 (2008).

\bibitem{Satz:2005hx}
H.~Satz,
J. Phys. G \textbf{32}, R25 (2006).

\bibitem{Mazumder:2022jjo}
S.~Mazumder, V.~Chandra and S.~K.~Das,
[arXiv:2211.06985 [hep-ph]].

\bibitem{Caristo:2021tbk}
F.~Caristo, M.~Caselle, N.~Magnoli, A.~Nada, M.~Panero and A.~Smecca,
JHEP \textbf{03}, 115 (2022).

\bibitem{Burnier:2016mxc}
Y.~Burnier and A.~Rothkopf,
Phys. Rev. D \textbf{95}, no.5, 054511 (2017).


\bibitem{Xing:2021xwc}
W.~J.~Xing, G.~Y.~Qin and S.~Cao,
Phys. Lett. B \textbf{838}, 137733 (2023).



 \end{thebibliography}
\end{document}